\documentclass[a4paper,11pt]{article}
\usepackage{jheppub}
\usepackage{graphicx}
\usepackage{array}
\usepackage{subcaption}
\usepackage{bm}
\usepackage{hyperref}
\usepackage{verbatim}
\usepackage{amssymb}
\usepackage{amsmath}
\usepackage{xspace}
\usepackage{longtable}
\usepackage[dvipsnames]{xcolor}
\definecolor{darkred}{rgb}{0.55, 0.0, 0.0}
\usepackage{lipsum}
\usepackage{physics}
\usepackage{simpler-wick}
\usepackage{mathtools}
\usepackage{dsfont}
\usepackage{float}
\usepackage{dblfloatfix}
\usepackage[section]{placeins}
\usepackage{enumerate}
\usepackage{tikz}
\usepackage{bbm}
\usepackage{ulem}

\begin{document}

\title{Generalized Free Cumulants for Quantum Chaotic Systems}
\author{Siddharth Jindal}
\author{and Pavan Hosur}
\affiliation{
 Department of Physics and Texas Center for Superconductivity, University of Houston, Houston, TX 77004
}
\emailAdd{siddharth@sidjindal.dev}

\abstract{
The eigenstate thermalization hypothesis (ETH) is the leading conjecture for the emergence of statistical mechanics in generic isolated quantum systems and is formulated in terms of the matrix elements of operators. An analog known as the ergodic bipartition (EB) describes entanglement and locality and is formulated in terms of the components of eigenstates. In this paper, we significantly generalize the EB and unify it with the ETH, extending the EB to study higher correlations and systems out of equilibrium. Our main result is a diagrammatic formalism that computes arbitrary correlations between eigenstates and operators based on a recently uncovered connection between the ETH and free probability theory. We refer to the connected components of our diagrams as generalized free cumulants. We apply our formalism in several ways. First, we focus on chaotic eigenstates and establish the so-called subsystem ETH and the Page curve as consequences of our construction. We also improve known calculations for thermal reduced density matrices and comment on an inherently free probabilistic aspect of the replica approach to entanglement entropy previously noticed in a calculation for the Page curve of an evaporating black hole. Next, we turn to chaotic quantum dynamics and demonstrate the ETH as a sufficient mechanism for thermalization, in general. In particular, we show that reduced density matrices relax to their equilibrium form and that systems obey the Page curve at late times. We also demonstrate that the different phases of entanglement growth are encoded in higher correlations of the EB. Lastly, we examine the chaotic structure of eigenstates and operators together and reveal previously overlooked correlations between them. Crucially, these correlations encode butterfly velocities, a well-known dynamical property of interacting quantum systems.
}
\maketitle
\section{Introduction}
\label{sec:intro}
There has been recent interest in the physics of thermalization in quantum many-body systems. Thermalization was historically established by Boltzmann's ergodic hypothesis, which states that systems uniformly sample the entire phase that is consistent with their macroscopic symmetries~\cite{schwabl_statistical_2006}. Despite its remarkable predictive power, the ergodic hypothesis is manifestly inconsistent with unitarity. A similar issue arises in the study of the black hole information paradox, where general relativity appears inconsistent with unitarity near an event horizon~\cite{hawking_breakdown_1976}. Nonetheless, quantum systems do thermalize~\cite{von_neumann_proof_2010, deutsch_quantum_1991, srednicki_chaos_1994, srednicki_approach_1999}, and understanding how they thermalize has led to a deeper understanding of quantum and statistical mechanics~\cite{popescu_entanglement_2006, linden_quantum_2009}. Since the field of quantum chaos was born partly to resolve the inconsistency between quantum and statistical mechanics, it is fitting that it has returned to prominence in attempts to solve the inconsistency between quantum mechanics and general relativity~\cite{page_information_1993,shenker_black_2014}.

Roughly 30 years ago, Deutsch and Srednicki introduced the eigenstate thermalization hypothesis (ETH) to justify the emergence of statistical mechanics for generic isolated quantum systems~\cite{deutsch_quantum_1991, srednicki_chaos_1994}. Since then, the ETH has been implicated in a variety of physical phenomena including holography~\cite{lashkari_eigenstate_2016}, quantum error correction~\cite{bao_eigenstate_2019}, scrambling~\cite{foini_eigenstate_2019, murthy_bounds_2019}, and transport~\cite{dymarsky_bound_2018}. Furthermore, the failure of the ETH has led to the discovery of a fascinating phase of matter known as many-body localization~\cite{nandkishore_many-body_2015}. Six years ago, motivated by contemporaneous work in the string theory community~\cite{maldacena_bound_2016}, Foini and Kurchan reformulated the ETH to account for higher order correlations~\cite{foini_eigenstate_2019}, and last year, along with Pappalardi, reinterpreted their results in terms of free probability theory~\cite{pappalardi_eigenstate_2022}.

The ETH purports that the emergence of statistical mechanics and thermalization in isolated quantum systems is the result of pseudorandomness in the matrix elements of operators. Formally~\cite{foini_eigenstate_2019},
\begin{align}
\label{eq:ETH}
    &\overline{X_{i_1i_2}\cdots X_{i_ni_1}}=e^{-(n-1)S(\bar{E})}f_n(\bar{E};\Vec{\omega}), \quad i_p\neq i_q\text{ for } p\neq q.
\end{align}
where $X$ is a local operator\footnote{One may consider correlations between $n$ distinct operators, too. For any sequence of operators or ``word", $W$, constructed from individual operators or ``letters", there is a unique associated $f_W$, up to cyclic permutations of the letters. For example, for operators $X,\,Y,\,Z$ and word $YZYX$, we have $\overline{Y_{i_1i_2}Z_{i_2i_3}Y_{i_3i_4}X_{i_4i_1}}=e^{-3S(\overline{E})}f_{YZYX}(\overline{E};\Vec{\omega})$ for distinct $i_1,i_2,i_3,i_4$.}, $\{\ket{i_m}\}$ are eigenstates of a chaotic Hamiltonian, $S$ is the microcanonical entropy, $f$ is an $\mathcal{O}(1)$ spectral function, and the overline denotes arithmetic averaging over a narrow energy band. More than justifying the results of equilibrium statistical mechanics, the ETH ensures that thermalization occurs in real time. Though systems which obey the ETH are considered \textit{chaotic}, it is not clear how the ETH interacts with other principles of quantum chaos. For instance, eq. (\ref{eq:ETH}) does not fundamentally harbor a notion of locality except via an implicit but vague restriction that it only applies to ``simple" operators. It also does not obviously capture the non-local entanglement structure of the Page curve.

A recent, state-based idea that addresses this issue is the ergodic bipartition (EB), an ansatz on the pseudorandom structure of chaotic eigenstates when split over two subsystems. Formally~\cite{deutsch_thermodynamic_2010, lu_renyi_2019, murthy_structure_2019},
\begin{align}
\label{eq:EB}
    &\bra{i}\ket{ab} \equiv c^{i}_{ab}, \quad \overline{\left|c^{i}_{ab}\right|^2} = e^{-S(E_i)}F(E_i-E_a-E_b)
\end{align}
where $\ket{i}$ is an eigenstate of the full Hamiltonian $H = H_A + H_B + H_{AB}$, $\ket{ab}$ is a product of eigenstates of subsystem Hamiltonians $H_A$ and $H_B$, and $H_{AB}$ is the interacting term that couples subsystems $A$ and $B$. $e^{S(E)}$ is the density of states of the system at energy $E$ and $F$ is a window function that ensures $E_i\approx E_a+E_b$. The moments of $F$ are roughly those of $H_{AB}$, $\int_\omega F(\omega)\omega^n \approx \expval{H_{AB}^n}_i$ (see appendix~\ref{subapp:normForm}). Therefore, if $H_{AB}$ lives on the boundary between $A$ and $B$, it may be regarded as a subextensive perturbation and $F$ is sharply peaked in its argument. The EB implies that the reduced density matrices of eigenstates are consistent with those of the microcanonical ensemble and obey a Page curve. Additionally, the tensor product structure allows it to distinguish local systems from nonlocal ones. Structurally, the EB is formulated in analogy with the ETH, but presently is only capable of computing static quantities for systems prepared in eigenstates.

Both the ETH and the EB are avatars of Berry's conjecture~\cite{berry_regular_1977}: the hypothesis that most eigenfunctions of chaotic potentials are, in essence, Gaussian random waves. Berry's conjecture is predated by Von Neumann's quantum ergodic theorem~\cite{von_neumann_proof_2010}, which states that the overwhelming majority of pure states in a many-body system approximate the local properties of microcanonical ensemble arbitrarily well. It is then natural to conjecture that the eigenstates of many-body systems also retain the properties of the microcanonical ensemble, if they can be treated as random waves. In this light, Berry's conjecture would be a manifestly quantum version of Boltzmann's ergodic hypothesis where uniform sampling of phase space has been supplanted by random vectors. 

In systems with few degrees of freedom, not all eigenstates can be treated as random~\cite{berry_regular_1977}. However, as one takes the thermodynamic limit, $\mathcal{V}\rightarrow\infty$, provided it is well-defined, two things happen simultaneously: (1) the level spacing of the system vanishes exponentially fast $\sim\mathcal{O}(e^{-S})$ and (2) the local physics of the system become insensitive to microscopic perturbations\footnote{This cannot strictly be true for eigenstates sufficiently close to the ground state of the system as both conditions are violated. An arbitrarily small relevant perturbation can modify IR physics and the density states is not exponentially large.} $\ll \mathcal O(1)$~\cite{deutsch_eigenstate_2018}. As such, any microscopic perturbation $\ll \mathcal O(1)$ added to the system can mix an exponentially large number of eigenstates $\sim\mathcal{O}(e^{S})$ without modifying any physics. Thus, if the thermodynamic limit is to be well-defined, \textit{all} pure states in a microscopically small energy window should have identical physical properties~\cite{dymarsky_bound_2018, dymarsky_new_2019}\footnote{In a system with multiple conservation laws, or in a symmetry broken phase~\cite{fratus_eigenstate_2015}, mixing states with distinct, non-energy, quantum numbers would generically require a perturbation that is $\sim\mathcal{O}(1)$ and capable of modifying some local physics. In  \textit{integrable} systems, which contain an extensive number of conservation laws, the number of states in the microscopic window that can be mixed by an integrability-preserving perturbation is not exponentially large, and the ETH cannot hold in the strong form we present~\cite{deutsch_eigenstate_2018}.}. This line of reasoning underlies early work on chaos in many-body systems~\cite{feingold_distribution_1986, deutsch_quantum_1991} and, more recently, investigations into \textit{emergent rotational symmetry} at small frequencies, that is, an invariance of physics to arbitrary norm-preserving\footnote{i.e. orthogonal, unitary, or symplectic} linear transformations of eigenstates that are sufficiently close in energy~\cite{foini_eigenstate_2019-1,wang_emergence_2023}. From it, we can conjecture that eigenstates of thermodynamically large systems retain the properties of random vectors and of the microcanonical ensemble.

In this paper, we formulate a many-body Berry's conjecture (MBBC): the hypothesis that chaotic eigenstates are essentially random vectors up to the symmetry constraints of the system, and show that it unifies the ETH and the EB. To our knowledge, this term was first used in ref.~\cite{lu_renyi_2019}, which studied entanglement entropies of chaotic eigenstates. We show that free probability theory naturally emerges from the MBBC and build on this insight to develop a diagrammatic formalism for calculating general correlations of eigenstates and operators, greatly generalizing the EB, in particular. With this formalism we are able to justify or improve several known results for chaotic eigenstates on a common footing. We show that chaotic eigenstates have thermal reduced density matrices consistent with the subsystem ETH conjecture and Von Neumann entanglement entropies consistent with the Page curve. We then show that both these properties are shared by non-equilibrium states at late times, indicating that chaotic systems thermalize in real time. We additionally study signatures of local dynamics: ballistic and diffusive entanglement growth and butterfly velocities.

\begin{figure}
    \centering
    \includegraphics[width = 0.75\linewidth, viewport=0 70 800 530, clip]{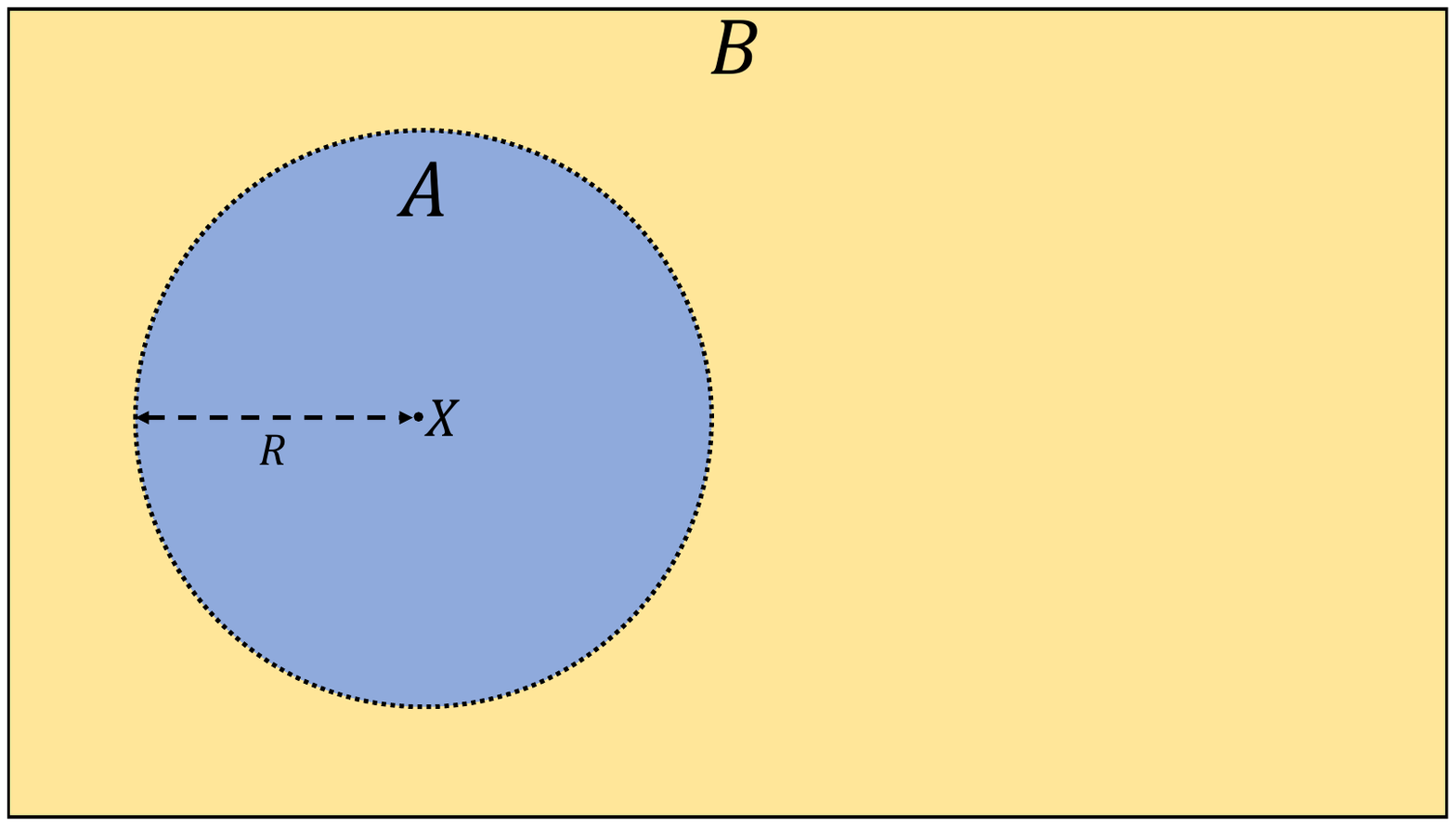}
    \caption{The system is split into two subsystems $A$ and $B$. An operator $X$ lives within $A$, a distance $R$ from the boundary.}
    \label{fig:system}
\end{figure}

In our setup, we consider a system split into two subsystems $A$ and $B$ and a local operator $X$ living deep within $A$, see figure \ref{fig:system}. The goal of doing so is to study the reduced density matrix on a single subsystem and the bipartite entanglement between the subsystems. However, our techniques should extend to any kind of partitioning of the system. 

The outline of this paper is as follows. In section \ref{sec:motivation}, we motivate our approach by studying the properties of random vectors without appealing to formal results of Haar measures or of free probability theory. The main result in this section, which illustrates the basic principle behind the MBBC, is that random vectors, and therefore chaotic eigenstates, have the same properties as equilibrium density matrices. 

Section \ref{sec:diagrams} is the technical backbone of the paper and contains our main results. It builds on section \ref{sec:motivation} to establish the framework of free probability theory and its connection to the EB and the ETH. Here, we introduce generalized free cumulants and describe the diagrammatic framework for computing arbitrary correlations between states and operators. This formalism is the main result of our paper. We compare our technique to analogous techniques in the literature and provide some numerical evidence for its validity. Furthermore, we connect the predictions of the MBBC to the predictions of rotationally invariant random matrix ensembles. We also discuss how the emergence of free probability in the ETH has unique consequences for the behavior of local observables over long times, connecting our observations to recent results in the literature. We argue that the ETH, eq. \eqref{eq:ETH}, the \textit{quantum butterfly effect} characterized by the decay of \textit{out-of-time-ordered correlators} (OTOCs)~\cite{maldacena_bound_2016, hosur_chaos_2016}, and the emergence of rotational symmetry at small frequencies~\cite{foini_eigenstate_2019-1, wang_emergence_2023} are all, in essence, equivalent definitions of chaos and avatars of free probability (we elaborate on these ideas in appendix~\ref{app:FPT}). To summarize, this section contains the mathematical and diagrammatic framework of free cumulants and a glimpse of implications that we explore in depth in the next two sections. 

In section \ref{sec:equilibrium}, we study the structure of chaotic eigenstates. Several previous works have been dedicated to this topic~\cite{deutsch_thermodynamic_2010, lu_renyi_2019, murthy_structure_2019, shi_local_2023}, but our discussion is more than a review. We provide new derivations that show disparate results as consequences of the MBBC. First, in section \ref{subsec:RDMEq}, we show that the reduced density matrix of chaotic eigenstates takes the form,
\begin{align}
\label{eq:RDMEqIntro}
    \rho^{i}_{aa'}\equiv \bra{a}\!\operatorname{Tr}_B\left[\ket{i}\!\!\bra{i}\right]\!\ket{a'} = e^{-S(E_i)+S_B(E_i-E_{aa'})}\left(\delta_{aa'}+\sqrt{e^{S_{\text{min}}(E_{aa'})}\tilde F(E_{aa'};\omega_{aa'})}R_{aa'}\right)
\end{align}
where $E_{aa'}\equiv \left(E_a+E_{a'}\right)/2$, $S_{\text{min}} \equiv \operatorname{min}\left[S_A(E_{aa'}),S_B(E_i-E_{aa'})\right]$, $\tilde F(E_{aa'},\omega_{aa'})$ contains a $\omega_{aa'} \equiv E_{a} - E_{a'}$ dependence narrowly peaked around zero, and $R_{aa'}$ is a nearly Gaussian random matrix that encodes higher correlations in the density matrix. Eq. \eqref{eq:RDMEqIntro} improves the result of \cite{murthy_structure_2019} which contained an unphysical suppression of off-diagonal elements and generalizes the \textit{state-averaging ansatz} of ref.~\cite{de_boer_principle_2023}. Next, we study the entanglement entropy of chaotic systems and show that we are able to reproduce the results of ref. \cite{murthy_structure_2019} exactly. In particular, we show that the Von Neumann entropy obeys a Page curve,
\begin{align}
\label{eq:PageEqIntro}
    S_1=\operatorname{min}\left[S_A(E_{iA}),S_B(E_{iB})\right]+\Delta S
\end{align}
where $E_{iA}$ and $E_{iB}$ are the microcanonically expected subsystem energies defined by $E_{iA}+E_{iB} = E_i$ and $S_A^\prime(E_{iA}) = S_B^\prime(E_{iB})$ and $\Delta S$ denotes subextensive corrections to the Page curve. Furthermore, we discuss an inherently free probabilistic structure in the replica calculation of entanglement entropy previously noticed in the gravitational path integral calculation of the Page curve for an evaporating black hole~\cite{liu_entanglement_2021,penington_replica_2022, wang_beyond_2023}. Finally, in section \ref{subsec:SETH}, we show that reduced density matrices of nearby eigenstates are exponentially close in trace distance, a hypothesis known as the subsystem ETH~\cite{dymarsky_subsystem_2018}. More precisely, we show,
\begin{align}
\label{eq:SubETHIntro}
    \mathcal{O}\left(e^{S_A/2-S/2}\right)\lesssim ||\rho^{i}_A-\rho^j_A||_1\lesssim \mathcal{O}\left(e^{S_{\text{min}}^{(\infty)}/2+S_A/2-S/2}\right)
\end{align}
where $||\cdots||_1$ is the Schatten 1-norm\footnote{For an operator $M$, the Schatten $p$-norm is defined as $||M||_p\equiv \sqrt[\leftroot{-2}\uproot{2}p]{\sum_i \abs{m_i}^p}$ where $m_i$ are the eigenvalues or singular values of $M$.} and $S_{\text{min}}^{(\infty)}$ is the microcanonical entropy of the smaller subsystem at infinite temperature\footnote{A stronger bound was derived in ref.~\cite{kudler-flam_distinguishing_2021} using similar methods.}.  

In section \ref{sec:thermalization}, we then study a system prepared out-of-equilibrium with no entanglement between the subsystems. We first verify that all states with the same initial energy, $E_0$, relax towards that same equilibrium at late times, which shares its properties with nearby chaotic eigenstates\footnote{This statement holds for all properties except for fluctuations of approximately conserved operators, namely the fluctuations of subsystem energies. We discuss this caveat in appendix~\ref{subapp:subvariances} and thank Tarun Grover for bringing it to our attention.}. Our approach reproduces the \textit{equilibrated pure state} formalism of ref.~\cite{liu_entanglement_2021}. Precisely, we show in section \ref{subsec:RDM} that
\begin{align}
\label{eq:RDMNeqIntro}
    \rho_{aa'}(t\rightarrow\infty) = e^{-S(E_0)+S_B(E_0-E_{aa'})}\left(\delta_{aa'}+\sqrt{e^{S_{\text{min}}(E_{aa'})}\tilde F(E_{aa'};\omega_{aa'})}R_{aa'}\right)
\end{align}
and in section \ref{subsec:Page} that
\begin{align}
\label{eq:PageNeqIntro}
    S_1(t\rightarrow\infty)=\operatorname{min}\left[S_A(E_{0A}),S_B(E_{0B})\right]+\Delta S
\end{align}
where $E_{0A},\, E_{0B}$ are the microcanonical subsystem energies for total energy $E_0$. The key insight behind this result is that the time-independent partitions of the diagrams for time-dependent states map uniquely to those of eigenstates. In section \ref{subsec:entanglementVelocities}, we study the different phases of entanglement growth~\cite{mezei_entanglement_2017, rakovszky_sub-ballistic_2019, huang_dynamics_2020} and find that distinct phases are encoded in distinct diagrams. Our results provide a novel organization to chaotic entanglement dynamics.

In section \ref{sec:locality}, we consider the correlations between operators and eigenstates. When an operator $X$ lives deep within subsystem $A$, it may be natural to assume it becomes uncorrelated with $c$ defined in eq. (\ref{eq:EB}). In section \ref{subsec:corr}, we show that this assumption strongly violates causality and that operator-state correlations are necessary for the ETH to apply in a system with many degrees of freedom. In essence, decorrelating the elements of $X$ with those of $c$ disconnects $X$ from the time evolution operator $e^{iHt}$ and trivializes the dynamics of the system. Thus, in section \ref{subsec:butterfly}, we exploit causality to constrain operator-state correlations. For $X$ living deep within subsystem $A$, time evolution under $H_A + H_B$ will be identical to that under $H$ for times shorter than $R/v_B$, where $v_B$ is the butterfly velocity of the system~\cite{roberts_lieb-robinson_2016} and $R$ is the distance to the boundary. Thus, local physics is necessarily encoded in eigenstate-operator correlations.

Some calculations and discussions in this paper are left to the appendices. Appendix~\ref{app:FPT} contains an informal introduction to the concepts of free probability theory we use in this paper and motivate why the subject plays a central role in quantum chaos. Appendix~\ref{app:saddlePoints} provides an introduction to the use of saddle-point approximations for making thermal approximations with the ETH. Appendix~\ref{app:width} discuss the non-zero width of the $F$ function that we mostly neglect in the body of this paper. Appendix~\ref{subapp:subvariances} focuses on corrections to the energy fluctuations of subsystems that are a finite fraction of the whole system, which are able to distinguish different pure states with the same energy density, even in equilibrium. Appendix~\ref{subapp:normForm} discusses the form of $F$ in eq.~\eqref{eq:EB} in the context of previously conjectured Gaussian  and Lorentzian forms for $F$~\cite{murthy_structure_2019, shi_local_2023}. Our analysis complements appendix A of ref.~\cite{shi_local_2023}. Appendix~\ref{app:OFSR} derives eq.~\eqref{eq:ETH} from the MBBC.

\section{Many-body Berry's conjecture}
\label{sec:motivation}

Our fundamental postulate is that eigenstates of a chaotic Hamiltonian behave \textit{for all intents and purposes} as random vectors up to the symmetry constraints of the system. This idea is what we refer to as the many-body Berry's conjecture and attempt to make precise in this section. We will assume conservation of energy and no other symmetries going forward. The purpose of this section is to develop the properties of random vectors while sidestepping a formal discussion of Haar measures or of free probability.

First, consider a $d$-dimensional Hilbert space and sample two normalized vectors randomly, $\ket{1}, \ket{2}$ from it. What is the expected value of their squared overlap, $\bra{1}\ket{2}\!\!\bra{2}\ket{1}$? An easy way to calculate this overlap is to rotate into an orthonormal basis, $\{\ket{1'}\}$, that contains $\ket{1}$. Then, $\ket{2}$, being chosen independently of $\ket{1}$, on average will have overlap $1/\sqrt{d}$ with each basis element. Thus,
\begin{align}
    &\overline{\bra{1}\ket{2}\!\!\bra{2}\ket{1}} = \frac{1}{d}
\end{align}
We may sample three vectors, $\ket{1}$, $\ket{2}$ and $\ket{3}$, and want to know the expected value of $\bra{1}\ket{2}\!\!\bra{2}\ket{3}\!\!\bra{3}\ket{1}$. One may be forgiven for guessing that each overlap contributes a factor of $d^{-\frac{1}{2}}$ yielding $d^{-\frac{3}{2}}$, but that is incorrect. The \textit{fluctuating} part of a single amplitude is indeed $d^{-\frac{3}{2}}$, but the noncommutativity of the projection operators $\ket{i}\!\!\bra{i}$ ensures that the 3 overlaps are correlated. Instead, consider again summing over the entire basis that contains $\ket{1}$. This procedure corresponds to 
\begin{align}
    &\sum_{1'}\bra{1'}\ket{2}\!\!\bra{2}\ket{3}\!\!\bra{3}\ket{1'} = \operatorname{Tr}\left[\ket{2}\!\!\bra{2}\ket{3}\!\!\bra{3}\right] = \bra{3}\ket{2}\!\!\bra{2}\ket{3}.
\end{align}
Linearity of averages implies 
\begin{align}
    &\operatorname{Tr}\left[\overline{\ket{2}\!\!\bra{2}\ket{3}\!\!\bra{3}}\right] = \overline{\bra{3}\ket{2}\!\!\bra{2}\ket{3}}.
\end{align}
Lastly, since the basis $\{\ket{1'}\}$ is essentially arbitrary, 
\begin{align}
    &\overline{\bra{1}\ket{2}\!\!\bra{2}\ket{3}\!\!\bra{3}\ket{1}} = \frac{1}{d}\sum_{1'}\bra{1'}\overline{\ket{2}\!\!\bra{2}\ket{3}\!\!\bra{3}}\ket{1'} = \frac{1}{d}\operatorname{Tr}\left[\overline{\ket{2}\!\!\bra{2}\ket{3}\!\!\bra{3}}\right] = \frac{1}{d}\overline{\bra{3}\ket{2}\!\!\bra{2}\ket{3}} = d^{-2}.
\end{align}
Following this inductive argument, we can now assert:
\begin{equation}
\label{eq:HaarLoop}
    \overline{\bra{1}\ket{2}\!\!\bra{2}\cdots\ket{n}\!\!\bra{n}\ket{1}}=d^{-(n-1)}.
\end{equation}
 Eq. (\ref{eq:HaarLoop}) applies to vectors sampled uniformly over their Hilbert space. Note that the mean $d^{-(n-1)}$ is smaller than the fluctuating part $d^{-n/2}$ for $n>2$.

We now want to understand how eigenstates of chaotic Hamiltonians can be understood as random vectors. First consider a pair of Hamiltonians: $H_1$ and $H_2 = H_1 + \lambda X$, where $X$ is some subextensive perturbation. Physical arguments~\cite{feingold_distribution_1986, deutsch_quantum_1991} and numerical evidence~\cite{rigol_quantum_2010} indicate that even for a very small\footnote{The precise expectation is that for $\lambda X\gtrsim \mathcal{O}(\mathcal V^{-\gamma})$, eigenstates of $H_2$ and $H_1$ within a small energy window are related essentially by a random rotation and cannot be computed from one another to any order in perturbation theory where $\mathcal{V}$ is the volume of the system, and $\gamma$ is an exponent that depends on the transport properties of the system~\cite{georgeot_quantum_2000, modak_finite_2014, dymarsky_bound_2018}.} $\lambda$ the eigenstates of $H_2$ are nearly orthogonal to those of $H_1$. Specifically, we expect that for eigenstates $\ket{i_1}$ of $H_1$ and $\ket{i_2}$ of $H_2$ with $E_{i_1}\approx E_{i_2}$,
\begin{align}
    \overline{\bra{i_1}\ket{i_2}\!\!\bra{i_2}\ket{i_1}} \sim e^{-S}.
\end{align}
Furthermore, even for $\lambda\sim\mathcal{O}(1)$, $\lambda X$ cannot mix eigenstates of $H_1$ that are very distant in energy. Our precise requirement is that $\lambda X$ is small enough that $H_1$ and $H_2$ retain the same entropy functions, ${S_2(E)}\approx {S_1(E)} \equiv {S(E)}$, which is expected to hold for any subextensive perturbation~\cite{shi_local_2023}. Lastly, $X$ may have some nontrivial energy dependence. Thus we hypothesize,
\begin{align}
    \overline{\bra{i_1}\ket{i_2}\!\!\bra{i_2}\ket{i_1}} = e^{-S(\bar E)}F(\bar E;\omega).
\end{align}
where $\overline{E}=\frac{1}{2}\left(E_{i_1}+E_{i_2}\right)$, $\Vec{\omega}=E_{i_1}-E_{i_2}$, and $F$ serves as a cutoff function that encodes the nontrivial energy dependences. Analogously, considering $n$ distinct Hamiltonians $H_1,\dots,\,H_n$ and an overlap of $n$ respective eigenstates $\ket{i_1},\dots,\,\ket{i_n}$, we hypothesize,
\begin{align}
\label{eq:stateCumulant}
    &\overline{\bra{i_1}\ket{i_2}\!\!\bra{i_2}\cdots\ket{i_n}\!\!\bra{i_n}\ket{i_1}}=e^{-(n-1)S(\bar{E})}F(\bar{E};\Vec{\omega})
\end{align}
where $\bar{E}=\frac{1}{n}\sum_mE_m$ and $\Vec{\omega}=(E_{i_1}-E_{i_2}, \cdots, E_{i_{n-1}}-E_{i_{n}})$. The crux of the MBBC is that the structure of eq.~\eqref{eq:stateCumulant} holds for the eigenstates of any set of chaotic Hamiltonians that are neither extremely close to one another nor extremely far from one another. In section~\ref{subsec:pH} we consider the situation where non-adjacent eigenstates in eq.~\eqref{eq:stateCumulant} are taken from the same Hamiltonian. 

To reiterate, we can understand eigenstates of physical Hamiltonians as random vectors by incorporating two constraints: mutual orthogonality and fixed energy. To handle the first constraint, we asserted that consecutive projection operators in eq.~\eqref{eq:stateCumulant} are eigenstates of distinct, but similar, Hamiltonians. For the second constraint, the Hilbert space dimension, $d$ was replaced with the density of states, $e^S$. Lastly, we inserted a cutoff function, $F(\bar{E};\Vec{\omega})$, that ensures correlations are local in energy.

It is useful to formulate the above argument in terms of the density matrix of the space from which the eigenstate was sampled. Consider, again, the overlap of $n$ random vectors sampled from the entire Hilbert space. The corresponding density matrix is simply the infinite temperature state, $\frac{1}{d}\mathbbm{1}$. Our result for random vectors can be stated as
\begin{align}
    &\overline{\bra{1}\ket{2}\!\!\bra{2}\cdots\ket{n}\!\!\bra{n}\ket{1}}=\operatorname{Tr}\left[\left(\frac{1}{d}\mathbbm{1}\right)^n\right]=d\cdot d^{-n}=d^{-(n-1)}
\end{align}
while for eigenstates,
\begin{align}
    &\overline{\bra{i_1}\ket{i_2}\!\!\bra{i_2}\cdots\ket{i_n}\!\!\bra{i_n}\ket{i_1}}\sim \operatorname{Tr}[\rho(\bar{E})^n] \sim e^{-(n-1)S(\bar{E})}
\end{align}  
where $\rho(\bar{E})$ is the microcanonical density matrix with energy $\bar{E}$. Hence, narrow band averaging imbues eigenstates with the properties of corresponding equilibrium ensembles.

We can also consider subsystems. Let the Hilbert space be a tensor product of subspaces $A$ and $B$ with dimensions $d_A$ and $d_B = d/d_A$, respectively. We want to know how to compute the expected value of an arbitrary overlap between vectors sampled on the subspaces and the full space. If the vectors are sampled uniformly, we will again assume they can be replaced with the corresponding density matrix. For example, consider two vectors from each space $\ket{1}$, $\ket{2}$, $\ket{1_A}$, $\ket{2_A}$, $\ket{1_B}$, $\ket{2_B}$ and the following overlap\footnote{An analogous calculation was performed in Appendix B of ref.~\cite{dymarsky_subsystem_2018}.},
\begin{align}
\label{eq:nontrivial}
    \overline{\bra{1}\ket{1_A1_B}\!\!\bra{1_A2_B}\ket{2}\!\!\bra{2}\ket{2_A2_B}\!\!\bra{2_A1_B}\ket{1}} &= d^{-2}\operatorname{Tr}\left[\overline{\mathbbm{1}\ket{1_A1_B}\!\!\bra{1_A2_B}\mathbbm{1}\ket{2_A2_B}\!\!\bra{2_A1_B}}\right] \nonumber \\ &= d^{-2}\overline{\bra{2_A}\ket{1_A}\!\!\bra{1_A}\ket{2_A}} \nonumber \\ &= d^{-2}d_A^{-1}.
\end{align}
An analogous overlap of eigenstates is expected to have weight $e^{-2S-S_A}$. We clarify this generalization in section \ref{subsec:diagrams2} and provide numerical evidence for it in section \ref{subsec:numerics}. 

Remarkably, eq. \eqref{eq:stateCumulant} has the same form as eq. \eqref{eq:ETH}. The fact that both changes-of-basis and local observables obey the same correlated structure may seem surprising and raise the question of precisely which objects are amenable to such an analysis. The answer is that eigenstates are the natural object of study of the ETH, where local operators and change-of-basis rotations serve equally well as scramblers (see appendix \ref{app:OFSR} for details). We claim that one can mix any combination of chaotic eigenstates, density matrices, and observables and retain such an expression. The remaining task of the section is to relate this observation to thermalization. 

We study an $n$-point \textit{cumulant} of an operator defined as the time-dependent sum over a product of operators that neglects repeated indices:
\begin{align}
\label{eq:connected}
    &\sum_{\left[i_2\cdots i_n\right]} e^{i\Vec{\omega}\cdot\Vec{t}}X_{i_1i_2}\cdots X_{i_ni_1}, \quad \Vec{t} \equiv \left(t_1,\dots,t_{n-1}\right), \quad \Vec{\omega} \equiv \left(\omega_{i_1i_2},\dots,\omega_{i_{n-1}i_n}\right)
\end{align}
where the braces $[\cdots]$ indicate that indices are not to be repeated in summation (i.e. $i_m\neq i_{m'}$). We will replace sums over sufficiently smooth correlations with integrals via the substitution
\begin{align}
    \sum_{i}\longrightarrow\int_{E_i}e^{S(E_i)}.
\end{align}
Then we replace the right-hand-side of eq. \ref{eq:connected} with an integral as
\begin{align}
\label{eq:crux}
    \sum_{\left[i_2\cdots i_n\right]} e^{i\Vec{\omega}\cdot\Vec{t}}X_{i_1i_2}\cdots X_{i_ni_1} &= \sum _{[i_2\cdots i_n]}e^{i\Vec\omega\cdot\Vec t - (n-1)S(\bar E)}f(\bar E; \Vec\omega) \nonumber \\
    &=\int_{E_{i_2}\cdots E_{i_{n}}}e^{i\Vec\omega\cdot\Vec t +\sum_{m=2}^n (S(E_{i_m})-S(\bar E))}f(\bar E; \Vec\omega) \nonumber \\
    &= \int_{\Vec\omega}e^{(i\Vec{t}-\beta\Vec{l})\cdot\Vec\omega}f(E_{i_1};\Vec\omega), \quad \Vec{l} = \left(\frac{n-1}{n},\dots,\frac{1}{n}\right) \nonumber \\
    &\equiv f(E_{i_1};\Vec t + i\beta \Vec l\,\,)
\end{align}
where $\beta\equiv S'(E_{i_1})$ is the thermodynamic temperature. To go from the second to the third line in eq. \eqref{eq:crux}, we have utilized the approximation $S(E_{i_m})-S(E_{i_{m+1}}) \approx \beta\omega_{i_mi_{m+1}}$, which holds so long as the heat capacity of the system is extensive (see appendix \ref{app:saddlePoints}) and we have approximated $f(\bar E;\Vec{\omega}) = f(E_{i_1} -\Vec{l}\cdot\Vec{\omega};\Vec{\omega})\approx f(E_{i_1};\Vec{\omega})$ since $f$ is a slow function of the total energy. 

It is clear that the time-dependent cumulant is simply a mixed Fourier--Laplace transform of $f$. For the case of $n=1$, the right-hand side of eq. \eqref{eq:crux} simply reduces to the time-independent microcanonical expectation value of $X$. For $n>1$, we instead have a complete set of thermal correlation functions which must decay in order for the system to equilibrate. These cumulants are more precisely operator \textit{free cumulants}, which we discuss in more detail in section \ref{subsec:cactus}. 

One should wonder when replacing sums with integrals over smooth components, as we did in eq. (\ref{eq:crux}), is safe. For example, the fluctuating component of $X_{i_1i_2}\cdots X_{i_ni_1}$ is larger than its smooth component, $e^{-nS/2}\nless e^{-(n-1)S}$. The answer is that it depends on the number of sums being performed. Let the number of sums being performed be $n^*$. While the smooth parts will grow with $e^{n^*S}$, the fluctuating parts are uncorrelated and will only grow with $e^{n^*S/2}$. Thus, replacing the sums with integrals is valid when
\begin{align}
\label{eq:comparison}
    e^{-nS/2+n^*S/2} < e^{-(n-1)S+n^*S} \implies n^*>n-2
\end{align}
In eq. \eqref{eq:crux}, we had $n^*=n-1$ and were safe by one factor of $e^S$. If $n^*=n-2$, then the fluctuating and smooth parts will have the same magnitude. If $n^*<n-2$, then the smooth part will be washed away. In general, one may consider more complex situations with different entropy factors that will need to be compared. 

Another difficulty with converting sums to integrals is the presence of \textit{level repulsion}\footnote{the suppressed likelihood of finding two eigenstates much closer in energy than the average level spacing in a chaotic system}. In chaotic systems, the eigenvalues of the Hamiltonian should be considered correlated (pseudo-)random variables~\cite{dyson_statistical_2004}. Thus when converting a sum over multiple indices to a multi-variable integral, in principle, one would have to introduce a joint density of states, e.g.,
\begin{align}
    \sum_{ij}\longrightarrow\int_{E_iE_j}e^{S^{(2)}(E_i,E_j)}
\end{align}
for a pair of energy indices~\cite{lal_mehta_random_2004}. However, to leading order in $e^{-S}$, the joint density of states factorizes $e^{S^{(2)}(E_i,E_j)} \approx e^{S(E_i)+S(E_j)}$. Furthermore, level spacing corrections only become relevant at unphysical late times order of the Heisenberg timescale $\tau_{\text{H}}\sim e^{S}$~\cite{srednicki_approach_1999}. In principle, level spacing corrections may be computed if one has e.g. a \textit{matrix model}\footnote{A matrix model is an integral over matrices (e.g. a probability distribution). For example, the matrix model of ref.~\cite{jafferis_matrix_2023} is a distribution over all Hamiltonians consistent with a given set of correlators. Within the validity of saddle-point integration, (single trace) correlators computed from a single Hamiltonian sampled from a matrix model are identical to those averaged over that matrix model to all orders in $e^{-S}$. The ETH holds for a system if there is a matrix model in the sense of ref.~\cite{jafferis_matrix_2023} that generalizes its Hamiltonian.}~\cite{jafferis_matrix_2023} for a system but a matrix model is not generally accessible in practice. More importantly, though we consider higher-order corrections in $e^{-S}$, we largely neglect the timescales necessary to observe level repulsion effects and will presume that joint densities of states factorize.

By the convolution theorem, the decay of time-dependent free cumulants, $f(\Vec{t})$, is equivalent to the fine resolution behavior of their spectral free cumulants, $f(\Vec{\omega})$. If the spectral free cumulant becomes smooth at over an energy scale $\varepsilon$, the associated time-dependent free cumulant will vanish to zero over a timescale $\varepsilon^{-1}$. In general, correlation functions of local operators are expected to have nontrivial behavior over short timescales that are associated with dissipation and scrambling and become trivial over longer timescales that are associated with transport and hydrodynamics~\cite{dymarsky_bound_2018}. Beyond those timescales, correlation functions are expected to vanish until the Heisenberg time when the level spacing of the system becomes relevant. Accordingly, a given spectral free cumulant should appear entirely smooth over energy scales that are smaller than those associated with hydrodynamics but are much larger than the level spacing of a system. 

\section{Generalized free cumulants}
\label{sec:diagrams}

Recently, it was shown that the structure of the ETH, eq. (\ref{eq:ETH}), is intimately related to free probability theory~\cite{pappalardi_eigenstate_2022}, which describes the statistics of highly non-commutative random variables~\cite{mingo_free_2017}. For correlated operators, eq. (\ref{eq:ETH}), turn out to be their \textit{free cumulants}, analogs of the classical cumulants for maximally non-commutative variables. When classical (commutative) random variables are sampled independently, their classical cumulants\footnote{A cumulant of degree $n$ is, in general, some polynomial in the moments of degree $m \leq n$. Both free and classical cumulants can be defined by their respective moment-cumulant relationship.} vanish. Analogously, when large $N\times N$ random matrices are sampled independently, their free cumulants vanish to leading order in $1/N$. In the context of the ETH, we think of physics as being local in energy space so the $1/N$ expansion is replaced by an $e^{-S}$ expansion and free probability becomes a good model in the thermodynamic limit.  We discuss free probability theory and its role in quantum chaos in more detail in appendix~\ref{app:FPT}.

While ref.~\cite{pappalardi_eigenstate_2022} clarifies the role of free cumulants for operators, an analogous structure is missing for states, even though the analysis of section \ref{sec:motivation} indicates that factors of $e^{-S}$ show up for correlations of states in the same manner as they do for operators. In this section, we show that correlations between states are described by \textit{generalized free cumulants} (GFCs) of which the EB, eq. (\ref{eq:EB}), is a simple example. Our generalizations account for the partitioning of our system (figure \ref{fig:system}) and correlations between eigenstates from distinct Hamiltonians. We introduce a diagrammatic representation of GFCs that will aid the investigation of state-based physics, such as evolution towards an equilibrium density matrix and the behavior of entanglement entropy without requiring a firm knowledge of free probability. In general, we are able to compute to arbitrary order in an $e^{-S}$ expansion. This formalism is the main result of the paper. 

To study ordinary free cumulants, refs.~\cite{foini_eigenstate_2019, pappalardi_eigenstate_2022, pappalardi_general_2023} utilize a similar diagrammatic formalism known as \textit{cactus diagrams} while standard references on free probability utilize so-called \textit{non-crossing partitions}~\cite{speicher_free_1997} and ref.~\cite{jafferis_matrix_2023} utilizes \textit{'t Hooft diagrams}. Existing diagrammatic approaches have substantial merit and we may have been able to modify them instead of introducing our own diagrams. However, we feel the diagrams we introduce will be more natural for our purposes. 

\subsection{Eigenstate correlations: perturbed Hamiltonian}
\label{subsec:pH}
To ease the introduction of GFCs, we first consider a Hamiltonian $H_0$ and a perturbed Hamiltonian $H$ without necessarily bipartitioning the system. Later, we will take $H_0$ to be $H_A+H_B$ and $H$ to be $H_A+H_B + H_{AB}$. We wish to compute, as an example, the diagonal elements with respect to $H_0$ of a system prepared in the canonical ensemble of $H_0$ but evolved under $H$. Let us again define the symbol $c^{\mu}_{\nu}\equiv \bra{\mu}\ket{\nu}$ for the change-of-basis tensor, $\ket{I}, \ket{J}$ as eigenstates of the initial Hamiltonian, and $\ket{i}, \ket{j}$ as eigenstates of the perturbed Hamiltonian. Then our expression is,
\begin{align}
\label{eq:example}
    &\rho_{II}(t) = \frac{1}{Z_\beta}\sum_{Jij}\bra{I}\ket{i}\!\!\bra{i}\ket{J}\!\!\bra{J}\ket{j}\!\!\bra{j}\ket{I}e^{-\beta E_{J}-i\omega_{ij}t}\equiv \frac{1}{Z_\beta}\sum_{Jij}c^I_ic^i_{J}c^{J}_{j}c^{j}_{I}e^{-\beta E_{J}-i\omega_{ij}t}.
\end{align}
One should expect that the typical magnitude of an element $\rho_{II}$ is of order $\mathcal{O}(e^{-S})$. The right-hand expression in eq. (\ref{eq:example}) appears to sum $e^{3S}$ terms while $Z_\beta$ is the canonical partition function and contributes a factor of $e^{-S}$, so the summand should have average weight $e^{-3S}$. We need the overall expression to have magntidue $e^{-S}$. If we compare the summand to eq. \eqref{eq:stateCumulant}, we see that a cyclic product of 4 overlaps has the necessary remaining factor of $e^{-3S}$.

However, we have neglected some terms. If $I=J$, the dominant contribution to the sum will be a product of 2 2-cycles, $\left(c^{I}_ic^{i}_I\right)\left(c^{I}_jc^j_I\right)$ and the summand will have magnitude $\left(e^{-(2-1)S}\right)^2=e^{-2S}$. But we have also dropped a sum and are now only summing over $e^{2S}$ terms, so our overall magnitude is the same. The same argument applies when $i=j$. We may also have $I=J$ and $i=j$ simultaneously. Indeed, there are 4 distinct index partitions to consider: 
\begin{align}
\label{eq:parts}
     Z_\beta\rho_{II}(t) &= \sum_{[Jij]}c^I_ic^i_{J}c^{J}_{j}c^{j}_{I}e^{-\beta E_{J}-i\omega_{ij}t} 
     +\sum_{[ij]}c^I_ic^i_Ic^{I}_{j}c^{j}_{I}e^{-\beta E_{I}-i\omega_{ij}t} \nonumber\\
     &\quad\quad+\sum_{[Ji]}c^I_ic^i_{J}c^{J}_{i}c^{i}_{I}e^{-\beta E_{J}} 
     +\sum_{[i]}c^I_ic^i_{I}c^{I}_{i}c^{i}_{I}e^{-\beta E_{I}}  \nonumber \\
     &=\int_{\omega_1\omega_2\omega_3} e^{-it(\omega_2+\omega_3)-\beta(E_I-\frac{1}{4}\omega_1-\frac{1}{2}\omega_2+\frac{1}{4}\omega_3)}F_{}(\omega_1,\omega_2,\omega_3) \nonumber \\
     &\quad + \int_{\omega_1\omega_3} e^{-it(-\omega_1+\omega_3)-\beta(E_I+\frac{1}{2}\omega_1+\frac{1}{2}\omega_3)}F_{(IJ)}(\omega_1,\omega_3) \nonumber \\
     &\quad + \int_{\omega_1\omega_2} e^{-\beta E_I}F_{(ij)}(\omega_1,\omega_2) \nonumber \\
     &\quad + \int_{\omega_1} e^{-S(E_I)-\beta E_I}F_{(IJ)(ij)}(\omega_1)
\end{align}
where we have defined window functions for contracted partitions, $F_{(\cdots)}$, and used the symbol $(\cdots)$ to label relevant index contractions. We have also dropped the slow dependence of $F$ on the average energy. Notice how the partition $(IJ)(ij)$ carries an extra factor of $e^{-S}$ ensuring that its contributions will be suppressed. Furthermore, notice that the only other time-independent term is $(ij)$. Therefore, if each $F$ has a finite resolution, then the long time value of $\rho_{II}$ will be entirely determined by the Laplace Transform of $F_{(ij)}$.

\begin{figure}[htbp]
    \centering
    \begin{subfigure}{0.49\textwidth}
        \centering
        \includegraphics[width=\linewidth, viewport=20 112 770 550, clip]{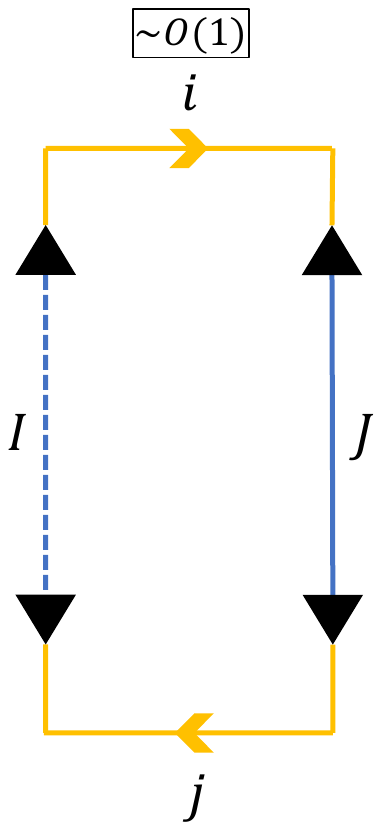}
        \caption{The full partition has 1 4-loop and 3 sums, therefore the overall weight is $\mathcal{O}(e^{3S-3S})\sim\mathcal{O}(1)$.}
        \label{subfig:PHpart1}
    \end{subfigure}
    \hfill
    \begin{subfigure}{0.49\textwidth}
        \centering
        \includegraphics[width=\linewidth, viewport=0 112 800 550, clip]{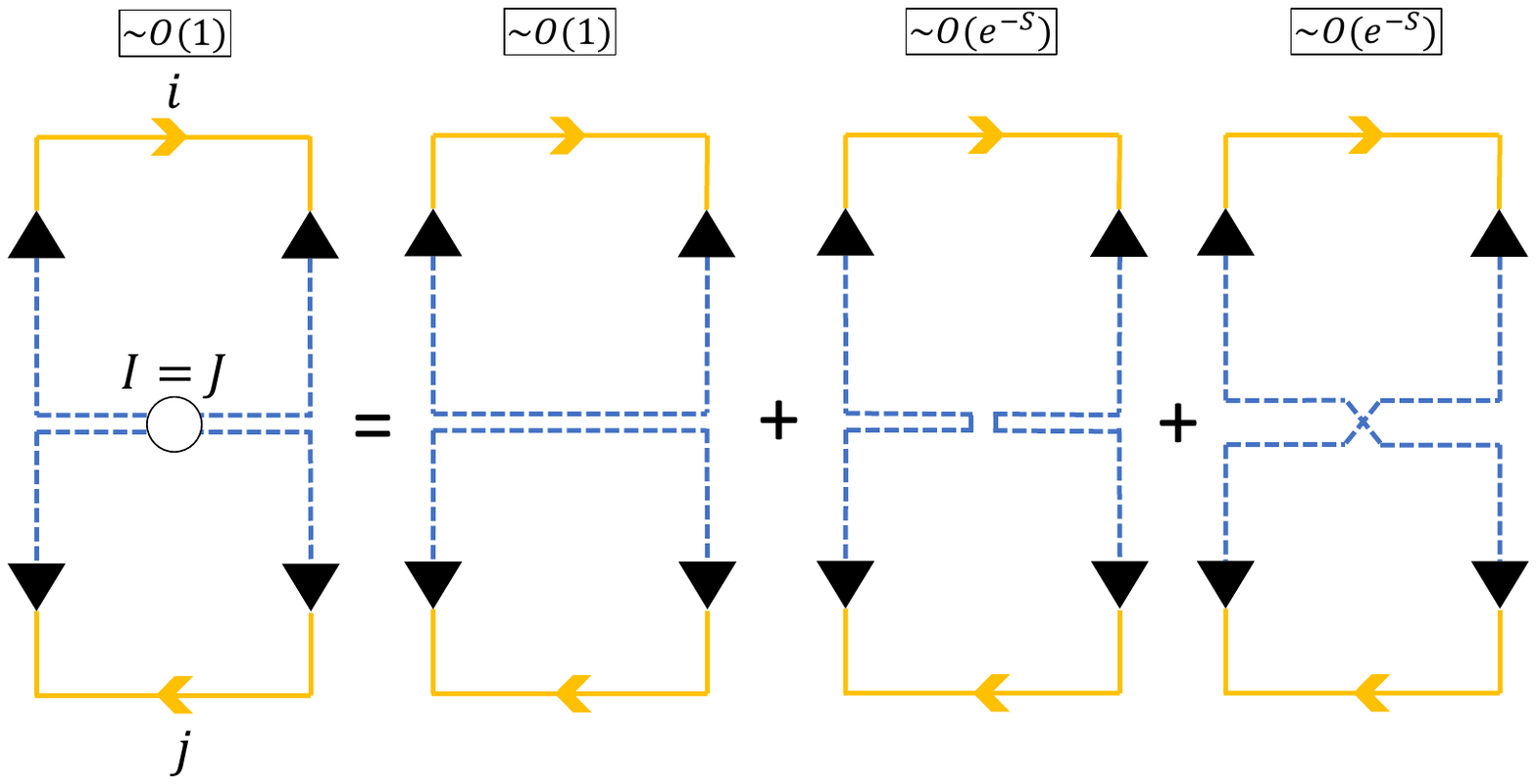}
        \caption{Partition $(IJ)$ is dominated by a product of 2 2-loops with 2 sums, therefore the overall weight is $\mathcal{O}(e^{2S-S-S})\sim\mathcal{O}(1)$.}
        \label{subfig:PHpart2}
    \end{subfigure}

    \vspace{0.1cm} 

    \begin{subfigure}{0.49\textwidth}
        \centering
        \includegraphics[width=\linewidth, viewport=0 112 800 550, clip]{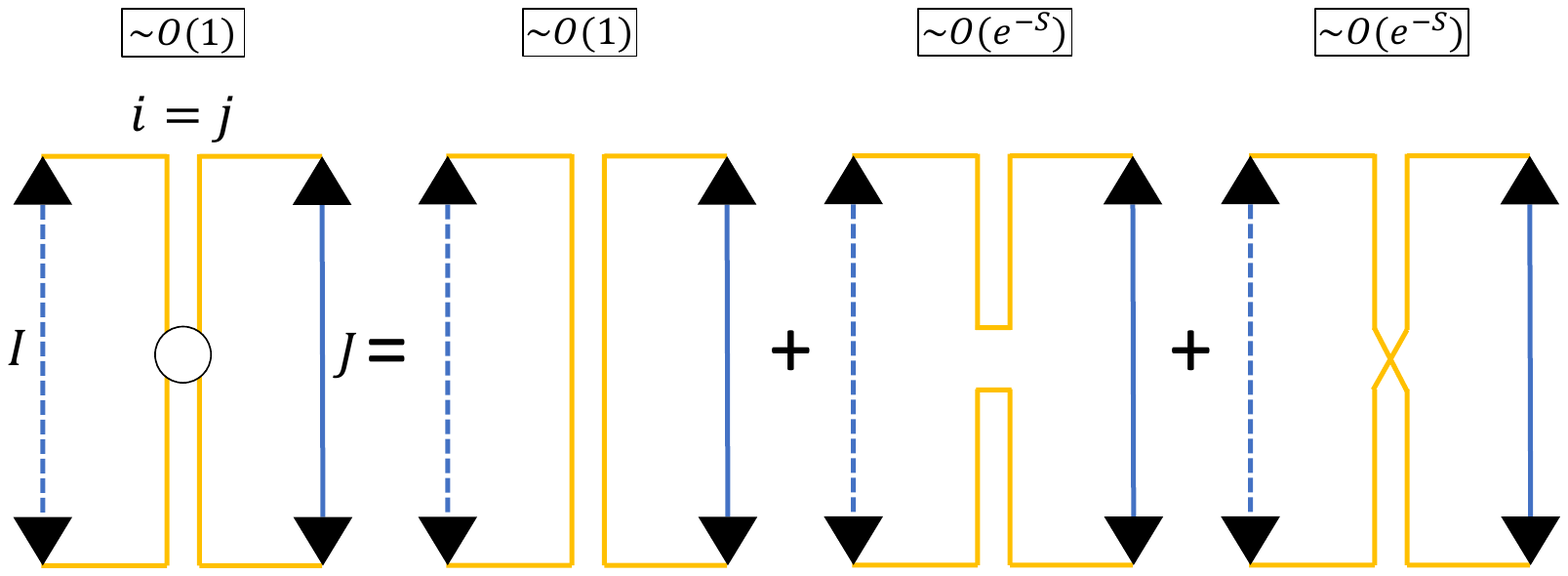}
        \caption{Partition $(ij)$ is dominated by a product of 2 2-loops with 2 sums, therefore the overall weight is $\mathcal{O}(e^{2S-S-S})\sim\mathcal{O}(1)$.}
        \label{subfig:PHpart3}
    \end{subfigure}
    \hfill
    \begin{subfigure}{0.49\textwidth}
        \centering
        \includegraphics[width=\linewidth, viewport=20 112 770 550, clip]{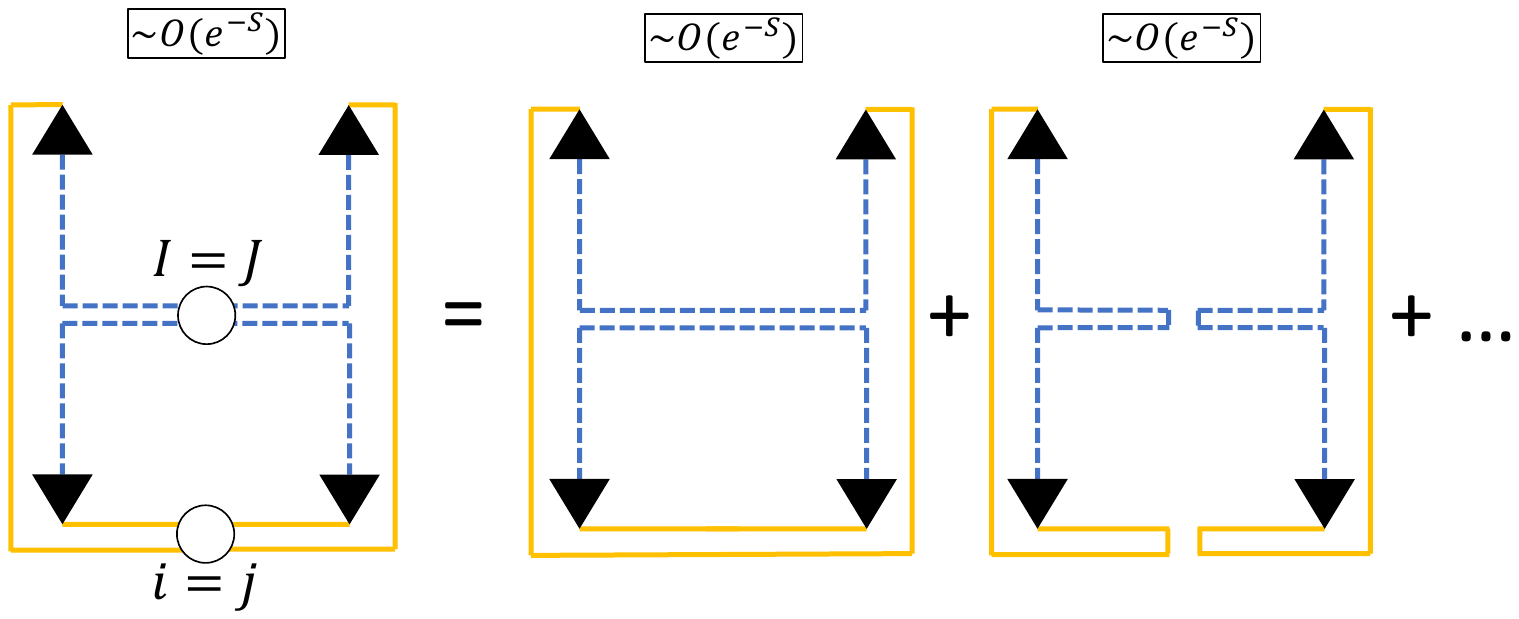}
        \caption{Partition $(IJ)(ij)$ is dominated by a product of 2 2-loops with 1 sum, therefore the overall weight is $\mathcal{O}(e^{S-S-S})\sim\mathcal{O}(e^{-S})$.}
        \label{subfig:PHpart4}
    \end{subfigure}

    \caption{The partitions in eq. (\ref{eq:parts}). We have used color to indicate the eigenstates of different Hamiltonians. Unless explicitly stated otherwise in the figure, indices are taken to be distinct in summations. When a partition contains a contraction, it can be expressed as a sum of products of connected components obtained from different ways of pairing off indices. In (b) and (c), there are 3 pairings only 1 of which is dominant. In (d), there are $3 \times 3 = 9$ pairings, of which we have drawn 2 of 3 dominant terms. In (c) and (d), notice that a contraction of indices with opposite time-dependences yields partitions that are time-independent. \textbf{Note:} Though we have removed the contraction symbol on the right-hand side of the above figures, the contractions still exist and need to be considered when counting the number of sums in a diagram.}
    \label{fig:PHpart}
\end{figure}

General expressions may have many more partitions, and counting every possible partition would be tricky. To simplify the process of identifying partitions, we introduce \textit{index diagrams} in fig. \ref{fig:PHpart}. The building blocks of {index diagrams} are labeled in table \ref{tab:tablePieces}. Let us state the rules of these diagrams.
\begin{enumerate}
    \item Each change-of-basis is represented by a triangle vertex, each index by a line. Operators are represented in labeled boxes. An arrow on a line indicates time evolution. Solid lines are summed over; dashed ones are fixed. 
    \item An index contraction occurs when two or more indices are equal, and is represented by an open circle. Attached to each contraction are $2p$ legs for some $p\geq 2$. The weight of the partition is given by the sum over distinct contributions from each of the $(2p-1)!!$ pairings of the $2p$ legs\footnote{ We show in section \ref{subsec:RMT} that the assumption that contractions decompose into pairs is equivalent to the assumption that random matrix terms are Gaussian to leading order, which is, in essence, a converse of Isserlis' (Wick's) theorem and a generic prediction of rotationally invariant random matrix ensembles.}. 
    \item Cycles appear as loops in the diagrams and contractions decompose large loops into smaller ones non-uniquely. However, there will be at least one decomposition with the largest overall weight that is, in general, the decomposition with the greatest number of component loops.
    \item The smooth weight assigned to each partition is dominated by the product of the weights of its component loops multiplied by an appropriate density of states factor for each sum. The weight of each loop is $e^{-(n-1)S(\bar{E})}$, where $n$ is the number of indices in each loop and $\bar{E}$ is the average energy of those indices. So long as we neglect the finite width of $F$, we can replace $\bar{E}$ with the energy of any index in the loop. Each sum contributes $e^{S(E_i)}$, where $E_i$ is the energy being summed over.
    \item The fluctuation assigned to each partition is dominated by the product of fluctuations from each wavefunction overlap multiplied by an appropriate factor for each sum. This term amounts to, in general, $e^{-(n-n^*)S(\bar{E})/2}$, where $n$ is the number of indices and $n^*$ is the number of indices being summed over.
    \item The arguments of the window function $F$ are the average energy of the connected part of the diagram and the energy differences across the triangle vertices. Per cycle, one such argument is redundant and omitted. We will not study the dependence of $F$ functions on average energy in this paper, so we will usually drop its dependence.
\end{enumerate}
These rules are intended to reproduce factors of $e^{S}$ in line with what one would expect for factors of $d$ that appear in uniform sampling of a Hilbert space. However, they will also allow us to introduce frequency-dependent structure into the smooth part of correlations that encode the nontrivial physical properties of systems.

\begin{table}[ht]
  \centering
  \begin{tabular}{|m{0.4\linewidth}|m{0.4\linewidth}|}
    \hline
    \textbf{Object} & \textbf{Symbol} \\
    \hline
    Summed index: $\sum_{i} \cdots\delta_{ii}\cdots$ & \includegraphics[width=0.8\linewidth, viewport = 55 400 225 450, clip]{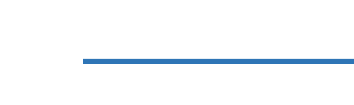} \\
    \hline
    Unsummed index: $\cdots\delta_{ii}\cdots$ & \includegraphics[width=0.8\linewidth, viewport = 295 290 465 340, clip]{TablePieces.pdf} \\
    \hline
    Summed index with time dependence: $\sum_{i} \cdots e^{iE_it}\delta_{ii}\cdots$ & \includegraphics[width=0.8\linewidth, viewport = 530 400 700 450, clip]{TablePieces.pdf} \\
    \hline
    Change-of-basis: $\cdots\bra{i}\ket{J}\cdots$ & \includegraphics[width=0.8\linewidth, viewport = 60 160 230 210, clip]{TablePieces.pdf} \\
    \hline
    Index contraction: $\cdots\delta_{ij}\cdots$ & \includegraphics[width=0.8\linewidth, viewport = 535 163 705 213, clip]{TablePieces.pdf} \\
    \hline
    Operator: $\cdots X_{ij}\cdots$ & \includegraphics[width=0.8\linewidth, viewport = 315 175 465 225, clip]{TablePieces.pdf} \\
    \hline
  \end{tabular}
  \caption{Building blocks of the index diagrams.}
  \label{tab:tablePieces}
\end{table}

First, recognize that in Fig. \ref{subfig:PHpart4}, the suppression of $(IJ)(ij)$ is related to a double constraint imposed by the two contractions. In general, partitions may have any number of contractions, however, if any two contractions are connected by more than two indices topologically, they will be suppressed by a relevant factor of the density of states. What we refer to as a generalized free cumulant is a single connected partition, without contractions, before summing, in which no two indices are equal. In this case, the GFC's are individual loops. For example, a single, $n=4$ generalized free cumulant is pictured in figure \ref{subfig:PHpart1}, whereas each of figures \ref{subfig:PHpart2}, \ref{subfig:PHpart3}, \ref{subfig:PHpart4} depict partitions which are dominated by a product (or products) of $n=2$ generalized free cumulants. GFCs are the irreducible building blocks of correlations in the $e^{-S}$ expansion.
\begin{table}[h]
  \centering
  \begin{tabular}{|c|c|c|}
    \hline
    \textbf{} & \textbf{Notation} & \textbf{Example} \\
    \hline
    (a) &$E_{ijk\cdots}$ & $E_{ij} = \frac{1}{2}(E_i+E_j)$ \\
    (b)&$\bar E$ & $\bar E = \frac{1}{n}\sum_{m=1}^nE_m$ \\
    (c)&$\omega_{ij}$ & $\omega_{ij} = E_i-E_j$ \\
    (d)&$(\cdots)$ & eq. \eqref{eq:parts} \\
    (e)&$[\cdots]$ & $\sum_{[ij]}X_{ij}X_{ji} = \sum_{i}\sum_{j\neq i}X_{ij}X_{ji}$ \\
    (f)&$f$ & $f_{XY}(E; \omega) = e^{-S(\bar E)}\sum_{[ij]}X_{ij}Y_{ji}\delta(E_i-(E+\omega/2))\delta(E_j-(E-\omega/2))$ \\
    (g) & $F$ & eq. \eqref{eq:parts} \\
    (h) &$\expval{\cdots}_i$ & $\expval{X(t)Y(0)X(t)Y(0)}_i = \bra{i}\!X(t)Y(0)X(t)Y(0)\!\ket{i}$ \\
    (i) & $\expval{\cdots}_\beta$ & $\expval{X(t)Y(0)X(t)Y(0)}_\beta=\sum_{ijkl}X_{ij}Y_{jk}X_{kl}Y_{li}e^{i(\omega_{ij}+\omega_{kl})t-\beta E_{ijkl}}/\sum_{i}e^{-\beta E_i}$ \\
    \hline
  \end{tabular}
  \caption{Table of various notation. All notation holds analogously for various indices we will use ($i,\,j,\,I,\,J,\,a,\,b$). (a) We use subscripts of $E$ to denote average energies over specified indices. (b) We use an overline to denote averaging the energy over an unspecified number of indices. (c) We use $\omega$ in general to denote energy differences, and subscripts to specify indices. (d) We use parentheses, $(\cdots)$, as a short hand to denote a set of indices that are contracted. Multiple contractions are written as products, $(\cdots)(\cdots)$. (e) Braces are used to indicate indices that are not equal to any other index in the expression. (f) We use the letter $f$ to denote operator free cumulants in time and frequency space. (g) We used the letter $F$ to denote state generalized free cumulants in time and frequency space. (h) We use $\expval{\cdots}_i$ to denote the ordinary expectation value evaluated on an individual eigenstate $i$. (i) We use $\expval{\cdots}_\beta$ to denote special thermally regulated correlators that are cyclically symmetric in their arguments and are natural objects in quantum chaos.}
  \label{tab:mytable}
\end{table}

Each GFC has an associated window function. In the present case, we can use $F_2(\omega)$ to represent the window function for the $n=2$ GFC and $F_4(\omega_1,\omega_2,\omega_3)$ for the $n=4$ GFC\footnote{We should point out here the $F$ depends, in general, on the Hamiltonians considered. Rather than name distinct functions for the various cases we consider throughout this paper, with some abuse of notation, we will reuse the label $F$ for general window functions and let its arguments and context specify the specific function under consideration.}. More precisely,
\begin{align}
\label{eq:GFCex}
    &\overline{c^I_{j}c^{j}_I} = e^{-S(E_{Ij})}F_2(E_I-E_j) \nonumber \\
    &\overline{c^I_{i}c^{i}_{J}c^{J}_{j}c^{j}_I} = e^{-3S(E_{IJij})}F_4(E_I-E_i, E_i-E_J, E_J-E_j),\quad I\neq J,\, i\neq j.
\end{align}
Then, by comparison to eq. (\ref{eq:parts}) or by inspecting figure \ref{fig:PHpart}, we can write down the window functions for the contracted terms to leading order, 
\begin{align}
    \label{eq:reducible}
    &F(\omega_1,\omega_2,\omega_3) = F_4(\omega_1,\omega_2,\omega_3) \nonumber \\
    &F_{(IJ)}(\omega_1, \omega_2) = F_2(\omega_1)F_2(\omega_2) \nonumber \\
    &F_{(ij)}(\omega_1, \omega_2) = F_2(\omega_1)F_2(\omega_2) \nonumber \\
    &F_{(IJ)(ij)}(\omega) = 3F_2(\omega)^2.
\end{align}
The general procedure for performing computations in this paper will be inspecting diagrams to acquire the correct factors of $e^S$ and the correct form of the window function, then evaluating integrals via saddle-points by neglecting the finite width of the window functions. We have included a summary of notation we use in this paper in table \ref{tab:mytable}.

\subsection{Operator correlations: OTOCs and freeness}
\label{subsec:cactus}
In eq. \eqref{eq:crux}, we showed that the sum over a product of operators neglecting repeated indices evaluated on an eigenstate reduced to the mixed Fourier-Laplace transform of the relevant $f$ function. We referred to this object as a free cumulant. Instead of evaluating the free cumulant on an eigenstate, we can evaluate it via a special thermal regulator,
\begin{align}
    \label{eq:reg}\frac{1}{Z_\beta}\sum_{[i_1\cdots i_n]}X_{i_1i_2}\cdots X_{i_ni_1} e^{i\Vec\omega\cdot\Vec t-\beta\bar E} &= e^{\beta E_\beta-S(E_\beta)}\int_{\bar E \Vec\omega} e^{i\Vec\omega\cdot\Vec t + (n - (n-1))S(\bar E) - \beta\bar E} f(\bar E; \Vec\omega) \nonumber \\ 
    &= \int_{\Vec\omega}e^{i\Vec\omega\cdot\Vec t} f(E_\beta; \Vec\omega) \nonumber \\ 
    &= f_n(E_\beta;\Vec t\,\,).
\end{align}
where the integral over $\bar E$ was evaluated via its saddle-point (see appendix \ref{app:saddlePoints}). The symmetry between indices has removed the shift vector $\Vec l$ which forced us to work in complex time. Thus we find it is more natural to work with the thermally regulated correlator which is conventional in the study of quantum chaos~\cite{maldacena_bound_2016, pappalardi_microcanonical_2023}.

\begin{figure}[htb]
    \centering
    \includegraphics[width = 0.8\linewidth, viewport = 0 100 800 500, clip]{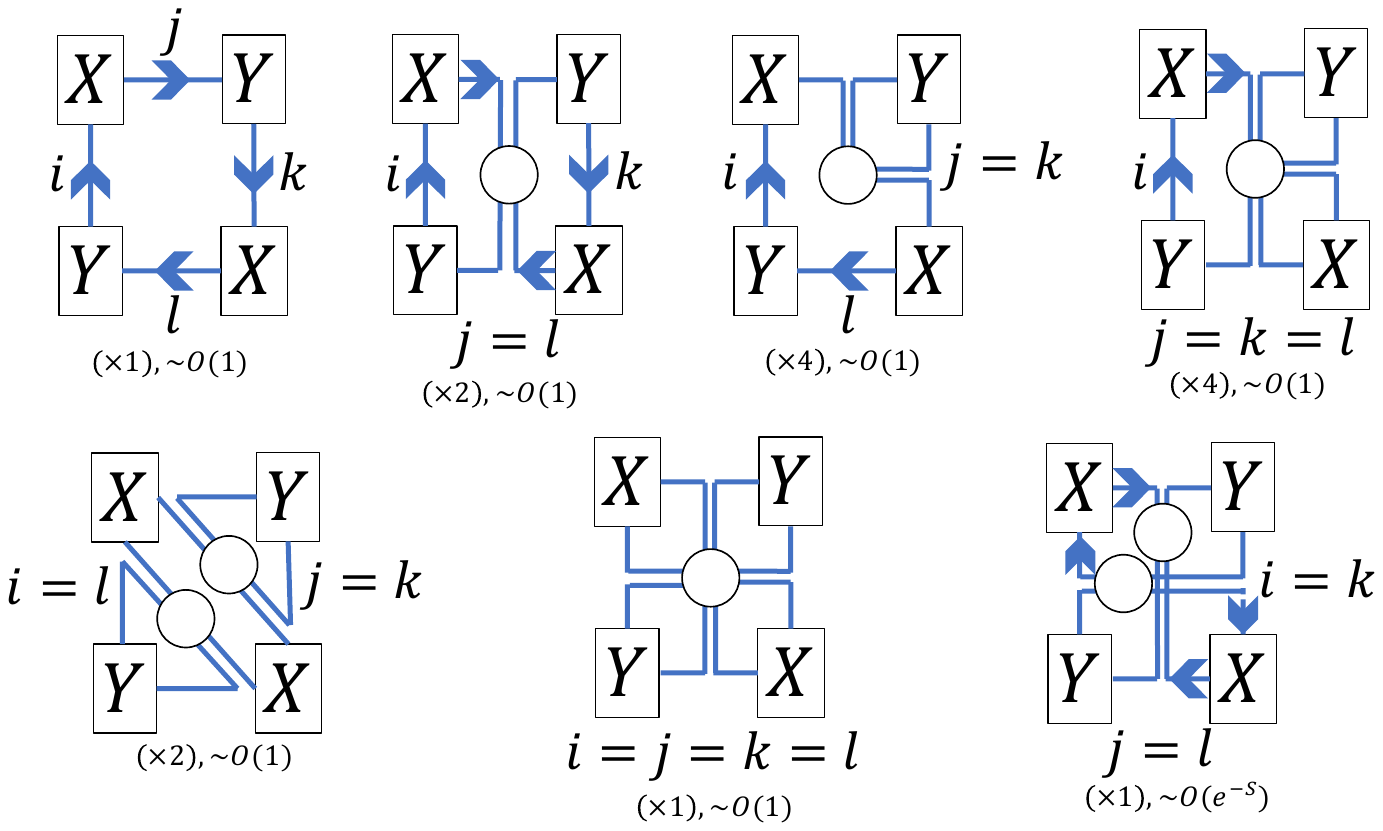}
    \caption{There are a total of 15 partitions from 7 different graphs. We have drawn one representative partition from each graph and labeled their multiplicities. Note that two different partitions with the same graph contribute differently, so the multiplicity is not a ``symmetry factor''. We have also labeled the overall weight of each partition divided by an implicit factor of $Z_\beta$.}
    \label{fig:cactus}
\end{figure}

Here, we study an out-of-time-ordered correlator (OTOC),
\begin{align}
\label{eq:OTOC}
    \expval{X(t)Y(0)X(t)Y(0)}_\beta = \frac{1}{Z_\beta}\sum_{ijkl}X_{ij}Y_{jk}X_{kl}Y_{li}e^{i\omega_{ij}t+i\omega_{kl}t-\beta E_{ijkl}}.
\end{align}
We have drawn all the possible contractions of eq. \eqref{eq:OTOC} in figure \ref{fig:cactus}. In this case, since any index can contract with any other index, the GFCs of our diagrams, which are the cumulants mentioned above, are the celebrated free cumulants of the operators $X(t)$ and $Y(0)$ and are represented by single loop diagrams and are associated to the $f$ functions.

We can isolate the free cumulants from eq. \eqref{eq:OTOC} using non-crossing \textit{operator} partitions inline as follows. For convenience we will assume that all $1$-point functions vanish, $\expval{X}_\beta = \expval{Y}_\beta = 0$. We define an operator partition as a symbol that represents the associated free cumulant of an ordered set of operators. A crossing of operator partitions results in an overwhelming suppression. Then,
\begin{align}
\label{eq:freeWick}
    \expval{X(t)Y(0)X(t)Y(0)}_\beta &\equiv \raisebox{-0.33em}{\begin{tikzpicture}
    \begin{scope}[xshift=-2.8em]
    \draw (0em,0) -- ++ (0,1.05em) -| (2em,0);
    \draw (2em,1.05em) -| (4em,0);
    \draw (4em,1.05em) -| (6em,0);
    \end{scope}
    \node[fill=white, inner sep=1pt] at (1em,0) {$\expval{X(t)Y(0)X(t)Y(0)}_\beta$};    
  \end{tikzpicture}} \nonumber \\
    & + \wick[sep = 0pt, offset = 1.3em]{\expval{\c1 X(t) \c1 Y(0) \c2 X(t) \c2 Y(0)}_\beta} + \wick[sep = 3pt, offset = 1em]{\expval{\c2 X(t) \c1 Y(0) \c1 X(t) \c2 Y(0)}_\beta} \nonumber \\
    &= f_{XYXY}(E_\beta;t,0,t) + 2f_{XY}(E_\beta;t)^2 \nonumber \\
    &= \int_{\omega_{1}\omega_2\omega_3}f_{XYXY}(E_\beta;\omega_{1},\omega_2,\omega_3)e^{it(\omega_1+\omega_3)} \nonumber \\
    & + 2\left(\int_{\omega}f_{XY}(E_\beta;\omega)e^{i\omega t}\right)^2   
\end{align}
where $E_\beta = \expval{H}_\beta$ and we have extracted 3 terms: a partition of all 4 operators, a product of the partition of the first two and last two operators, and a product of the partition of the first and last operator and the second and third operator. To a non-crossing partition of $n$ operators is associated their $n$-point free cumulant. From eq. (\ref{eq:freeWick}) we can see that the $f$-functions of the ETH are the regularized free cumulants that have a special significance in quantum chaos. 

Partitions whose contractions cross exist, but are suppressed. In this case, we have one:
\begin{align}
\label{eq:freeWickCross}
    \wick[sep = 3pt, offset = 1em]{\expval{\c1 X(t) \c2 Y(0) \c1 X(t) \c2 Y(0)}_\beta} &= \frac{1}{Z_\beta}\left(2f_{XY}(E_\beta;t)^2 + f_{XX}(E_\beta;0)f_{YY}(E_\beta;0)\right)\sim \mathcal{O}(e^{-S}).
\end{align}
One may wonder why 2 copies of $f_{XY}(E_\beta;t)^2$ appeared in eq. (\ref{eq:freeWickCross}) when we appeared to have only contracted the pair of $X$'s and the pair of $Y$'s. One can find all 3 terms from the last partition in figure \ref{fig:cactus}. But it is not obvious how to derive them directly from the left-hand expression in eq. \eqref{eq:freeWickCross}. However, it is also natural to utilize non-crossing index partitions. In table \ref{tab:constrasts}, we compare our own diagrammatic approach to correlations to others in the literature, namely, cactus diagrams, non-crossing partitions of indices, and 
't Hooft diagrams. All approaches fundamentally contain the same content and are united in their connection to free probability. Redoing the above calculations but taking into account 1- and 3-point functions will generate all 15 partitions discussed in figure \ref{fig:cactus}. The full decomposition into non-crossing partitions is given in appendix~\ref{app:FPT}.
\begin{table}[ht]
  \centering
  \begin{tabular}{|c|c|c|c|}
  \hline
    \textbf{Index} & \textbf{Cactus} & \textbf{Non-crossing} & \textbf{'t Hooft} \\
    \hline
    \includegraphics[width=0.14\linewidth, viewport = 25 315 140 490, clip]{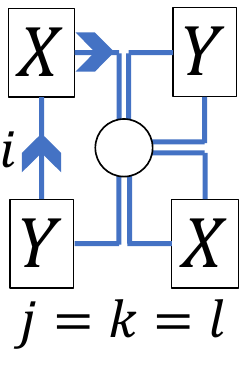} &
    \includegraphics[width=0.24\linewidth, viewport = 140 300 400 502, clip]{Table2.pdf} &
    \includegraphics[width=0.2\linewidth, viewport = 410 320 590 525, clip]{Table2.pdf} &
    \includegraphics[width=0.2\linewidth, viewport = 590 335 750 500, clip]{Table2.pdf} \\
    \hline
    \includegraphics[width=0.2\linewidth, viewport = 30 105 203 285, clip]{Table2.pdf} &
    \includegraphics[width=0.24\linewidth, viewport = 145 90 400 310, clip]{Table2.pdf} &
    \includegraphics[width=0.2\linewidth, viewport = 410 110 590 315, clip]{Table2.pdf} &
    \includegraphics[width=0.2\linewidth, viewport = 590 125 750 280, clip]{Table2.pdf} \\
    \hline
  \end{tabular}
  \caption{Comparison of diagrammatic techniques. Column 1 contains two partitions from figure \ref{fig:cactus}. Column 2 redraws these partitions using the formalism of refs.~\cite{foini_eigenstate_2019, pappalardi_eigenstate_2022}. Whether a diagram is leading order is equivalent to whether it resembles a cactus (essentially a tree with round branches) in the following sense. Each loop is a cactus pad; to any cactus pad, any number of other pads may be connected. However, to travel from any pad to any other by hopping to adjacent pads, there can only ever be one route without crossing the same contraction twice. This feature of cactus diagrams is shared by the loops in index diagrams so long as we are not considering subsystems (which we do in the next section). We also note that these are essentially the properties of tadpole diagrams which govern the Hartree approximation in large $N$ vector systems~\cite{fradkin_quantum_2021}. Column 3 utilizes so-called non-crossing partitions of indices whose order is determined by the number of crossings. In eq.~\eqref{eq:freeWick} we utilized non-crossing partitions of operators. Non-crossing partitions of operators and of indices enjoy a \textit{dual} relationship and both function to describe free cumulants~\cite{pappalardi_eigenstate_2022}. The black circumscribing circle represents the trace and the internal lines represent index contractions. The top row contains no crossings, while the bottom row contains 1. Column 4 depicts analogous 't Hooft diagrams, which play a role in large $N$ gauge theories and whose order is closely related to the genus of the surface on which the diagram may be drawn without crossings~\cite{hooft_planar_1974}. As for non-crossing partitions, the black circumscribing circle represents the trace and the internal lines represent index contractions. The solid circles explicitly represent the appropriate free cumulants, often referred to in this context as \textit{planar connected Green's functions}~\cite{hooft_planar_1974, gopakumar_mastering_1994}, and as such the bottom row only encodes the first term on the right hand side of eq. \eqref{eq:freeWickCross} and not all terms of order $e^{-S}$. 't Hooft diagrams are the natural method of evaluating matrix models and main technology employed by ref.~\cite{jafferis_matrix_2023} to study the ETH. Both 't Hooft and index diagrams utilize a type of double-line notation, where the former uses open circle to denote index contractions, and the latter uses solid circles to denote operator contractions. In this way, index and 't Hooft diagrams share a dual relationship similar to the one shared by non-crossing partitions of operators and of indices.}
  \label{tab:constrasts}
\end{table}

As discussed in section \ref{sec:motivation}, the thermalization of a system follows from the decay of these free cumulants at late times. The equilibrium value of a correlation function can be obtained by dropping all time-dependent partitions or, equivalently, all diagrams that contain arrows. An equivalent formulation of this statement is that operators that satisfy the ETH become \textit{freely independent}, or \textit{free}, at long time separations. Free independence, or simply \textit{freeness}, is the free probabilistic analog of classical independence~\cite{mingo_free_2017}. For example, whereas a sum of many classically independent random variables acquires a Gaussian probability distribution function, a sum of many freely independent random variables acquires a semicircle spectrum, a famous signature of Gaussian matrix ensembles~\cite{lal_mehta_random_2004}. We can draw conclusions about physical operators from this observation.

Consider an operator $X$ restricted to a narrow energy band,
\begin{align}
\label{eq:banded}
     X_{ij}^{(\omega_c)} \equiv X_{ij}\theta(\omega_c-\abs{\omega_{ij}}).
\end{align}
Ref.~\cite{richter_eigenstate_2020} showed numerically on a spin chain that as $\omega_c \rightarrow 0$, $X^{(\omega_c)}$ obtains a semicircular spectrum and interpreted this observation as the onset of \textit{random matrix theory}. We can interpret this observation as a consequence of free independence and, thus, a necessary consequence of the ETH. Consider the following identity,
\begin{align}
    \label{eq:freeTime}
    X^{(\omega_c)}_{ij} \equiv X_{ij}\theta(\omega_c-\abs{\omega_{ij}}) = \frac{\omega_c}{\pi}\int_{-\infty}^{\infty} X_{ij}(t)\operatorname{sinc}(\omega_ct)dt
\end{align}
which allows us to see $X^{(\omega_c)}$ as an average of $X(t)$ over a time period of $\omega_c^{-1}$. If $X(t)$ becomes freely independent from $X(0)$ for $t$ greater than some $t_{\text{free}}$, then for $\omega_c \ll t_{\text{free}}^{-1}$, $X^{(\omega_c)}$ must obtain a semicircular spectrum. This has been more recently interpreted in terms of the onset of an \textit{emergent rotational symmetry} in the energy eigenspace of ETH satisfying systems at asymptotically small frequencies~\cite{wang_emergence_2023}.

We can draw a more direct line between freeness and the quantum butterfly effect by defining the freeness of sublagebras. Consider two subalgebras, $\mathcal{X}$ and $\mathcal{Y}$, of the algebra of operators that act of the Hilbert space of our system. These subalgebras may be understood, for example, to be the Pauli algebras of two possibly identical sites in a spin chain at possibly different times. $\mathcal{X}$ and $\mathcal{Y}$ are freely independent if and only if for every $X_i\in\mathcal{X}$ and $Y_i\in\mathcal{Y}$ with $\expval{X_i}=\expval{Y_i}=0$,
\begin{align}
    \label{eq:freeIndependence}
    \expval{X_1Y_1\cdots X_qY_q} = 0,\,\forall q \geq 1.
\end{align}
Shifting $\mathcal{X}$ forward in time\footnote{by time evolving each element of $\mathcal{X}$}  $\mathcal{X} = e^{iHt}\mathcal{X}(0)e^{-iHt}$, we can see that eq. (\ref{eq:freeIndependence}) contains the statement that all $q$-OTOCs of an element of $\mathcal{X}$ with an element of $\mathcal{Y}$ have decayed. Thus OTOCs are themselves a measure of freeness. This observation was first made in ref.~\cite{fava_designs_2023}, where a connection was drawn between OTOCs, freeness, and unitary designs. The idea we wish to convey is that satisfying the ETH, the quantum butterfly effect, and the emergence of rotational symmetry at small frequencies are, in essence, equivalent definitions of chaos and avatars of free probability (see appendix~\ref{app:FPT} for an elaboration on these ideas). 

\subsection{Random matrices, replicas, and higher moments}
\label{subsec:RMT}
One may also be interested in partitions where some legs are left open, as these partitions represent fluctuations with a random tensor term, e.g. $\bra{i}\ket{J} = \sqrt{e^{-S(E_{iJ})}F(\omega_{iJ})} R^{i}_{J}$ or $X_{ij} = f_1(E_{ij})\delta_{ij} + \sqrt{e^{-S(E_{ij})}f_2(\omega_{ij})} R_{ij}$, where $R$ represents random matrix terms with mean $0$ and variance $1$. The variance of such a term can be computed by replicating the diagram and connecting the open legs of the original to the open legs of its conjugate, e.g., $|\bra{i}\ket{J}|^2\propto R^{i}_{J}R_{i}^{J} \sim 1$. In general, one may compute higher moments from suitably many replicas and apply the above-stated rules for the partitions that are generated. This procedure is key to our computations of entanglement entropy in sections~\ref{subsec:PageEq} and~\ref{subsec:Page}.

\begin{figure}
    \centering
    \includegraphics[width = 0.6\linewidth, viewport=0 150 800 430, clip]{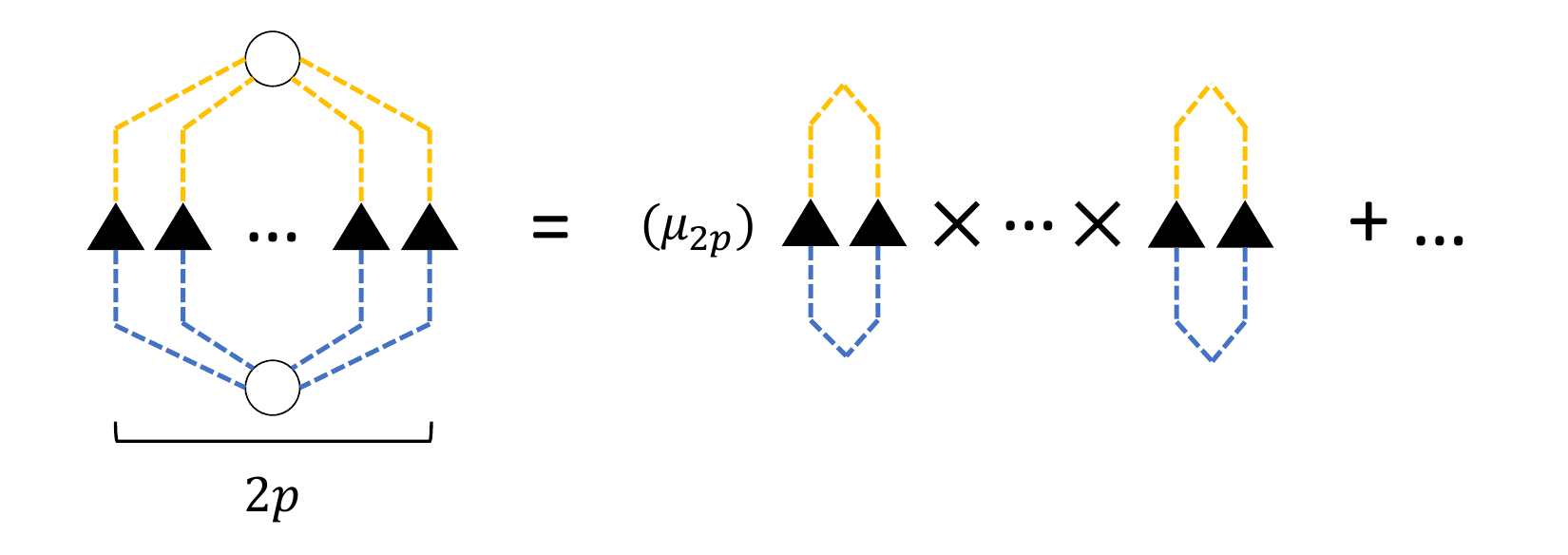}
    \caption{Decomposition of a contraction of $2p$ replicas into $\mu_{2p}$ products of $p$ 2-loops. This decomposition is not special to the case shown but a general result on the moments of random matrix terms. Where the triangle is in the above figure, any arbitrarily complicated diagram with two external legs may be placed instead but the consequence of the replica calculation will be the same: random matrix terms are Gaussian distributed to leading order in $e^{-S}$.}
    \label{fig:Gaussian}
\end{figure}

A simple consequence of the rules we have presented is that random matrix terms are Gaussian distributed, to leading order in $e^{-S}$. For example, we can show that $R^{i}_J$ is Gaussian distributed $\sim \mathcal{N}(0,1)$ by recognizing that (a) odd moments of $R$ have open indices and vanish while (b) even moments will be dominated by the decompositions with the largest number of loops: products of pairs (see figure \ref{fig:Gaussian}). Then, $\overline{|R^i_J|^{2p+1}} = 0$ while $\overline{|R^i_J|^{2p}} = \mu_{2p} (\overline{R^{i}_{J}R_{i}^{J}})^p + \mathcal{O}(e^{-S}) = (2p-1)!!$, where $\mu_{2p} = (2p-1)!!$ is the number of ways to pair up $2p$ copies of $|R^i_J|$ and are the central moments of the Gaussian distribution. An analogous argument was made for matrix elements of operators in~\cite{srednicki_approach_1999} and follows concretely from \eqref{eq:ETH}. Another result which can readily obtained is that variance of the diagonal elements of operators is exactly twice the variance of nearby off-diagonal operators~\cite{mondaini_eigenstate_2017, foini_eigenstate_2019-1}, as the number of ways to pair indices of $2p$ copies of a diagonal element $R_{ii}$ is instead $2^{2p}(2p-1)!!$. These properties are expected features of matrices sampled from rotationally invariant ensembles~\cite{lal_mehta_random_2004, collins_second_2006}.

The emergence of a Gaussian distribution originated from the decomposition of contractions into pairs of indices (rule 2). This derivation is essentially a converse of Isserlis' (Wick's) theorem. For the case of pure operator correlations, the validity of this decomposition is implicitly assumed throughout the literature and in key works~\cite{foini_eigenstate_2019, pappalardi_eigenstate_2022} and is supported by numerics on spin chains~\cite{steinigeweg_eigenstate_2013}. This assumption is justified so long as we imagine random matrix terms are consistent with a rotationally invariant random matrix ensemble. The matrix model of ref.~\cite{jafferis_matrix_2023} formalizes this idea for pure operator correlations, but heuristically it should hold for general eigenstate or operator-eigenstate correlations as well. To be more precise, as we discuss in appendix~\ref{app:FPT}, eq.~\eqref{eq:ETH} is an extension of the general form for the free cumulants of random matrices. But for the terms in eq.~\eqref{eq:ETH} to retain the combinatorial properties of free cumulants (i.e. generate non-crossing partitions), the pairwise decomposition must hold and as shown in ref.~\cite{jafferis_matrix_2023}, eq.~\eqref{eq:ETH} implies that non-Gaussian terms in the corresponding matrix model are suppressed by appropriate density of states factors. Thus the suppression of non-Gaussianities should be understood as a generic feature of the ETH.

The procedure of using replicas to compute higher moments also holds for partitions without open legs. In general, computing moments will reduce to counting contractions between replicas. One will find that partitions with sufficiently many summed indices (in the sense of eq. (\ref{eq:comparison})) will have higher central moments suppressed by factors of $e^{-S}$, as contractions inevitably cost sums without yielding sufficiently advantageous factorizations. Thus, these partitions will be very sharply peaked around their mean value. In contrast, partitions with very few summed indices will lose no sums from contractions but will gain advantageous factorizations and thus will be widely distributed from their means. The situation with open legs discussed in the previous paragraphs is a special case where random matrix terms acquire a Gaussian distribution with $O(1)$ variance about their means.

\subsection{Eigenstate correlations: interacting subsystems}
\label{subsec:diagrams2}
\begin{figure}[t!]
  \centering
  \includegraphics[width=0.5\linewidth, viewport = 10 130 770 520, clip]{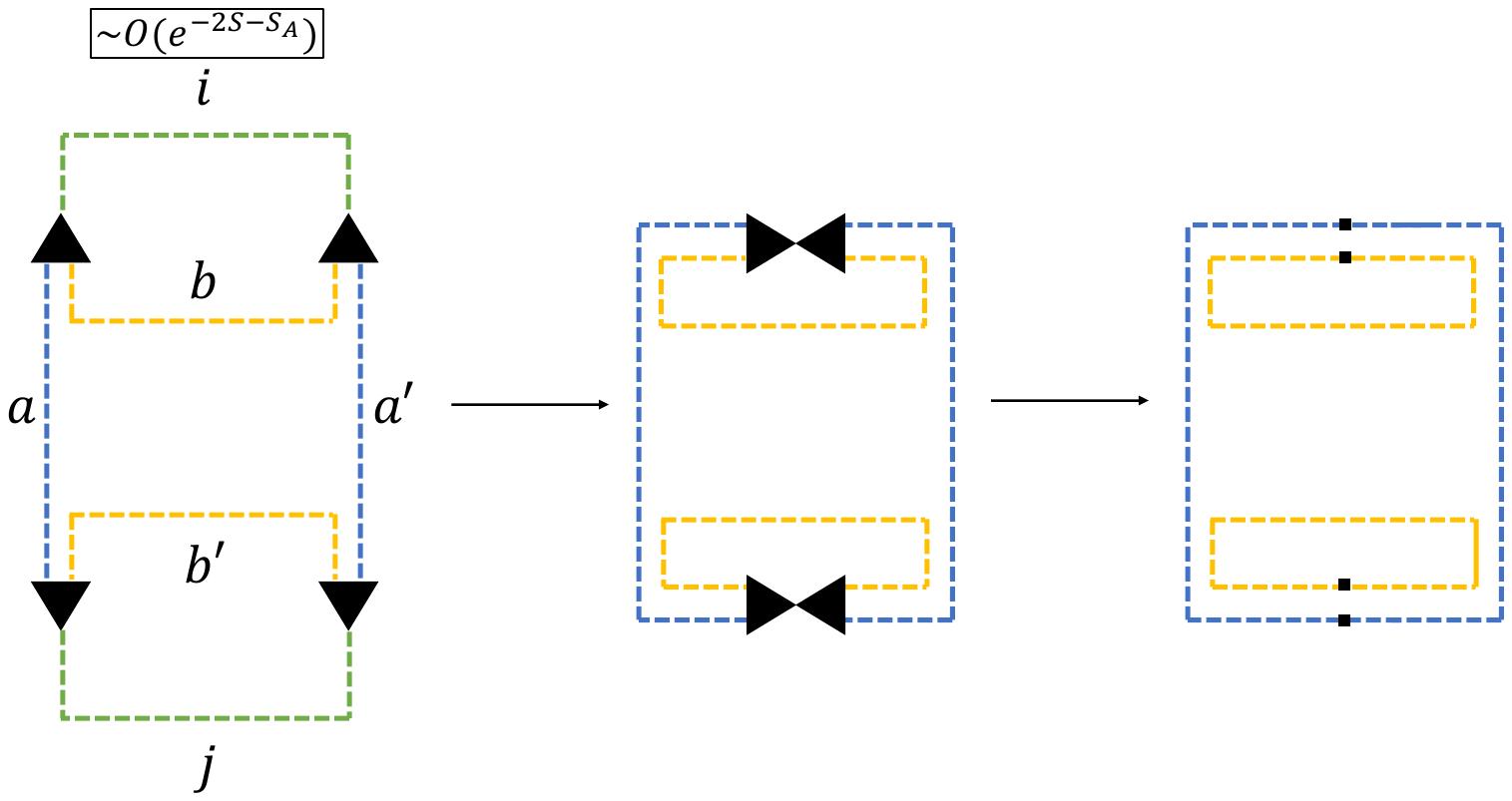}
  \caption{Schematic calculation of the weight associated with the uncontracted partition of eq. (\ref{eq:ISexample}). Whole system indices are removed, each carrying a factor of $e^{-S}$, and then subsystem indices contribute a total of $e^{-S_A}$ from what is left. Note that we are neglecting fluctuations which will vanish under summation.}
  \label{fig:schematic}
\end{figure}
We now wish to generalize our formalism to the situation of two interacting subsystems $A$ and $B$. We take $H_0 = H_A + H_B$ and $H=H_A+H_B+H_{AB}$. Take $\ket{i}$ and $\ket{j}$ as possibly identical eigenstates of $H$. We study, as an example of the various subtleties we will encounter, the Hilbert--Schmidt inner product of their reduced density matrices on $A$:
\begin{align}
\label{eq:ISexample}
    \big(\rho^{i}_A\big|\rho^j_A\big)\equiv \operatorname{Tr}_A\left[\rho^{i}_A\rho^j_A\right] = \sum_{aa'bb'}c^{ab}_ic^{i}_{a'b}c^{a'b'}_jc^{j}_{ab'}.
\end{align}
First, take $i$ and $j$ as distinct indices. We recognize that without sums, the term in eq. (\ref{eq:ISexample}) is the eigenstate analog of equation (\ref{eq:nontrivial}) which we conjectured to have unsummed weight $e^{-2S-S_A}$. The spirit of the calculation in eq. (\ref{eq:nontrivial}) was that we can remove indices on the whole system and collect a factor of $e^{-S}$ for each removed index until we are left with a product of partitions on the subsystem. We represent this calculation schematically in figure \ref{fig:schematic}. The remaining subsystem partitions contribute their factors of the subsystem densities of states. In this case we get $e^{-S}$ from each of $i$ and $j$, and a factor of $e^{-S_A}$ from the remaining 2 index cycles from $a$, $a'$. We neglect the orthogonality of $\bra{a}\ket{a'}$ so long as $a$ and $a'$ are not adjacent in the original diagram.
\begin{figure}[t!]
    \centering
    \begin{subfigure}{0.49\textwidth}
        \centering
        \includegraphics[width=0.9\linewidth, viewport=20 112 770 550, clip]{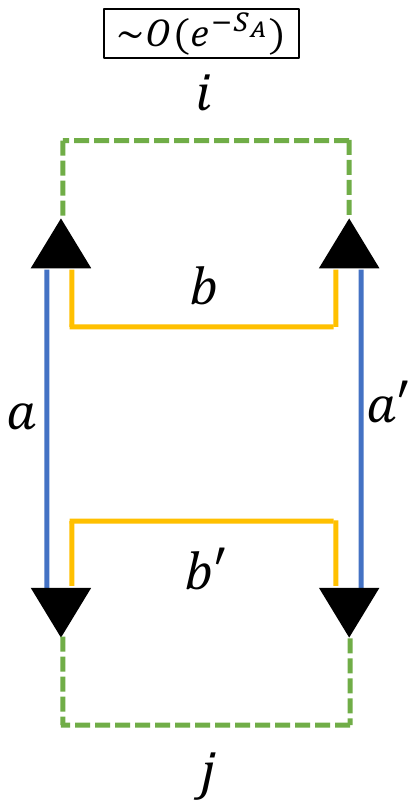}
        \caption{The full partition from eq. (\ref{eq:ISexample}) has overall weight $\mathcal{O}(e^{-S_A})$.}
        \label{subfig:ISij1}
    \end{subfigure}
    \hfill
    \begin{subfigure}{0.49\textwidth}
        \centering
        \includegraphics[width=0.9\linewidth, viewport=20 112 770 550, clip]{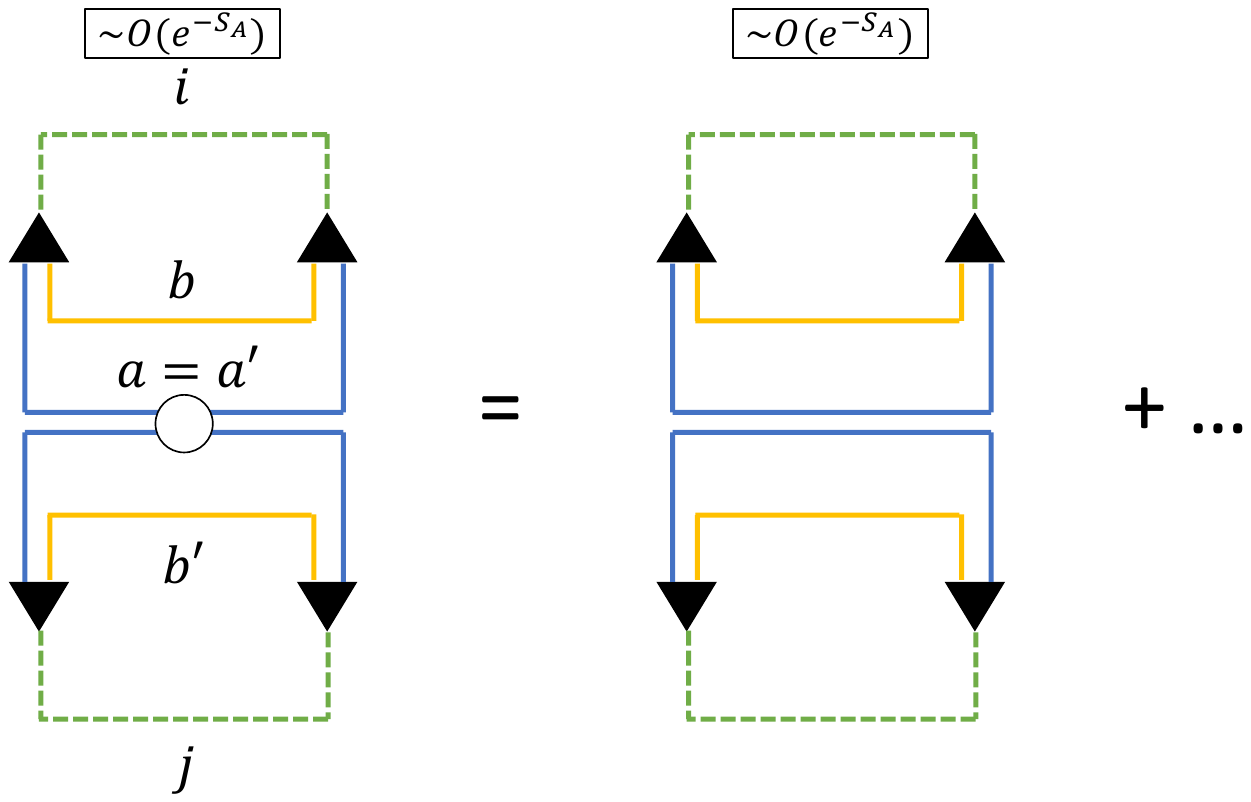}
        \caption{Partition $(aa')$ is dominated by a product of 2 2-cycles with 2 sums on subsystem $B$ and 1 sum on subsystem $a$. Therefore the overall weight is $\mathcal{O}(e^{2S_B+S_A-S-S})\sim\mathcal{O}(e^{-S_A})$.}
        \label{subfig:ISij2}
    \end{subfigure}

    \vspace{0.1cm} 

    \begin{subfigure}{0.49\textwidth}
        \centering
        \includegraphics[width=0.9\linewidth, viewport=20 112 770 550, clip]{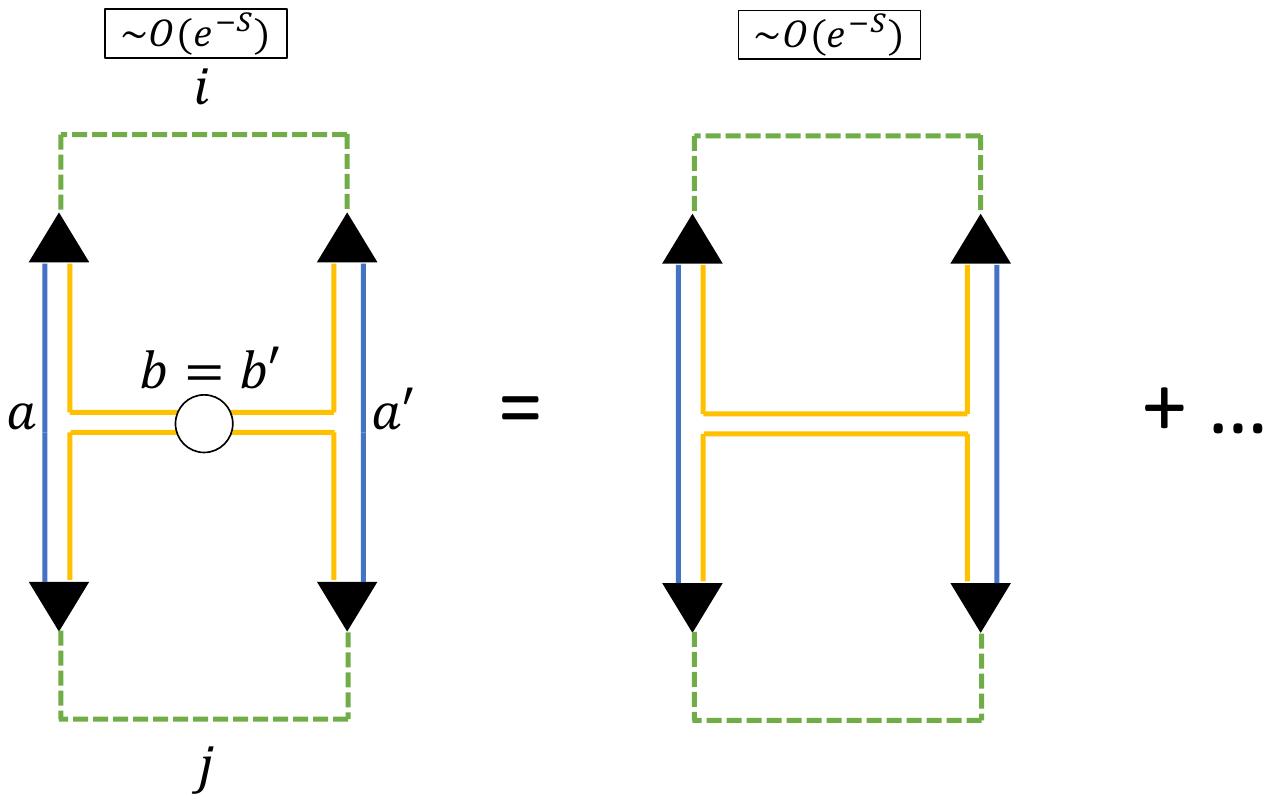}
        \caption{Partition $(bb')$ is dominated by the same partition as in figure \ref{subfig:ISij1}, but with 1 fewer sum on subsystem $B$, so the overall weight is $\mathcal{O}(e^{-S_A-S_B})=\mathcal{O}(e^{-S})$.}
        \label{subfig:ISij3}
    \end{subfigure}
    \hfill
    \begin{subfigure}{0.49\textwidth}
        \centering
        \includegraphics[width=0.9\linewidth, viewport=20 112 770 550, clip]{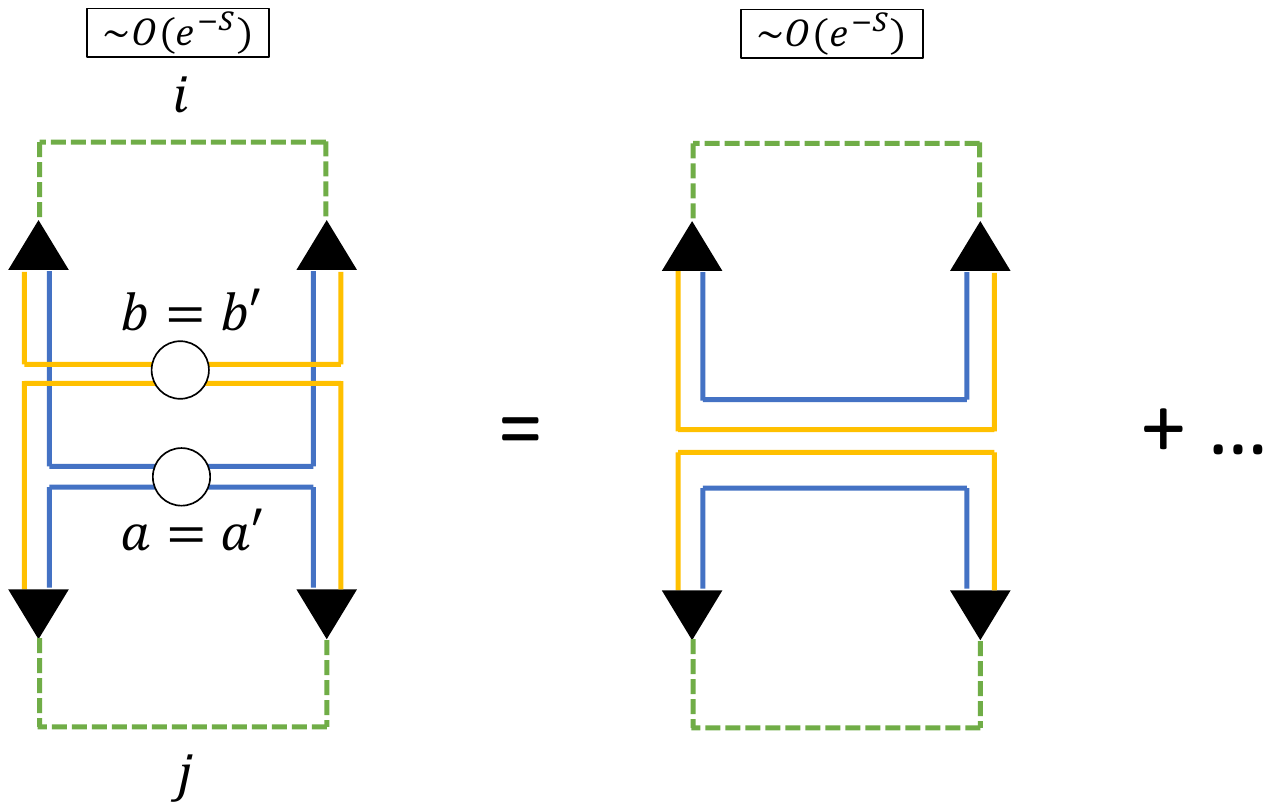}
        \caption{Partition $(aa')(bb')$ is dominated by the same partition as in figure \ref{subfig:ISij2} but with 1 fewer sum on subsystem $B$, therefore the overall weight is $\mathcal{O}(e^{-S})$.}
        \label{subfig:ISij4}
    \end{subfigure}

    \caption{The partitions of eq. (\ref{eq:ISexample}). For each diagram, leading order contributions are shown. It can be seen that the partition in figure \ref{subfig:ISij1} will share its window function with that of \ref{subfig:ISij3} and that the partition in figure \ref{subfig:ISij2} will share its window function with that of figure \ref{subfig:ISij4} while in each case having different weights.}
    \label{fig:ISij}
\end{figure}

However, we can also read the correct weights off from the full diagram directly. Restricted to subsystem $B$, the left-hand-side of figure \ref{fig:schematic} appears as a product of 2 2-loops (yellow and green lines), each of which would contribute a factor of $e^{-S_B}$. Restricted to subsystem $A$, it appears as a single 4-loop (blue and green lines) which would contribute $e^{-3S_A}$. For connected partitions that have potentially distinct behavior over different subsystems, the correct weight is acquired by combining terms of $S_A$ and $S_B$ as $S_A + S_B \rightarrow S$, leaving leftover factors of $S_{A(B)}$. In this case, we get $e^{-2S_B-3S_A}\rightarrow e^{-2S-S_A}$. 
\begin{figure}[t!]
    \centering
    \begin{subfigure}{0.49\textwidth}
        \centering
        \includegraphics[width=0.9\linewidth, viewport=20 112 770 550, clip]{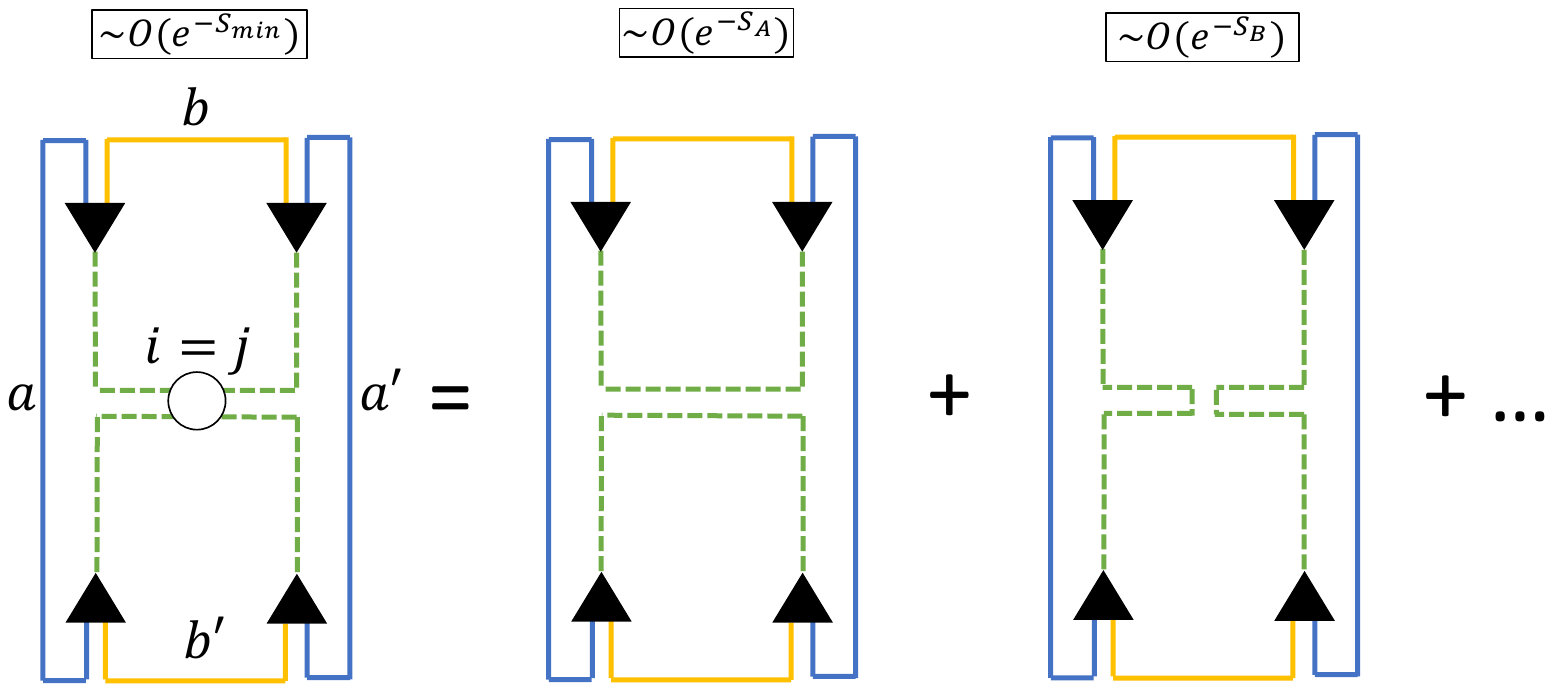}
        \caption{The partition $(ij)$ receives a contribution from two competing factors with respective weights $e^{-S_A}$ and $e^{-S_B}$. We define $S_{\text{min}}$ as  $\operatorname{min}(S_A,S_B)$ in this context.}
        \label{subfig:ISii1}
    \end{subfigure}
    \hfill
    \begin{subfigure}{0.49\textwidth}
        \centering
        \includegraphics[width=0.9\linewidth, viewport=20 112 770 550, clip]{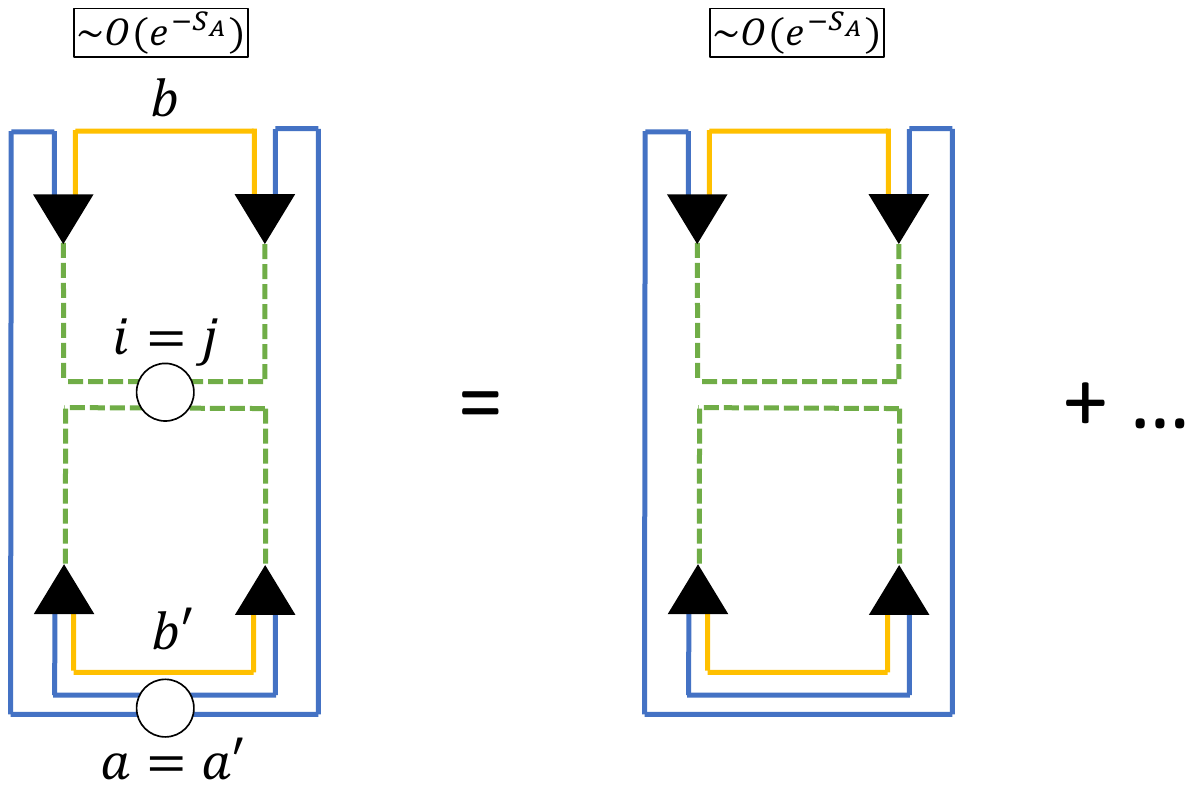}
        \caption{Partition $(ij)(aa')$ is dominated by the same term as the diagrma in figure \ref{subfig:ISij2}, which is 1 of 2 partitions which contribute to the diagram in figure \ref{subfig:ISii1}, and carries the same weight, $\mathcal{O}(e^{-S_A})$.}
        \label{subfig:ISii2}
    \end{subfigure}

    \vspace{0.1cm} 

    \begin{subfigure}{0.49\textwidth}
        \centering
        \includegraphics[width=0.9\linewidth, viewport=20 112 770 550, clip]{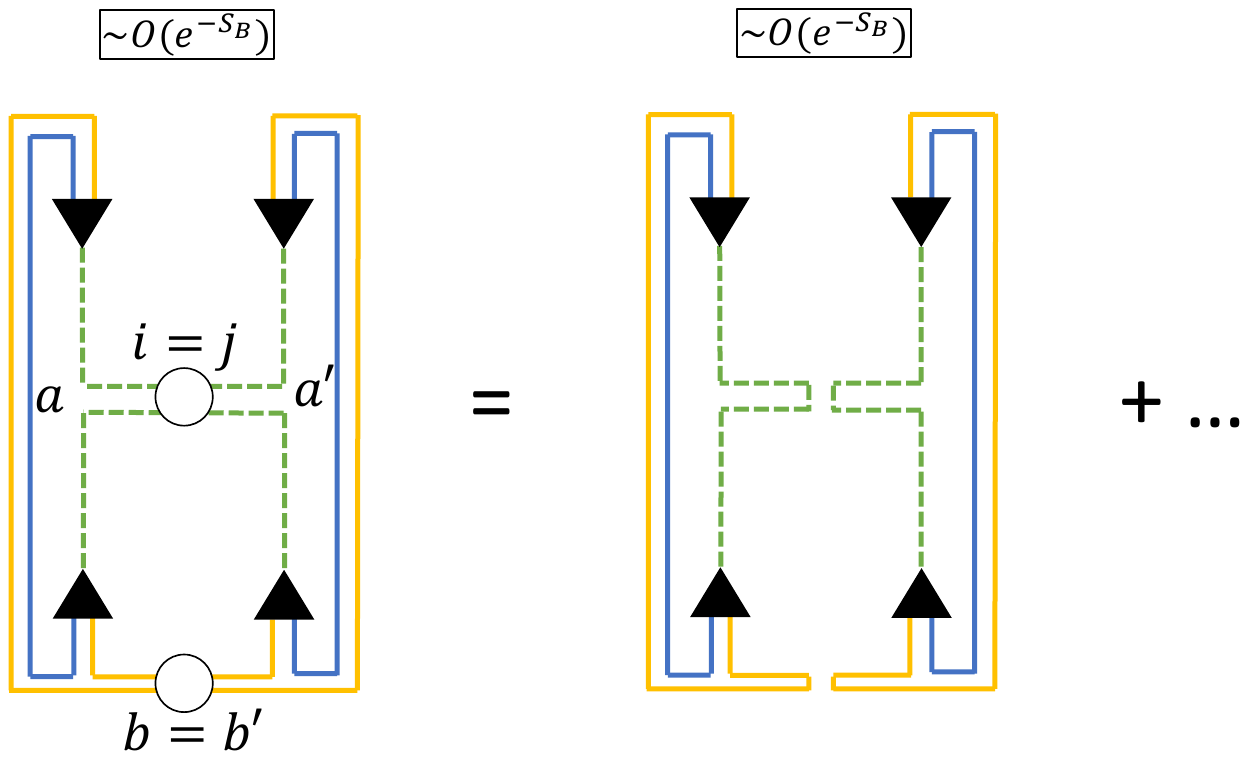}
        \caption{Partition $(ij)(bb')$ is dominated by 1 of 2 partitions that contribute to the diagram in figure \ref{subfig:ISii1} and carries weight $\mathcal{O}(e^{-S_B})$.}
        \label{subfig:ISii3}
    \end{subfigure}
    \hfill
    \begin{subfigure}{0.49\textwidth}
        \centering
        \includegraphics[width=0.9\linewidth, viewport=20 112 770 550, clip]{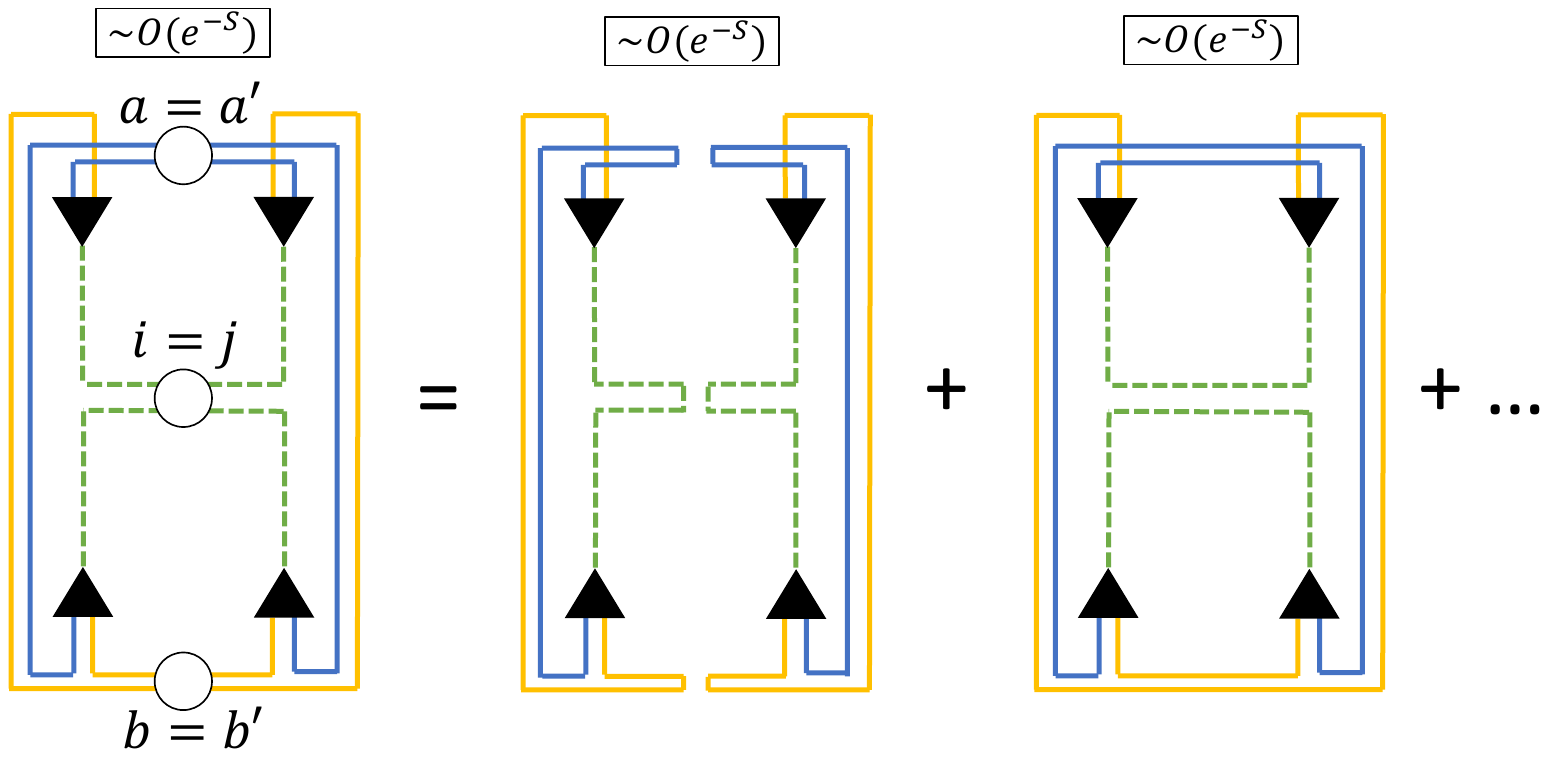}
        \caption{Partition $(ij)(aa')(bb')$ is dominated by the same partition as in figure \ref{subfig:ISij4} and one other enabled by the contraction $(ij)$ and carries overall weight $\mathcal{O}(e^{-S})$.}
        \label{subfig:ISii4}
    \end{subfigure}

    \caption{The partitions of eq. (\ref{eq:ISexample}) for $i=j$. Since there is no summation over $i$ or $j$ their contraction has not cost any factors of $e^{S}$, however, their contraction has enabled advantageous decompositions when $b=b'$ in figures \ref{subfig:ISii1}, \ref{subfig:ISii3}, and \ref{subfig:ISii4} not available in figures \ref{subfig:ISij1}, \ref{subfig:ISij3}, and \ref{subfig:ISij4}.}
    \label{fig:ISii}
\end{figure}

Next, we discuss the consequences of index contractions.  When summing over $a$, $a'$, $b$, $b'$, there will be terms where $a=a'$, or $b=b'$, or both. For $a=a'$, we can see from figure \ref{subfig:ISij2} that the diagram factorizes into two pieces, each with weight $e^{-S}$. However, for $b=b'$, we see from figure \ref{subfig:ISij3} that the contraction does not yield any advantageous factorization. Previously, in the case without subsystems, every contraction yielded an advantageous factorization. This is one way in which GFC's diverge from ordinary free cumulants. In the case of both $a=a'$ and $b=b'$, the best factorization is the same one as when $a=a'$ alone.

Now, take $i=j$, as in figure \ref{fig:ISii}. First, note that unlike the case $i\neq j$, there is a choice in how to decompose the contraction $(ij)$. This results in two possible weights from the partition, $e^{-S_A}$ or $e^{-S_B}$, depending on the smaller of $S_A$ and $S_B$ (see figure \ref{subfig:ISii1}). In other words, the contraction $(ij)$ has introduced a symmetry between the subsystems. Next, we consider again the effects of additional index contractions from $a=a'$ or $b=b'$. The contraction $(aa')$ works as before factorizing the partition into 2 simple partitions (see figure \ref{subfig:ISii2}). The contraction of $(bb')$, however, can now take advantage of the new symmetry and factor the partition into 2 simple partitions as well (see figure \ref{subfig:ISii3}). The contraction $(aa')(bb')$ also receives an additional contribution (see figure \ref{subfig:ISii4}). Thus, we have seen that index contractions on partitions with subsystems can lead to more diverse behavior than on partitions without subsystems.

An analogous technique for calculating such partitions was utilized in ref.~\cite{kudler-flam_distinguishing_2021} which treated $c^i_{ab}$ as a narrow banded Wishart matrix on its lower indices and then performed ensemble averages to calculate distinguishability measures via a version of 't Hooft double line notation\footnote{We thank Jonah Kudler-Flam for bring these calculations to our attention.}. Where both techniques apply, calculations are identical. However, our technique has the conceptual advantage of treating eigenstates of the full system on the same footing as eigenstates of the subsystem, while interpreting eigenstates as individual samples in the ensemble average. This conceptual advantage is particularly useful for introducing and discussing time-dependent physics as we do in sections~\ref{sec:thermalization} and~\ref{sec:locality}.

\subsection{Operator-eigenstate correlations}
\label{subsec:diagrams3}
\begin{figure}[t!]
    \centering
    \includegraphics[width = 0.5\linewidth, viewport=0 100 800 530, clip]{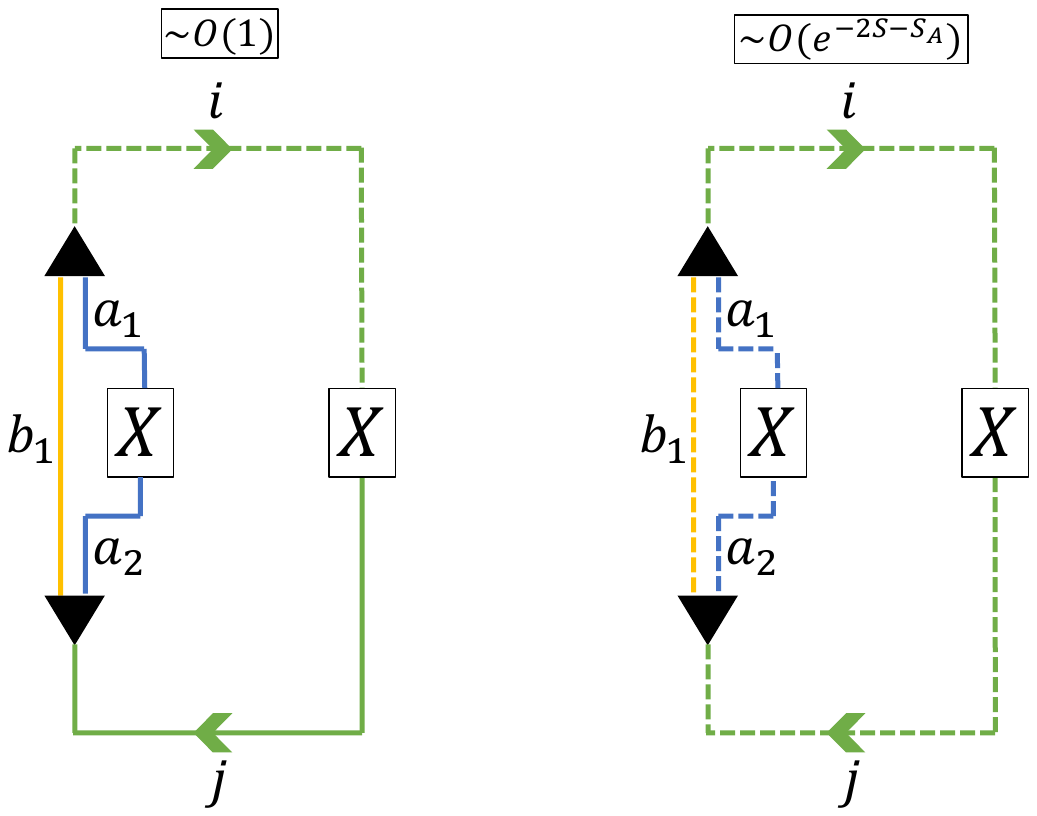}
    \caption{The summed and unsummed partitions representing eq. \eqref{eq:mixed}.}
    \label{fig:mixed}
\end{figure}

Lastly, we consider partitions which contain operators and changes-of-basis. Take an operator $X$ deep within subsystem $A$. $X$ can be understood in terms of its matrix elements in both the $H_A$ and $H$ eigenbases. We propose that either form of $X$ can be used in GFC's. For instance, consider the $2$-point correlator of $X$,
\begin{align}
\label{eq:mixed}
    \expval{X(t)X}_i = \sum_{jaa'b}X_{ij}c^j_{ab}X_{aa'}c^{a'b}_ie^{i\omega_{ij}t}.
\end{align}
We have represented the right-hand side of eq. \eqref{eq:mixed} in figure \ref{fig:mixed}. One may wonder if there is any nontrivial relationship between the matrix elements $X_{aa'}$ and $X_{ij}$. We explore such a relationship in section \ref{sec:locality}.

\subsection{Numerical evidence}
\label{subsec:numerics}
In this section, we provide some numerical evidence for our conjectured scalings via exact diagonalization of a quantum spin chain. We consider the non-integrable Ising model for $N$ spins with periodic boundary conditions, augmented by unequal local fields on the first and the last site to break translational and reflection symmetries:
\begin{align}
    H = h_{z1}\sigma^z_1+h_{zN}\sigma^z_N+\sum_{r}\left[J\sigma^z_r\sigma^z_{r+1}+h_z\sigma^z_r+h_z\sigma^x_r\right]
\end{align}
We choose the parameters $J=1.0$, $h_z=0.5$, $h_x=-1.05$, $h_{z1}=-0.45, h_{zN}=0.15$.

We split the system into two subsystems $A$ and $B$ with $N_A$ and $N_B=N-N_A$ spins, respectively, and define subsystem Hamiltonians,
\begin{align}
    &H_A = h_{zN}\sigma^z_N+\sum_{r=N_B+1}^N\left[J\sigma^z_r\sigma^z_{r+1}+h_z\sigma^z_r+h_z\sigma^x_r\right], \nonumber \\
    &H_B = h_{z1}\sigma^z_1+\sum_{r=1}^{N_B}\left[J\sigma^z_r\sigma^z_{r+1}+h_z\sigma^z_r+h_z\sigma^x_r\right].
\end{align}

\begin{figure}[t!]
  \centering
    \includegraphics[width=0.17\linewidth, viewport = 275 110 500 520, clip]{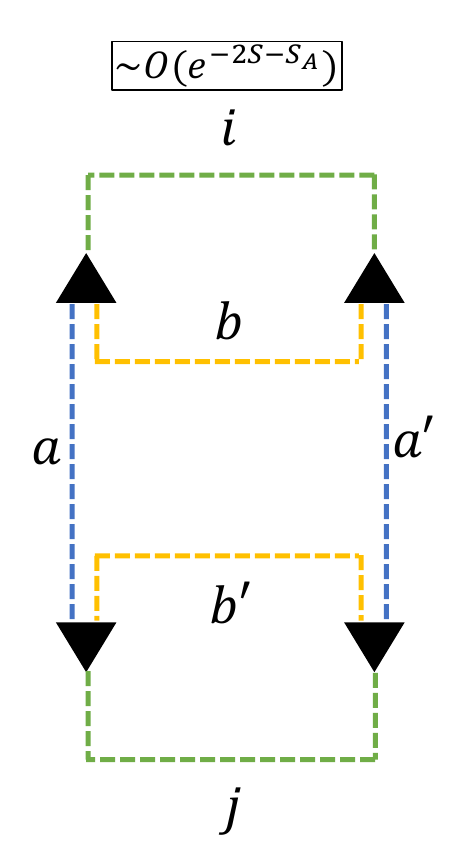}
  \caption{Diagrammatic representation of $C$ (eq. (\ref{eq:Cdef})), up to averaging. Averaging ensures that fluctuations are suppressed.}
  \label{fig:numFig}
\end{figure}

In this setup, we study the object we previously examined in figure \ref{fig:schematic} and redrawn in figure \ref{fig:numFig} for convenience. We had argued that $C$ should scale as $\mathcal{O}(e^{-2S-S_A})$ based on correlations under the implicit overline averaging. Now, we explicitly implement the overline averaging via a Gaussian weight function over a narrow energy band of width $\varepsilon$, $\Delta_\varepsilon^{(iab)} \equiv e^{-\frac{(E_i-E_a-E_b)^2}{2\varepsilon^2}}$. Thus, we compute
\begin{align}
\label{eq:Cdef}
    C\equiv\overline{c^i_{ab}c^{ab'}_{j}c^{j}_{a'b'}c^{a'b}_i} \equiv \frac{\sum_{[ijaa'bb']}\Delta_\varepsilon^{(iab)}\Delta_\varepsilon^{(jab')}\Delta_\varepsilon^{(ja'b')}c^i_{ab}c^{ab'}_{j}c^{j}_{a'b'}c^{a'b}_i}{\sum_{[ijaa'bb']}\Delta_\varepsilon^{(iab)}\Delta_\varepsilon^{(jab')}\Delta_\varepsilon^{(ja'b')}}
\end{align}

\begin{figure}[b!]
  \centering
  \begin{subfigure}{0.49\linewidth}
    \centering
    \includegraphics[width=0.75\linewidth, viewport = 245 150 550 450, clip]{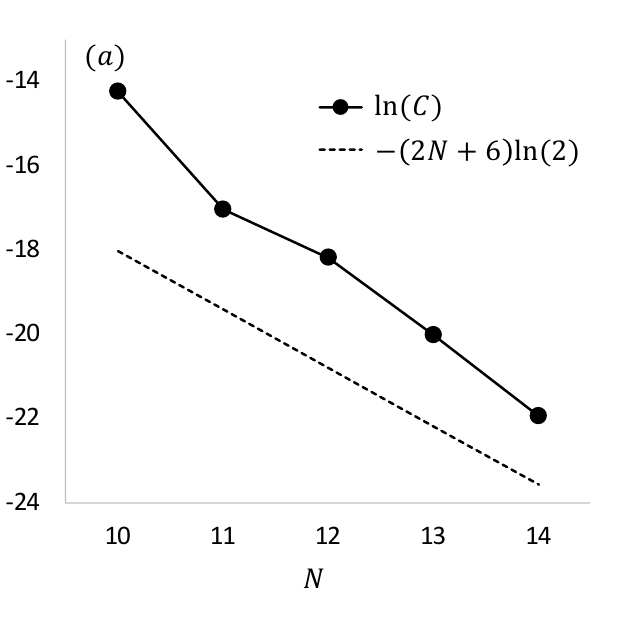}
  \end{subfigure}
  \hfill
  \begin{subfigure}{0.49\linewidth}
    \centering
    \includegraphics[width=0.75\linewidth, viewport = 245 150 550 450, clip]{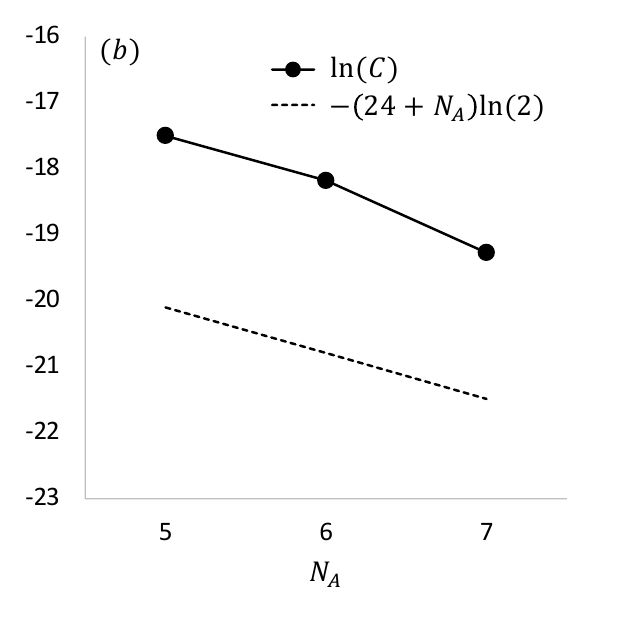}
  \end{subfigure}

  \caption{Scaling of $C$ (eq. (\ref{eq:Cdef})) vs. (a) $N$ for fixed $N_A=6$, and (b) $N_A$ for fixed $N=12$. We compare are numerical calculations for an Ising spin chain (solid dots) to the asymptotic scaling associated with the Hilbert space dimensions (dashed line). We find qualitatively good agreement in both figures (a) and (b) for the slope of the curve.}
  \label{fig:data}
\end{figure}

Our results are presented in figure \ref{fig:data}. We check the scaling with respect to $N$ by fixing $N_A=6$ and varying $N$ from 10 to 14. We also check the scaling of $N_A$ by fixing $N=12$ and varying $N_A$ from 5 to 7. In all cases, we find $\varepsilon = 0.4$ to be adequate. We are significantly limited by finite size corrections associated with the subsystem sizes which require us to discard data points for $N_A,\, N_B < 5$. Nonetheless, we find good agreement between the slopes of $\ln(d^{-2}d_A^{-1}) = (-2N-N_A)\ln(2)$  and $\ln(C)$ for accessible system sizes, which supports our analytical expectation of the scaling of $C$ with $N$ and $N_A$.

\section{Eigenstate correlations I: the structure of chaotic eigenstates}
\label{sec:equilibrium}
In this section, we consider an eigenstate $\ket{i}$ of the full system Hamiltonian $H$. We focus on two aspects of thermalization: the reduced density matrix and entanglement entropy. In section \ref{subsec:RDMEq}, we compute the on and off-diagonal elements of the reduced density matrix on subsystem A, improving the calculation of ref.~\cite{murthy_structure_2019} and generalizing that of ref.~\cite{de_boer_principle_2023}. In section \ref{subsec:PageEq}, we reproduce best-known results for the entanglement entropy of chaotic eigenstates from our formalism and discuss subextensive corrections. In doing so we remark on a qualitative resemblance of our calculation to the gravitational path integral calculation of the entanglement entropy of an evaporating black hole that is inherently free probabilistic~\cite{penington_replica_2022, liu_entanglement_2021, wang_beyond_2023}. Lastly, in section \ref{subsec:SETH}, we show that reduced density matrices of nearby eigenstates are exponentially close in trace distance, a hypothesis known as the subsystem ETH~\cite{dymarsky_subsystem_2018}.

\subsection{Reduced density matrix}
\label{subsec:RDMEq}

\begin{figure}[htb]
    \centering
    \includegraphics[width = 0.75\linewidth, viewport=0 180 800 470, clip]{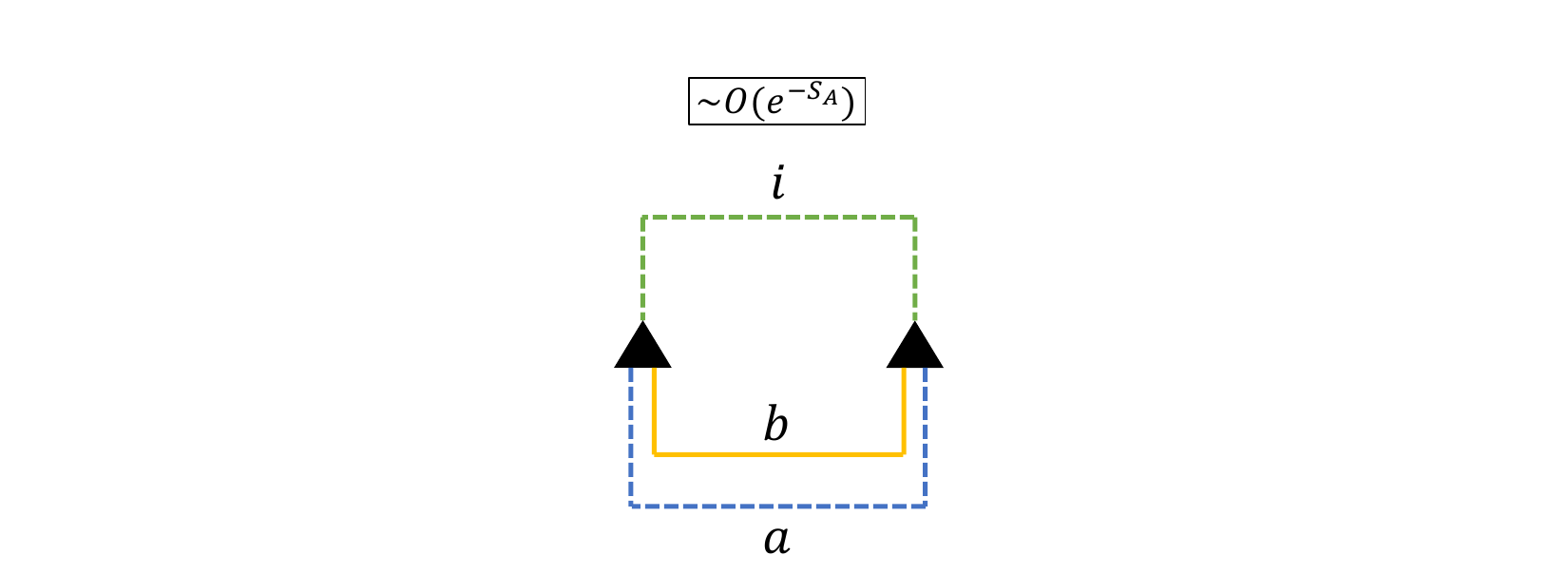}
    \caption{The diagonal elements of a reduced density matrix are represented by the simplest eigenstate cumulant for a bipartitioned system.}
    \label{fig:diagonal}
\end{figure}

We wish to study the matrix elements of the reduced density matrix of subsystem $A$,
\begin{equation}
    \rho^i_{aa'} = \bra{a}\!\operatorname{Tr}_{B}\left[\ket{i}\!\!\bra{i}\right]\!\ket{a'}=\sum_{b}c^{ab}_ic^{i}_{a'b}.
\end{equation}
Focusing on the diagonal elements first, we find a simple expression,
\begin{align}
\label{eq:diageq}
    \rho^i_{aa}=\sum_{b}c^{ab}_ic^{i}_{ab} = \int_{E_b}e^{S_B(E_b)-S(E_i)}F(E_i-E_a-E_b) = e^{-S(E_i)+S_B(E_i-E_a)}
\end{align}
which reduces to a Gibbs state when $A$ is much smaller than $B$. This term is represented in figure \ref{fig:diagonal}. We evaluated eq. \eqref{eq:diageq} at the trivial saddle-point $E_i-E_a-E_b=0$ associated with neglecting the finite width of $F$. If we neglect the width of window functions we are able to take the arguments of $F$ functions (other than the overall energy) as saddle-points. However, including a nonzero width of $F$ will slightly modify eq.~\eqref{eq:diageq} and the outcomes of the various saddle-point integrals we take in this paper. However, such a modification will only be felt by the fluctuations of approximately conserved operators (i.e. the fluctuations of subsystem energies) and only when subsystem $A$ is a finite fraction of the whole system (see appendix~\ref{subapp:subvariances}).

\begin{figure}[htb]
    \centering
    \begin{subfigure}{0.49\textwidth}
        \centering
        \includegraphics[width=\linewidth, viewport=0 100 800 520, clip]{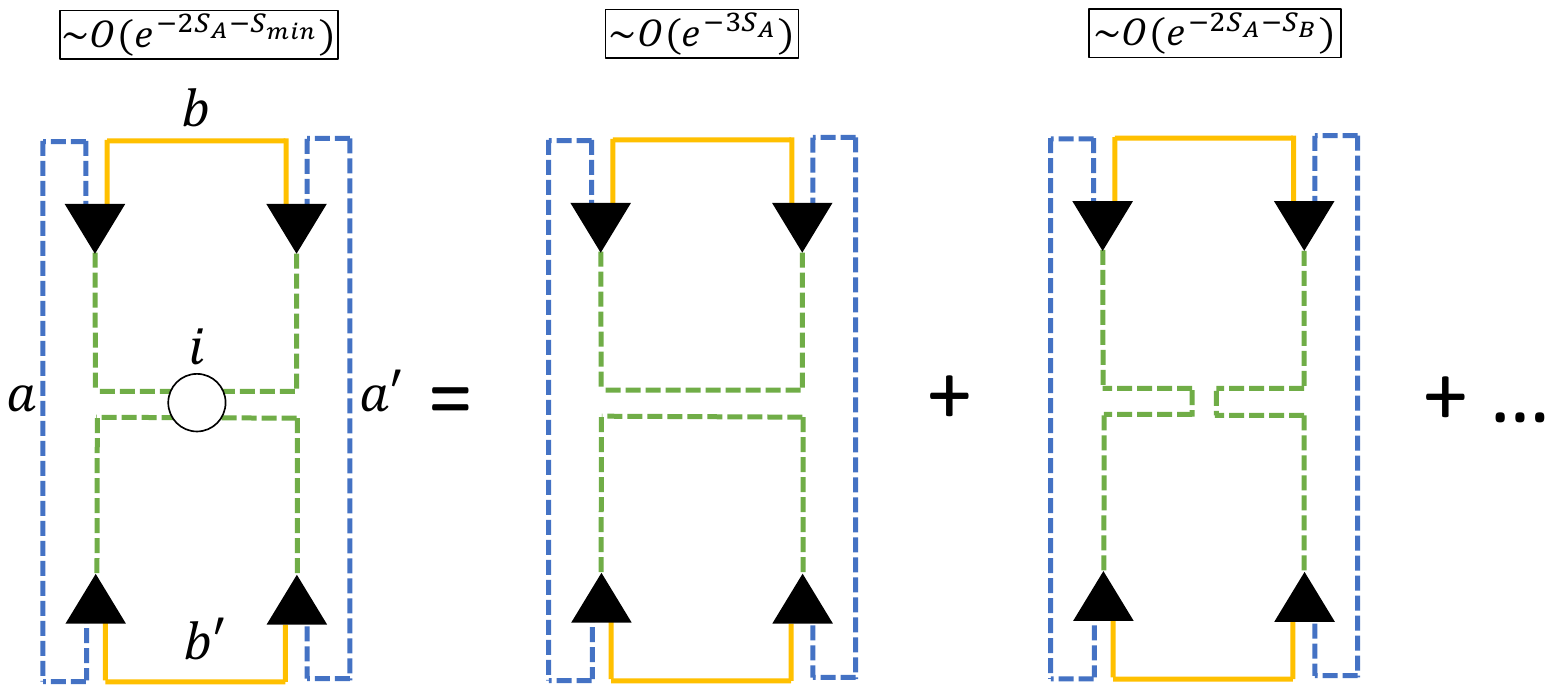}
        \caption{A true 4-point contribution which captures sharp behavior in system size when $S_{A}\sim S_B$. Note the similarity to figure \ref{subfig:ISii1}.}
        \label{subfig:offDiagonal4}
    \end{subfigure}
    \hfill
    \begin{subfigure}{0.49\textwidth}
        \centering
        \includegraphics[width=\linewidth, viewport=0 80 800 500, clip]{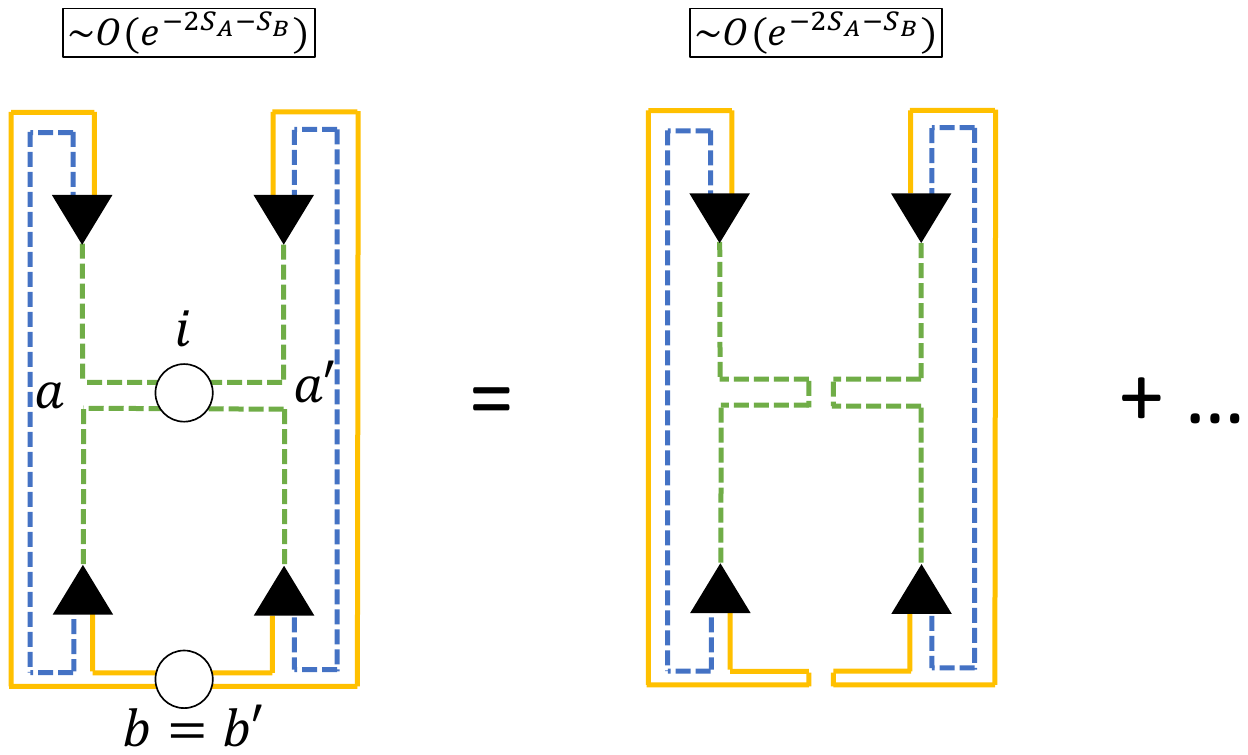}
        \caption{Contribution that factors into 2-point functions and only contributes at leading order when $S_A\gtrsim S_B$. Note the similarity to figure \ref{subfig:ISii3}.}
        \label{subfig:offDiagonal2}
    \end{subfigure}

    \caption{Diagrams which contribute to the off-diagonal matrix elements of reduced density matrices.}
    \label{fig:offDiagonal}
\end{figure}

For off-diagonal elements, we can immediately see that $\rho^i_{aa'}$ contains uncontracted indices and must have mean zero. We study instead, the variance, 
\begin{align}
\label{eq:offdiageq1}
    \rho^i_{aa'}\rho^i_{a'a}=\sum_{bb'}c^{ab}_ic^{i}_{a'b}c^{a'b'}_ic^{i}_{ab'}.
\end{align}
We represent eq. \eqref{eq:offdiageq1} in figure \ref{fig:offDiagonal}. Notably, the leading order contribution to eq. \eqref{eq:offdiageq1} depends on the smaller of $S_A(E_a)$ and $S_B(E_i-E_a)$. We define $S_{\text{min}}(E_a, E_b) \equiv \operatorname{min}\left[S_A(E_a),S_B(E_b)\right]$ and $S_{\text{min}}(E_a) \equiv \operatorname{min}\left[S_A(E_a),S_B(E_i-E_a)\right]$. Then, we can read off leading order terms from figure \ref{fig:offDiagonal} and evaluate via saddle-points,
\begin{align}
\label{eq:offdiageq2} 
    |\rho^i_{aa'}|^2 &= \sum_{[bb']}c^{ab}_ic^{i}_{a'b}c^{a'b'}_ic^{i}_{ab'} + \sum_{b}c^{ab}_ic^{i}_{a'b}c^{a'b}_ic^{i}_{ab} \nonumber \\
    &=\int_{E_bE_{b'}}e^{2S_B(E_{bb'})-2S(E)-S_{\text{min}}(E_{aa'}, E_{bb'})}F(\cdots)+\int_{E_b}e^{S_B(E_{bb'})-2S(E)}F_{(bb')}(\cdots) \nonumber \\
    &= e^{-2S(E)+2S_B(E_i-E_{aa'})-S_{\text{min}}(E_{aa'})}\tilde F(E_{aa'};\omega_{aa'})
\end{align}
where $\tilde F(E_{aa'};\omega_{aa'})$ has been defined to capture a leftover $\omega_{aa'}$ dependence that suppresses correlations away from the diagonal. Combining eqs. \eqref{eq:offdiageq2} and \eqref{eq:diageq}, we obtain for the reduced density matrix, 
\begin{align}
\label{eq:RDMEq}
    \rho^i_{aa'} = e^{-S(E_i)+S_B(E_i-E_{aa'})}\left(\delta_{aa'}+\sqrt{e^{-S_{\text{min}}(E_{aa'})}\tilde F(E_{aa'};\omega_{aa'})}R_{aa'}\right)
\end{align}
where $R_{aa'}$ is an approximate Gaussian random matrix with mean zero and variance one that encodes higher correlations in the reduced density matrix. Eq. \eqref{eq:RDMEq} is a generalization of the {state-averaging ansatz} of ref.~\cite{de_boer_principle_2023} to the case of arbitrary subsystem sizes. Ref.~\cite{de_boer_principle_2023} further discusses higher correlations in $R_{aa'}$ which we discuss implicitly in the next section in the context of entanglement entropy.

Eq. \eqref{eq:RDMEq} also improves the result of ref.~\cite{murthy_structure_2019} which neglected the terms in figure \ref{subfig:offDiagonal4} while keeping the term in figure \ref{subfig:offDiagonal2}. As a result, instead of our factor of $S_{\text{min}}$, they had a factor of $S_B(E_i-E_{aa'})$ in our notation. Our improvement implies a much smaller suppression of off-diagonal elements of the reduced density matrix below the critical energy. This improvement is physically necessary on the following grounds. The suppression by $e^{-S_B}$ would imply that the eigenstates of $H_A$ are exponentially close to the eigenstates of $\rho^i_A$ via $\bra{a}\!\left(\rho^i_A\right)^2\!\ket{a} = (\rho^i_{aa})^2\left(1 + \mathcal{O}(e^{-S_B})\right)$. However, $H_A$ is only defined up to an arbitrary area scaling term on the boundary of the subsystem and any two definitions of $H_A$ would likely share no eigenstates. In contrast, $\rho^i_A$ is defined unambiguously and cannot have eigenstates that are simultaneously exponentially close to eigenstates of all definitions of $H_A$. Another way of seeing the same problem is to consider preparing the system in an eigenstate of the total Hamiltonian $H$ and then switching off the interaction $H_{AB}$ thereby perturbing subsystem $A$ along its boundary. If the off-diagonal elements of $\rho^i_A$ are suppressed by $e^{-S_B}$, the timescale for diagonal elements of $\rho^i_A$ to evolve is exponentially long $\sim\mathcal O(e^{S_B-S_A})$. Physically, however, we should only expect the timescale to be the time it takes for information from the boundary of $A$ to reach the rest of the subsystem.

\subsection{Entanglement entropies and the Page curve}
\label{subsec:PageEq}
\begin{figure}
    \centering
    \includegraphics[width = 0.75\linewidth, viewport=0 180 800 480, clip]{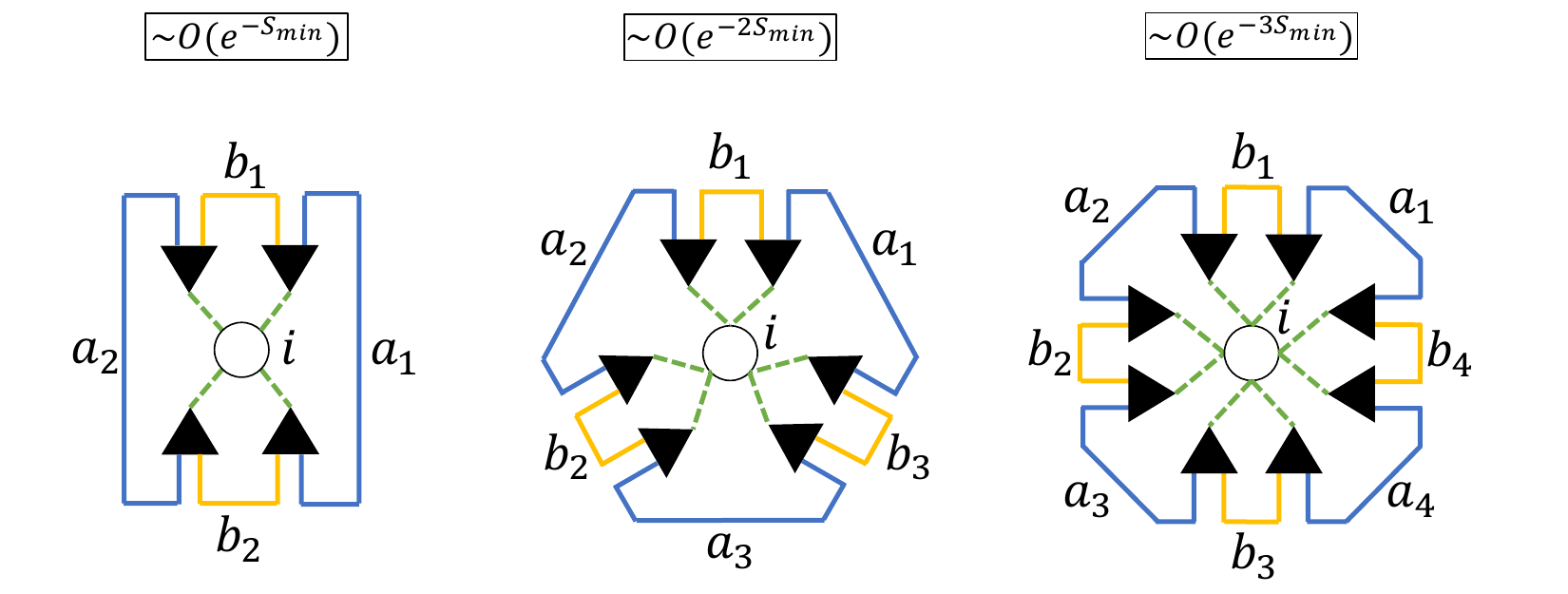}
    \caption{Diagrams which compute the second, third, and fourth moments of the reduced density matrix contained inside the log in eq. \eqref{eq:RenyiEq}. For general Renyi index $\alpha$, the diagram will appear as an $\alpha$-fold rotationally symmetric version of the diagrams above.}
    \label{fig:RenyiEq}
\end{figure}

\begin{figure}
    \centering
    \includegraphics[width = 1.0\linewidth, viewport=0 150 800 470, clip]{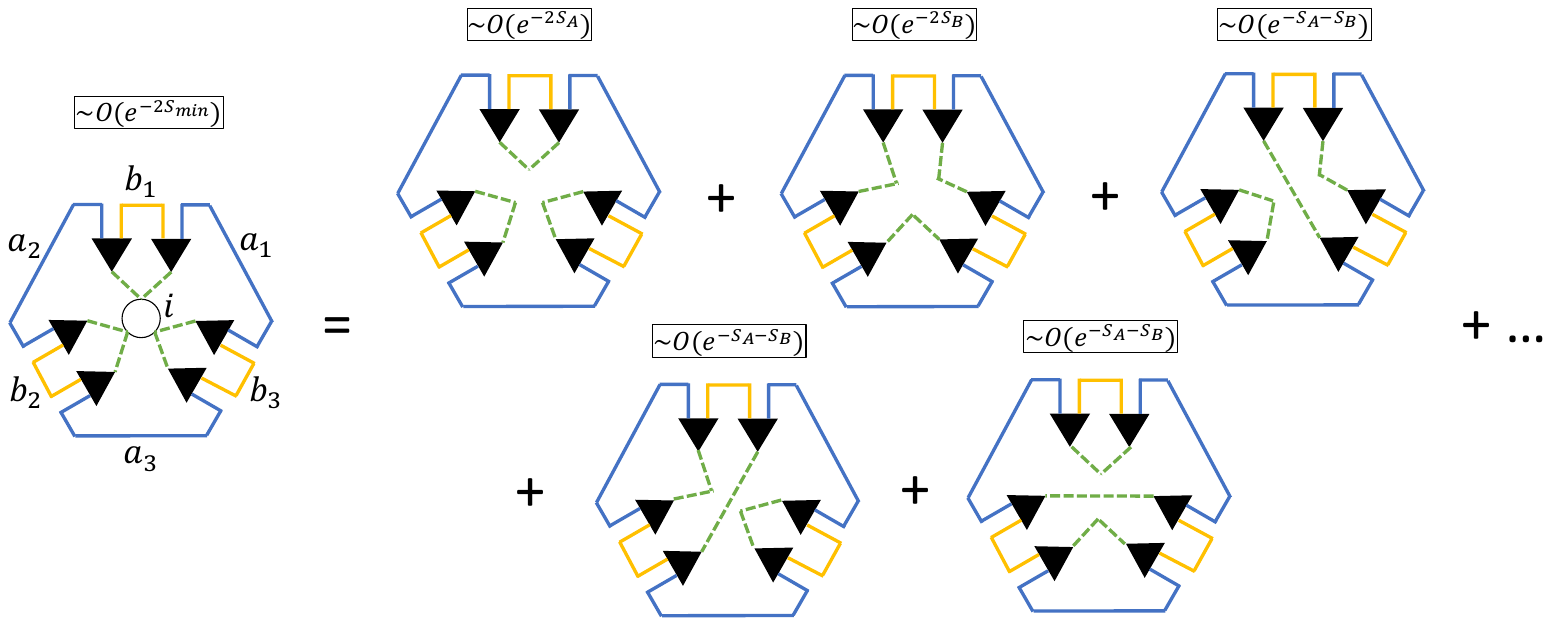}
    \caption{Pictured are the 5 five leading order contributions to eq. \eqref{eq:RenyiEq} for $\alpha = 3$, which come from the non-crossing pair partitions of the $i$ index. There is be a term associated to each pair partition of $2\alpha$ $i$ indices. The non-crossing condition is not universally true of index contractions in index diagrams but are a consequence of the ordinary rules presented in section \ref{subsec:pH} applied to the Renyi entropy diagrams, figure \ref{fig:RenyiEq}. We focus only on the first 2 terms as the remaining 3 only contribute near the isolated point $S_A(E_a)=S_B(E_b)$. Not drawn are contractions of $a$ and $b$ indices that will also contribute near the critical energy so long as they do not create any additional crossings.}
    \label{fig:PlanarEq}
\end{figure}

To compute the entanglement entropies we first focus on the $\alpha$-Renyi entropies for $\alpha \geq 2$,
\begin{align}
\label{eq:RenyiEq}
    S_\alpha \equiv \frac{1}{1-\alpha}\ln\left(\operatorname{Tr}_A\left[(\rho^i_A)^\alpha\right]\right) = \frac{1}{1-\alpha}\ln(\sum_{\{ab\}}c^{i}_{a_1b_1}c^{a_2b_1}_{i}\cdots c^{i}_{a_\alpha b_\alpha}c^{a_1b_\alpha}_i).
\end{align}
The diagrammatics for eq. (\ref{eq:RenyiEq}) are represented in figure \ref{fig:RenyiEq} for $\alpha = 2,\, 3,\, 4$. To leading order, we can ignore additional contractions in the sum and focus only on the full diagrams. The leading order terms will come from different ways of decomposing the contraction of $i$ indices into pairs. When subsystem $A(B)$ is far smaller than its complement, there is a single leading partition with weight $e^{-\alpha S(E_i)-(\alpha-1)S_{A(B)}(E_{a(b)})}$. However, when the subsystems are similar in size, there will be a critical energy $E^*_{A(B)}$ such that $S_{A(B)}(E_{A(B)}^*) = S_{B(A)}(E_i-E_{A(B)}^*)$ which denotes a crossover between leading partitions. Near $E_{A(B)}^*$, there will also be contributions from all other \textit{non-crossing} pairings of $i$ indices, and we draw examples of these partitions in figure \ref{fig:PlanarEq}. However, these partitions will, at best, contribute an area law term to entanglement entropy for $\alpha < 1$ and will be neglected other than to match terms to the time-dependent case we consider in section~\ref{sec:thermalization}.

Then we can read off for general $\alpha$,
\begin{align}
\label{eq:RenyiEqInt}
    S_\alpha = \frac{1}{1-\alpha}\ln(\int_{E_aE_b\{\omega\}}e^{\alpha S_A(E_a)+\alpha S_B(E_b)-\alpha S(E_i) - (\alpha-1)S_{\text{min}}(E_a,E_b)}F(\cdots))
\end{align}
and taking $\alpha\rightarrow 1$,
\begin{align}
\label{eq:VNEqInt}
    S_{1} = \int_{E_aE_b}e^{S_A(E_a)+S_B(E_b)-S(E_i)}S_{\text{min}}(E_a,E_b)F(E_i-E_a-E_b).
\end{align}
Eqs. (\ref{eq:RenyiEqInt}) and (\ref{eq:VNEqInt}) are equivalent to expressions derived in ref.~\cite{murthy_structure_2019} and our results are equivalent as well. In taking the $\alpha \rightarrow 1$ limit, we have neglected the implicit dependence of $F$ on $\alpha$. This neglect is valid as long as we also neglect the non-zero width of $F$.

We can then evaluate the integrals via saddle-points. The first saddle-point condition is given by $F$ which will enforce that $E_a+E_b=E_i$. The second saddle-point condition is given by maximizing the exponents in eqs. (\ref{eq:RenyiEqInt}) and (\ref{eq:VNEqInt}). For $\alpha\neq1$,
\begin{align}
\label{eq:RenyiEqResult}
    S_\alpha = \frac{1}{1-\alpha}\left\{\alpha\left[S_A(\mathcal{E
}_A)+S_B(\mathcal{E}_B)-S(E_i)\right]-(\alpha-1)S_{\text{min}}(\mathcal{E}_A,\mathcal{E}_B)\right\}
\end{align}
where $\mathcal{E}_{A}$ and $\mathcal{E}_B$ are the subsystem energies that dominantly contribute to the Renyi entropies and are defined by the saddle-point conditions of eq. (\ref{eq:RenyiEqInt}): 
\begin{align}
\label{eq:saddlePoints}
    &\text{(I)}\quad \mathcal{E}_{A}+\mathcal{E}_{B}=E_i\quad \& \quad
    \begin{cases}
      \text{(II.1)} \quad S_A'(\mathcal{E}_A) = \alpha S_B'(\mathcal{E}_B)\\
      \text{(II.2)} \quad S_B'(\mathcal{E}_B) = \alpha S_A'(\mathcal{E}_A)\\
      \text{(II.3)} \quad S_A(\mathcal{E}_A)=S_B(\mathcal{E}_B)
    \end{cases} 
\end{align}
Here, condition {(II)} is determined by which of the three given saddle-points minimizes $S_\alpha$. For $\alpha > 1$, the saddle point is given by condition {(II.1)} if $S_A < S_B$ and condition {(II.2)} if $S_A > S_B$. For $\alpha < 1$, the same result holds except at high temperatures and when the systems are of similar size, where condition {(II.3)} can hold, in which case $\mathcal E_A = E^*_A$. For $\alpha = 1$, we derive
\begin{align}
\label{eq:VNEqResult}
    S_1 = \operatorname{min}\left[S_A(E_{i,A}), S_B(E_{i,B})\right] + \Delta S
\end{align}
where the saddle-point conditions of \eqref{eq:VNEqInt} are given by,
\begin{align}
    \label{eq:saddlePoint}
    \text{(I)}\quad E_{i,A}+E_{i,B}=E_i \quad \& \quad \text{(II)} \quad S_A'(E_{i,A}) = S_B'(E_{i,B})
\end{align}
and $\Delta S < 0$ denotes subextensive corrections to $S_1$ we will discuss shortly.

First, we note that eq. (\ref{eq:VNEqResult}) recovers a finite temperature version of the Page curve, noted for its linear slopes as a function of subsystem size~\cite{page_average_1993}. However, for $\alpha \neq 1$, eq. (\ref{eq:RenyiEqResult}) does not obtain this form. In particular, $S_{\alpha>1}$ is superadditive, and $S_{\alpha<1}$ is subadditive and thus are convex and concave functions of system size, respectively. This convexity/concavity is referred to as the ``failure of the Page curve''~\cite{lu_renyi_2019}. We can interpret this failure in the following manner. The subsystems of a chaotic eigenstate $\ket{i}$ obtain an effective thermodynamic inverse temperature from $S(E_i)$, $\beta \equiv S'(E_i)$, which determines all aspects of local physics in finite energy density states. However, for $\alpha\neq 1$, Renyi entropies of subsystems that are a finite fraction of the whole system have access to highly \textit{nonlocal} information that encodes physics at different temperatures \cite{garrison_does_2018}. Thus, as the size of a small subsystem is increased, the Renyi entropies gain access to more information from different parts of the spectrum and bend accordingly. For $\alpha > 1$ ($\alpha < 1$), the Renyi entropies are dominated by low (high) temperature physics and as $\alpha\to\infty$ ($\alpha\to 0$) the entropy will approach that of the ground (infinite-temperature) state.

We now discuss the corrections to the Von Neumann entropy, $\Delta S$. One generally expects area law corrections to entanglement entropy. To acquire them, one would need to carefully consider the structure of the $F$ functions in the $\alpha \rightarrow 1$ limit. As the $F$ functions are generally system-dependent, so are the area law corrections, but we do not preclude the possibility that some generic structure may exist. There is another correction from contributions away from the saddle point in eq. (\ref{eq:VNEqInt}). These contributions are controlled by the heat capacities of the subsystems and, in homogeneous systems, contribute a correction to entanglement entropy that is order square root in system size. This term is derived in ref.~\cite{murthy_structure_2019}, and we direct readers to their calculation rather than repeat it. This is the leading order correction in $D<2$ dimensional systems, whereas the area law correction generally is leading order in $D>2$ dimensions. For $D=2$ dimensions both terms have equal order.

Despite the fact that previous studies of eigenstate entanglement~\cite{dymarsky_subsystem_2018, huang_universal_2019, huang_universal_2021, lu_renyi_2019, murthy_structure_2019} did not have a general way to compute higher eigenstate correlations, refs.~\cite{lu_renyi_2019, murthy_structure_2019} still successfully computed the Renyi and Von Neumann entropies\footnote{aside from the fact that~\cite{lu_renyi_2019} missed the possible saddle-point condition (II.3) in \eqref{eq:saddlePoints} and neglected subleading corrections}. Ref.~\cite{lu_renyi_2019} utilized similar assumptions to our own and computed the moments of the reduced density matrix averaged over all states consistent with the form in eq. \eqref{eq:EB}. Ref.~\cite{murthy_structure_2019}, on the other hand, computed entanglement entropies by guessing a distribution of eigenvalues of the reduced density matrix. Since the moments of the reduced density matrix are entirely determined by its eigenvalues, it is not necessarily surprising that this calculation could be done without direct computation of higher correlations.

The non-crossing constraint depicted in figure \ref{fig:PlanarEq} is similar to one found in the gravitational path integral computation for the Page curve of an evaporating black hole in JT gravity and discussed in refs.~\cite{liu_entanglement_2021, penington_replica_2022, wang_beyond_2023}. A connection was drawn to the non-crossing partitions of free probability theory in ref.~\cite{penington_replica_2022} and explicated in ref.~\cite{wang_beyond_2023} in terms of so-called Kreweras complements and free multiplicative convolution. To make the analogy precise, we model subsystem $B$ as an evaporating black hole and subsystem $A$ as its radiation. At early times, when $B$ is much larger than $A$, the Page curve is dominated by the first partition on the right-hand side of figure \ref{fig:PlanarEq} where pairings of $i$-indices disconnect the replicas of $B$. At late times, when most of $B$ has evaporated, the Page curve is dominated by the second partition where pairings of $i$-indices connect all replicas of $B$ together. Where in our calculation arise simple index contractions, the gravitational path integral predicts semiclassical wormholes that (dis)connect the black hole replicas (contrast our figures \ref{fig:RenyiEq} and \ref{fig:PlanarEq} with figure 2 of ref.~\cite{wang_beyond_2023}). Ref. ~\cite{de_boer_page_2023} finds a similar analogy between index contractions and replica wormholes in a toy model for entanglement dynamics.

The common thread between all calculations is an implicit ensemble averaging over correlations of wavefunction overlaps that connect originally disconnected replicas and suppress crossings by an appropriate density-of-states factor. The validity of such an averaged calculation in systems without explicit ensemble averaging is referred to as the \textit{factorization problem}~\cite{de_boer_page_2023}. The ETH may provide a solution. In our calculation, the ensemble average emerged as the smooth portion of the GFCs that dominate over fluctuating portions without any explicit ensemble average. Similar logic is summoned in ref.~\cite{de_boer_principle_2023} to justify the appearance of semiclassical wormholes in evaluating gravitational path integrals. 

\subsection{Subsystem ETH}
\label{subsec:SETH}
The expression for the density matrix given in eq. \eqref{eq:RDMEq} holds up to polynomial corrections in the system size. In contrast, ref.~\cite{dymarsky_subsystem_2018}, conjectured that reduced density matrices for a given subsystem of nearby\footnote{defined as having a difference in energy at the order of the level spacing} eigenstates of the full system should be \textit{exponentially} close in trace distance when the subsystem is smaller than one-half the system. More precisely,
\begin{align}
    \label{eq:SubETH}
    ||\rho^i_A-\rho^{j}_A||_1 \equiv \operatorname{Tr}_A\left[|\rho^i_A-\rho^{j}_A|\right] \sim \mathcal{O}(e^{-S/2}).
\end{align}
for $E_i = E_j$ up to irrelevant corrections of order the level spacing. The significance of this definition is that it implies the existence of an equilibrium reduced density matrix, $\rho^{\text{ETH}}_{A}$ that specifies the thermal properties of subsystem $A$ to a far greater degree than that of the canonical ensemble. 

Our diagrammatic formalism does not directly compute the Schatten 1-norm $||\cdots||_1$. However, we can compute the Schatten 2-norm and utilize the following bound\footnote{It was brought to our attention that under the same assumptions a replica calculation of fidelities yields a stronger bound on the trace distance. This calculation was performed in ref.~\cite{kudler-flam_distinguishing_2021} (eq. (197)).},
\begin{align}
\label{eq:CauchySchwarz1}
    ||M||_2 \leq ||M||_1 \leq \sqrt{\operatorname{rank}(M)}||M||_2
\end{align}
which holds for an arbitrary operator $M$ and is a corollary of the Cauchy--Schwarz inequality. First, let us expand, 
\begin{align}
\label{eq:diagramdiff}
    ||\rho^i_A-\rho^{j}_A||_2^2\equiv\operatorname{Tr}_A\left[\left(\rho^i_A-\rho^j_A\right)^2 \right] &= -2\operatorname{Tr}_A\left[\rho^i_A\rho^j_A \right] + \operatorname{Tr}_A\left[(\rho^i_A)^2 \right] +\operatorname{Tr}_A\left[(\rho^j_A)^2 \right] \nonumber \\
    &\approx 2\left(\operatorname{Tr}_A\left[(\rho^i_A)^2 \right] - \operatorname{Tr}_A\left[\rho^i_A\rho^j_A \right]\right) \nonumber \\ 
    &\equiv 2\left(\big(\rho^{i}_A\big|\rho^i_A\big) - \big(\rho^{i}_A\big|\rho^j_A\big)\right) 
\end{align}
where we the error in the approximation $\operatorname{Tr}_A\left[(\rho^i_A)^2 \right] \approx \operatorname{Tr}_A\left[(\rho^j_A)^2\right]$ is suppressed by $e^{-S}$ when $\ket{i}$ and $\ket{j}$ are nearby eigenstates\footnote{the given suppression can be justified by recognizing the gradual dependence of the 2-Renyi entropy on total energy and its small fluctuations between nearby eigenstates}. 

We can now recognize that the diagrammatics of $\big(\rho^{i}_A\big|\rho^i_A\big)$ and $\big(\rho^{i}_A\big|\rho^j_A\big)$ are given in figures \ref{fig:ISii} and \ref{fig:ISij}, respectively. By inspection, the leading order difference comes from the terms in figures \ref{subfig:ISii1} and \ref{subfig:ISii3} that are absent in \ref{subfig:ISij1} and \ref{subfig:ISij3}, where the contraction of index $i$ allowed diagrams of order $e^{-S_B}$ to contribute. Thus,
\begin{align}
    \big(\rho^{i}_A\big|\rho^i_A\big) - \big(\rho^{i}_A\big|\rho^j_A\big) &= \int_{E_bE_{b'}E_aE_{a'}}e^{2S_A(E_{aa'})+S_B(E_{bb'})-2S(E_{i})}F_{(ij)}(\cdots) \nonumber \\ &\quad +\int_{E_bE_aE_{a'}}e^{2S_A(E_{aa'})+S_B(E_b)-2S(E_{i})}F_{(ij)(bb')}(\cdots)  \nonumber \\ &\sim e^{-S(E_i)+S_A(E_{iA})}.
\end{align}
Lastly, recognizing that $\ln[\text{rank}(\rho_A)]$ is the $0$-Renyi entropy, per eq. (\ref{eq:RenyiEqResult}), we can recognize that the reduced density matrices are full rank (up to a polynomial correction in subsystem sizes): $\operatorname{rank}(\rho^{i(j)}_A) \sim e^{S^{(\infty)}_{\text{min}}}$, where $e^{S^{(\infty)}_{\text{min}}}$ is the microcanonical entropy of the smaller subsystem at infinite temperature. Then, ${\operatorname{rank}(\rho^i_A-\rho^{j}_A)}\lesssim e^{S^{(\infty)}_{\text{min}}}$ and eq. (\ref{eq:CauchySchwarz1}) becomes
\begin{align}
\label{eq:SETHbound}
\mathcal{O}(e^{S_A/2-S/2})\lesssim ||\rho^i_A-\rho^{j}_A||_1 \lesssim \mathcal{O}(e^{S^{(\infty)}_{\text{min}}/2+S_A/2-S/2})
\end{align}
which establishes the subsystem ETH as a consequence of the MBBC by its $e^S$ dependence. The remaining factor crucially dictates the subsystem sizes for which the trace distance is suppressed~\cite{dymarsky_subsystem_2018, kudler-flam_distinguishing_2021}. However, this factor remains to be determined and we leave a direct computation of the trace distance to future work.

\section{Eigenstate correlations II: thermalization of a non-equilibrium initial state}
\label{sec:thermalization}

In this section, we will show how a system that is not prepared in an eigenstate reaches thermal equilibrium. The general result reproduces the \textit{equilibrated pure state} formalism of ref.~\cite{liu_entanglement_2021}. Specifically, we show that reduced density matrices relax towards a form associated with their energy density, and entanglement entropies will relax towards their equilibrium values. The key insight that provides our general result in this section is that (a) time-dependent partitions vanish and (b) the time-independent partitions factor into a product of an ``outer'' partition that is identical to that of a system prepared in equilibrium and an ``inner'' partition that integrates out. In the special cases we consider, this factorization will reproduce the partitions we saw in section \ref{sec:equilibrium}.

We initialize our system in a product of subsystem eigenstates $\ket{ab}\textit{ s.t. }H_A^n H_B^m\ket{ab} = E_a^nE_b^m\ket{ab}$, but we stress that our main results do not depend on this choice. Our state has initial energy $E_0=E_a+E_b$, and all states with initial energy $E_0$ will relax to the same equilibrium so long as we neglect the width of $F$. In appendix~\ref{subapp:subvariances}, we show how fluctuations of the subsystem energy are sensitive to the specific form of $F$ and do not exactly obtain the properties of eigenstates, even in equilibrium, when subsystem $A$ is a finite fraction of the system. Ultimately, this is because the energy of a thermodynamically large subsystem is an approximately conserved quantity. 

Since dynamics are generally system-dependent, the specific forms of time-dependent partitions that encode dynamics will be system-dependent as well. However, that does not preclude generic features in the time-dependent partitions. Indeed, in section \ref{subsec:entanglementVelocities}, we find that the growth of entangelement entropy has an intriguing diagrammatic organization. We find that distinct diagrams contain the ballistic growth, diffusive growth, and saturation of entanglement entropy.

\subsection{Reduced density matrix}
\label{subsec:RDM}
Consider the matrix elements of the reduced density matrix on subsystem A over time given the initial state $|ab\rangle$:
\begin{align}
    \rho_{a'a''}(t) &= \bra{a'}\!\operatorname{Tr}\left[e^{-iHt}\ket{ab}\!\!\bra{ab}e^{iHt}\right]\!\ket{a''} \\
        &= \sum_{b'ij}\bra{a'b'}\ket{i}\!\!\bra{i}\ket{ab}\!\!\bra{ab}\ket{j}\!\!\bra{j}\ket{a''b'}e^{-i(E_i-E_j)t} \\
        &\equiv \sum_{b'ij}c^{a'b'}_ic^{i}_{ab}c^{ab}_{j}c^{j}_{a''b'}e^{-i\omega_{ij}t}.
\end{align}
We will once again be interested in both the diagonal elements $\rho_{a'a'}$ and the variance of off-diagonal elements $\left|\rho_{a'a''}\right|^2$. Let us first consider the diagonal elements,
\begin{align}
\label{eq:DiagonalNeq}
    \rho_{a'a'}(t) &= \sum_{b'ij}c^{a'b'}_ic^{i}_{ab}c^{ab}_{j}c^{j}_{a'b'}e^{-i\omega_{ij}t}.
\end{align}
For $\rho$ to relax to its equilibrium form in general, two things must be true: (1) the time-independent part of $\rho$ must equal its equilibrium form, and (2) the time-dependent part must vanish. We show the former in this section. The latter condition is contained in the smoothness assumption of the ETH -- the belief that there is a finite energy scale below which no structure can be seen in matrix elements. Given these conditions, in the $t\to\infty$ limit the oscillatory terms vanish and the time-independent value of $\rho$ is obtained.

\begin{figure}[htbp]
    \centering
    \begin{subfigure}{0.49\textwidth}
        \centering
        \includegraphics[width=\linewidth, viewport=0 100 800 500, clip]{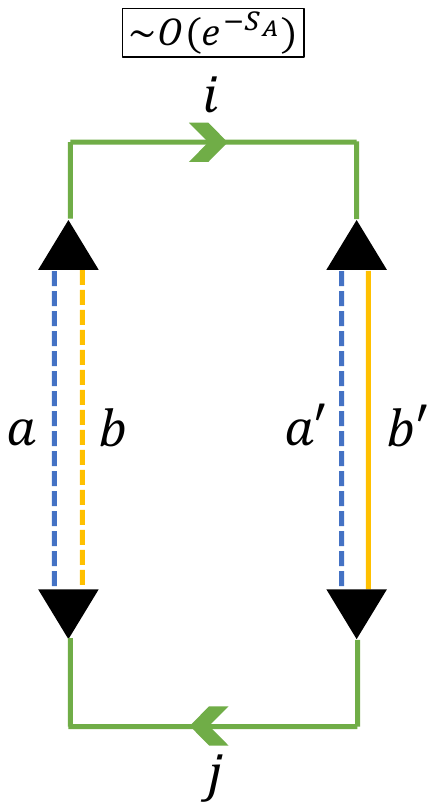}
        \caption{Uncontracted partition. Several contractions exist, however, only ones which cancel that time dependences on $i$ and $j$ will be time-independent.}
        \label{subfig:DiagonalNeq1}
    \end{subfigure}
    \hfill
    \begin{subfigure}{0.49\textwidth}
        \centering
        \includegraphics[width=\linewidth, viewport=0 115 800 470, clip]{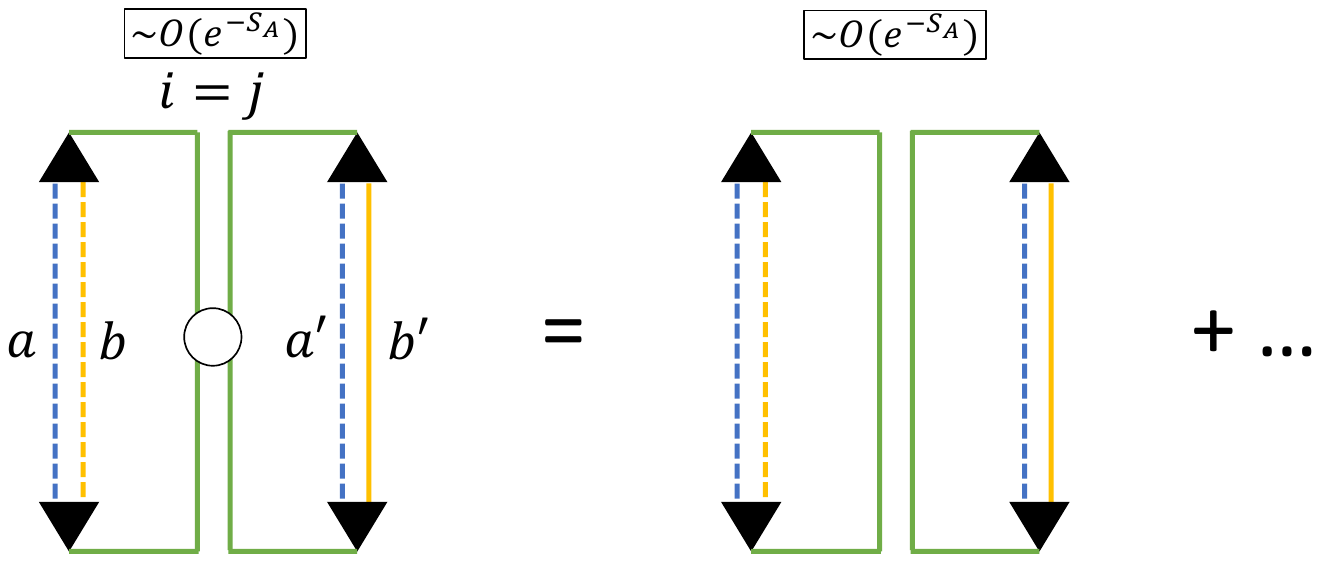}
        \caption{Time independent contribution to the reduced density matrix. Note, there is only one sum over index $i$.}
        \label{subfig:DiagonalNeq2}
    \end{subfigure}

    \caption{In (a) we have placed the uncontracted partition corresponding to eq. \eqref{eq:DiagonalNeq} for clarity. Time dependence ensures that it will decay as long as its corresponding $F$-function is smooth. In (b), we have placed the leading order time-independent partition. Time-independence is achieved by contracting indices with opposite time dependences. Further time-independent partitions can be generated by contracting $a=a'$ or $b=b'$. However, the former case will only contribute a factor of $2$ for the matrix element $\rho_{aa}$ while the latter case will suppressed by a factor of $e^{-S_B}$, so we neglect both cases. Focusing on the right hand side of the equation in (b) on the left, we have a partition that fully integrates out when summing over index $i$; we refer to this as the ``inner" partition. On the right, we have an ``outer'' partition that is identical to the partition pictured in figure \ref{fig:diagonal} and describes the diagonal elements of an equilibrium reduced density matrix. When we consider higher correlations the ``inner'' and ``outer'' labels will become more visually apparent.}
    \label{fig:DiagonalNeq}
\end{figure}

Diagrammatically, time-independent partitions are found by contracting indices with opposite time-dependences, thereby removing oscillatory terms. The expression in eq. (\ref{eq:DiagonalNeq}) has several unique partitions, but only 1 is both nontrivial and time-independent, so we will restrict our attention to it (see figure \ref{fig:DiagonalNeq}). We compute,
\begin{align}
\label{eq:integrateOut}
    \rho_{a'a'}(\infty) &= \sum_{b'i}\left|c_{a'b'}^i\right|^2\left|c^{i}_{ab}\right|^2 \nonumber \\ 
    &= \int_{E_i E_{b'}}e^{S(E_i)+S_B(E_{b'})-2S(E_i)}F(E_i-E_a-E_b)F(E_i-E_{a'}-E_{b'}) \nonumber \\
    &= e^{-S(E_0)+S_B(E_0-E_{a'})}
\end{align}
which is what we expect from the equilibrium case considered previously, though the second $F$ function will modify the subsystem energy variance expression (see appendix~\ref{subapp:subvariances}). We can see from this example that the recovery of the equilibrium result came from a factorization of the time-independent partition into a partition equivalent to the one considered for eigenstates, and a partition that was integrated out in eq. (\ref{eq:integrateOut}). This correspondence between time-independent partitions and equilibrium partitions is general.

\begin{figure}[htbp]
    \centering
    \begin{subfigure}{0.49\textwidth}
        \centering
        \includegraphics[width=\linewidth, viewport=0 50 800 550, clip]{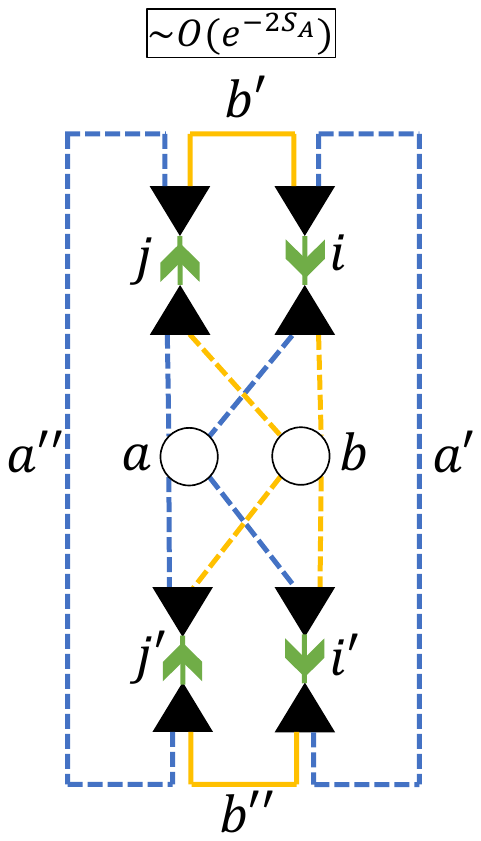}
        \caption{Full partition representing the time-dependent off-diagonal elements of the reduced density matrix on subsystem $A$.}
        \label{subfig:odn1}
    \end{subfigure}
    \hfill
    \begin{subfigure}{0.49\textwidth}
        \centering
        \includegraphics[width=\linewidth, viewport=0 50 800 550, clip]{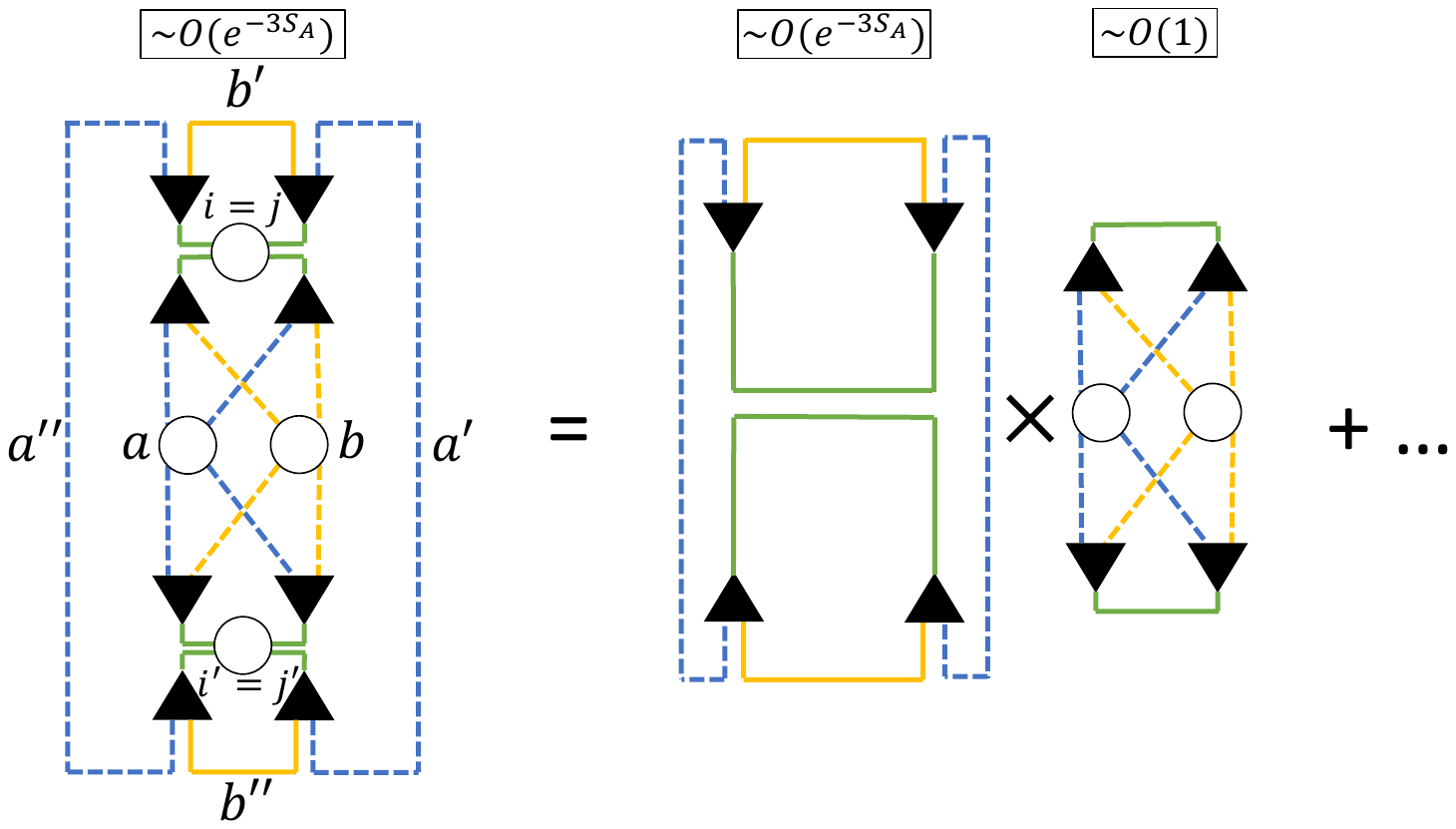}
        \caption{Time-independent partition $(ij)(i'j')$ that contributes to the off-diagonal elements of the reduced density matrix on subsystem $A$.}
        \label{subfig:odn2}
    \end{subfigure}
    \vspace{0.1cm}
    \centering
    \begin{subfigure}{0.49\textwidth}
        \centering
        \includegraphics[width=\linewidth, viewport=0 50 800 550, clip]{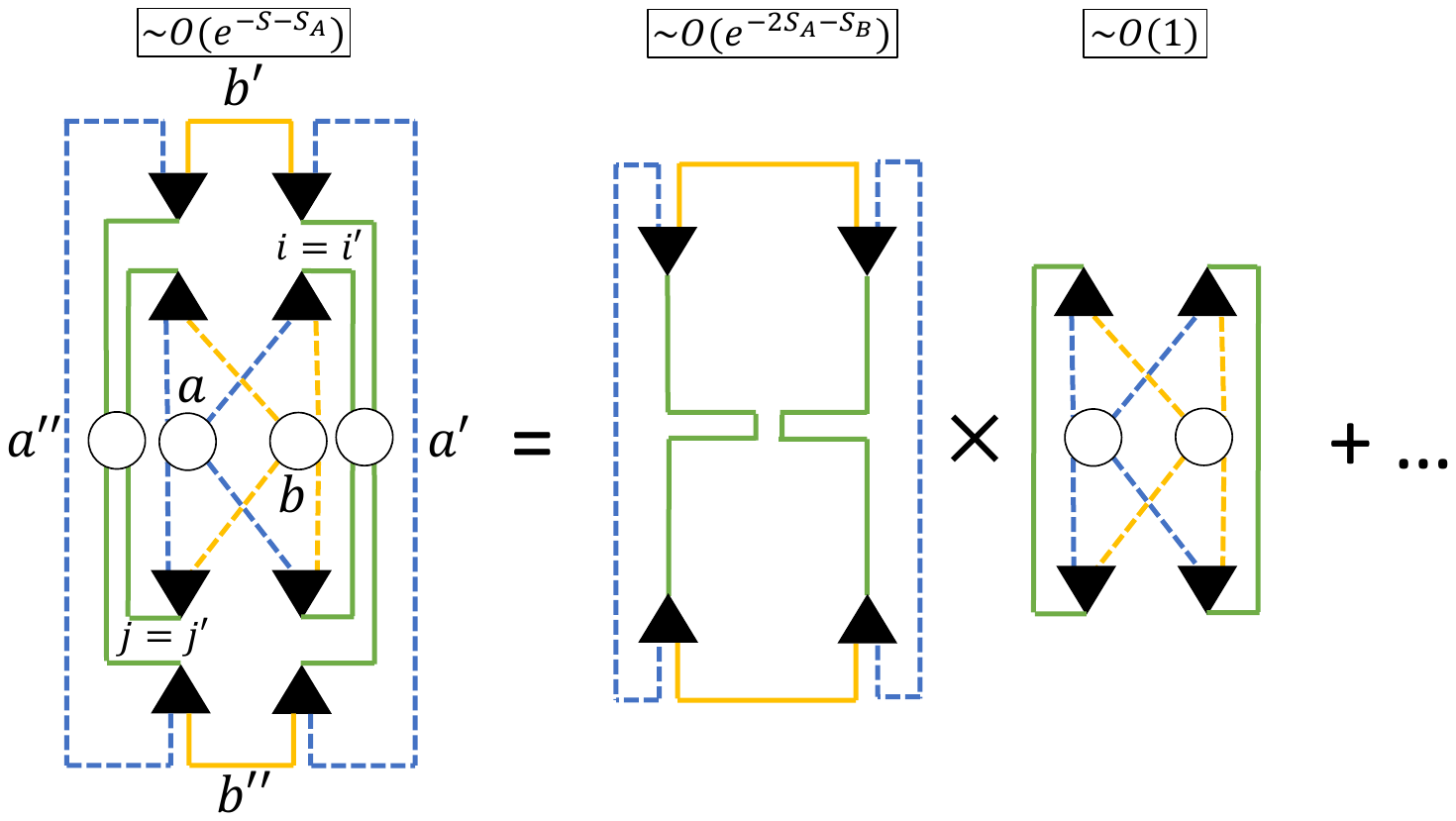}
        \caption{Time-independent partition $(ii')(jj')$ that contributes to the off-diagonal elements of the reduced density matrix on subsystem $A$.}
        \label{subfig:odn3}
    \end{subfigure}
    \hfill
    \begin{subfigure}{0.49\textwidth}
        \centering
        \includegraphics[width=\linewidth, viewport=0 50 800 550, clip]{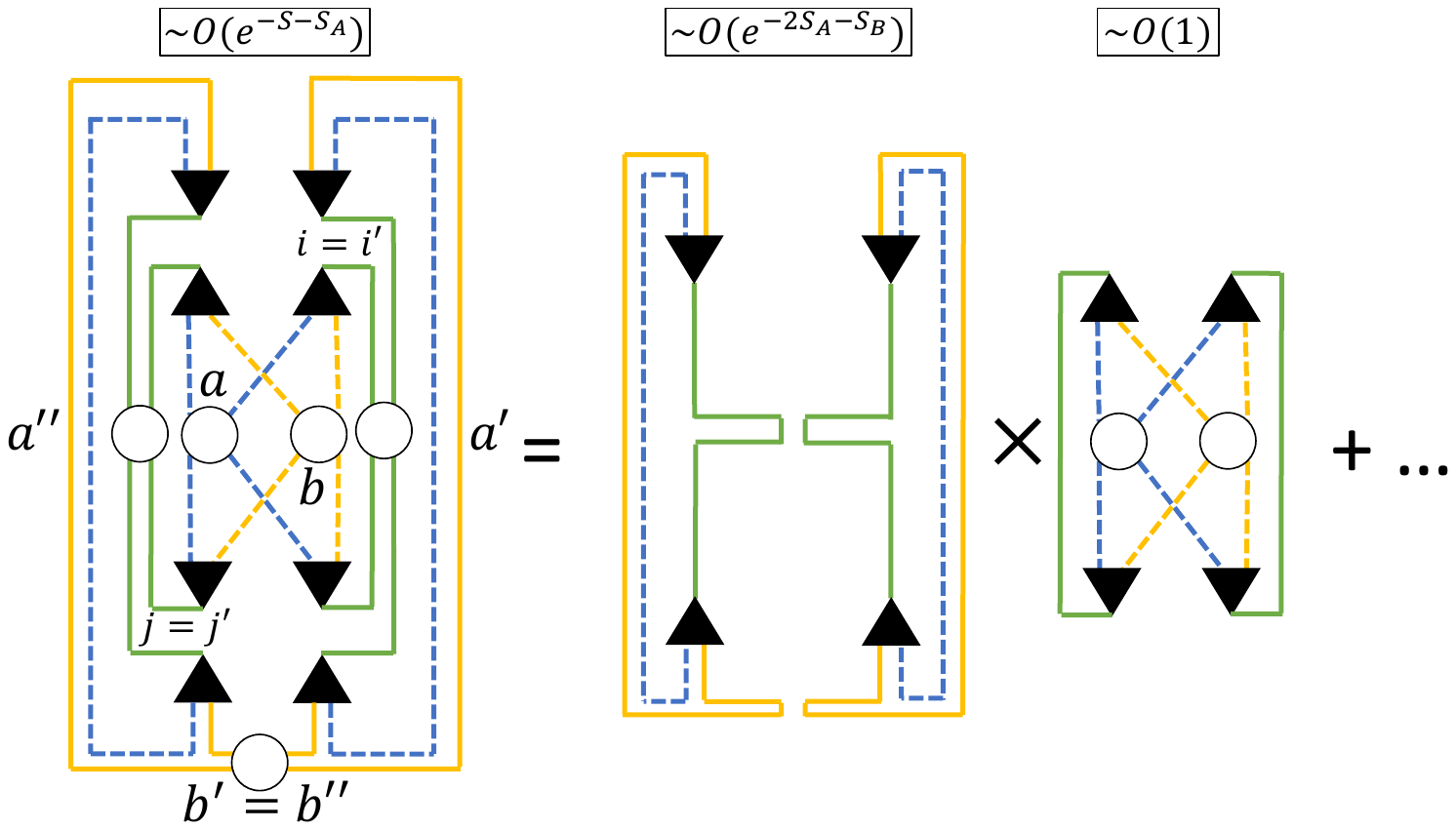}
        \caption{Time-independent partition $(ii')(jj')(b'b'')$ that contributes to the off-diagonal elements of the reduced density matrix on subsystem $A$.}
        \label{subfig:odn4}
    \end{subfigure}
    \caption{Diagrams that contribute to the off-diagonal matrix elements of the reduced density matrix on subsystem $A$. Subfigures (b-d) picture the time-independent terms and the factorizations that dominantly contribute. Each factorization contains an inner partition that integrates out and an outer partition that reduces to the partitions in figure \ref{fig:offDiagonal}. Note that one should not overcount the number of integrals on the right-hand side of each diagrammatic equation.}
    \label{fig:offDiagonalNeq}
\end{figure}

Next, we address the off-diagonal elements,
\begin{align}
    \rho_{a'a''}(t)\rho_{a''a'}(t) = \sum_{b'b''iji'j'}c^{a'b'}_{i}c^{i}_{ab}c^{ab}_{j}c^{j}_{a''b'}c^{a''b''}_{i'}c^{i'}_{ab}c^{ab}_{j'}c^{j'}_{a'b''}e^{-it(E_i-E_j-E_{i'}+E_{j'})}.
\end{align}
There are several dozen independent partitions; however, we will focus on just 4: the full partition and the 3 largest time-independent partitions (see figure \ref{fig:offDiagonalNeq}). The full partition drawn in figure \ref{subfig:odn1}, is given by
\begin{align}
\label{eq:tDeath}
    F(E_{a'a''};\omega_{a'a''};t) &= \sum_{[b'b''iji'j']}c^{a'b'}_{i}c^{i}_{ab}c^{ab}_{j}c^{j}_{a''b'}c^{a''b''}_{i'}c^{i'}_{ab}c^{ab}_{j'}c^{j'}_{a'b''}e^{-it(E_i-E_j-E_{i'}+E_{j'})}\nonumber \\ &= e^{-2S(E_a+E_b)+2S_B(E_a+E_b-E_{a'a''})}F(\omega_{a'a''};t)
\end{align}
where $F(\omega_{a'a''};t)$ is a smooth term fated to die off in $t$, whatever its precise form. There are no other partitions that contribute at the same order. This partition will be studied in the context of the $2$-Renyi Entropy in section \ref{subsec:entanglementVelocities}. 

Moving on to the time-independent partitions, $(ij)(i'j')$, $(ii')(jj')$, $(iji'j')$, we find that the partitions in figures \ref{subfig:odn2}, \ref{subfig:odn3}, \ref{subfig:odn4} factor into inner partitions that integrate out and outer partitions that resemble those in figures \ref{subfig:offDiagonal4} and \ref{subfig:offDiagonal2}. Then again, we see that time-independent partitions reduce to equilibrium partitions,
\begin{align}
\label{eq:integrateOut2}
    |\rho_{a'a''}(\infty)|^2 &=\int_{E_b'E_{b''}E_{i}E_{i'}}e^{2S_B(E_{b'b''})-2S(E_0)-S_A(E_{a'a''})}F_{(ij)(i'j')}(\cdots) \nonumber \\ 
    &+ \int_{E_{b'}E_{b''}E_{i}E_{i'}}e^{2S_B(E_{b'b''})-2S(E_0)-S_B(E_{b'b''})}F_{(ii')(jj')}(\cdots) \nonumber \\
    &+ \int_{E_{b'}E_{b''}E_{i}E_{i'}}e^{S_B(E_{b'b''})-2S(E_0)}F_{(ii')(jj')(b'b'')}(\cdots) + \cdots \nonumber \\
    &= e^{-2S(E_0)+2S_B(E_0-E_{a'a''})-S_{\text{min}}(E_{a'a''})}\tilde F(\omega_{a'a''})
\end{align}
which is the same expression derived in eq. (\ref{eq:offdiageq2}).

\subsection{Entanglement entropies and the Page curve}
\label{subsec:Page}
\begin{figure}[htbp]
    \centering
    \begin{subfigure}{0.49\textwidth}
        \centering
        \includegraphics[width=\linewidth, viewport=0 50 800 550, clip]{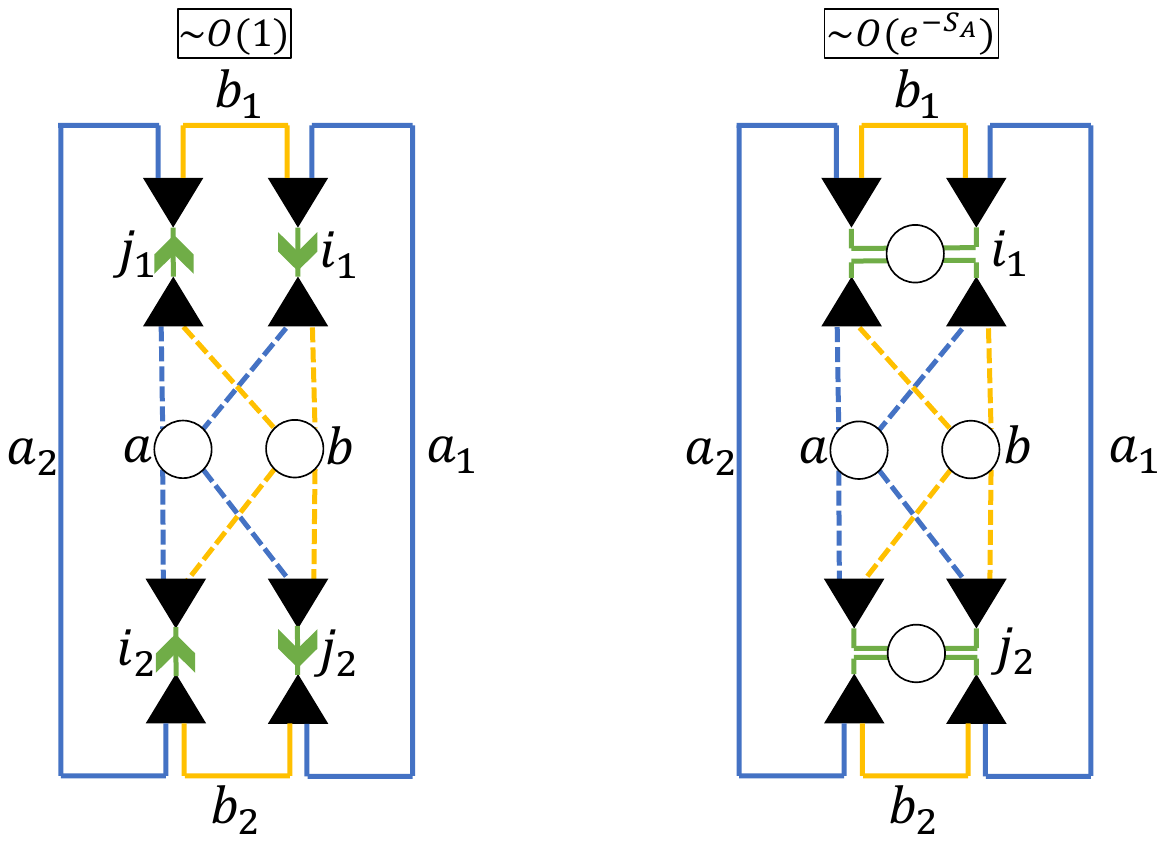}
        \caption{Time dependent 2\textsuperscript{nd} Renyi entropy.}
        \label{subfig:2RenyiNeq}
    \end{subfigure}
    \hfill
    \begin{subfigure}{0.49\textwidth}
        \centering
        \includegraphics[width=\linewidth, viewport=0 50 800 550, clip]{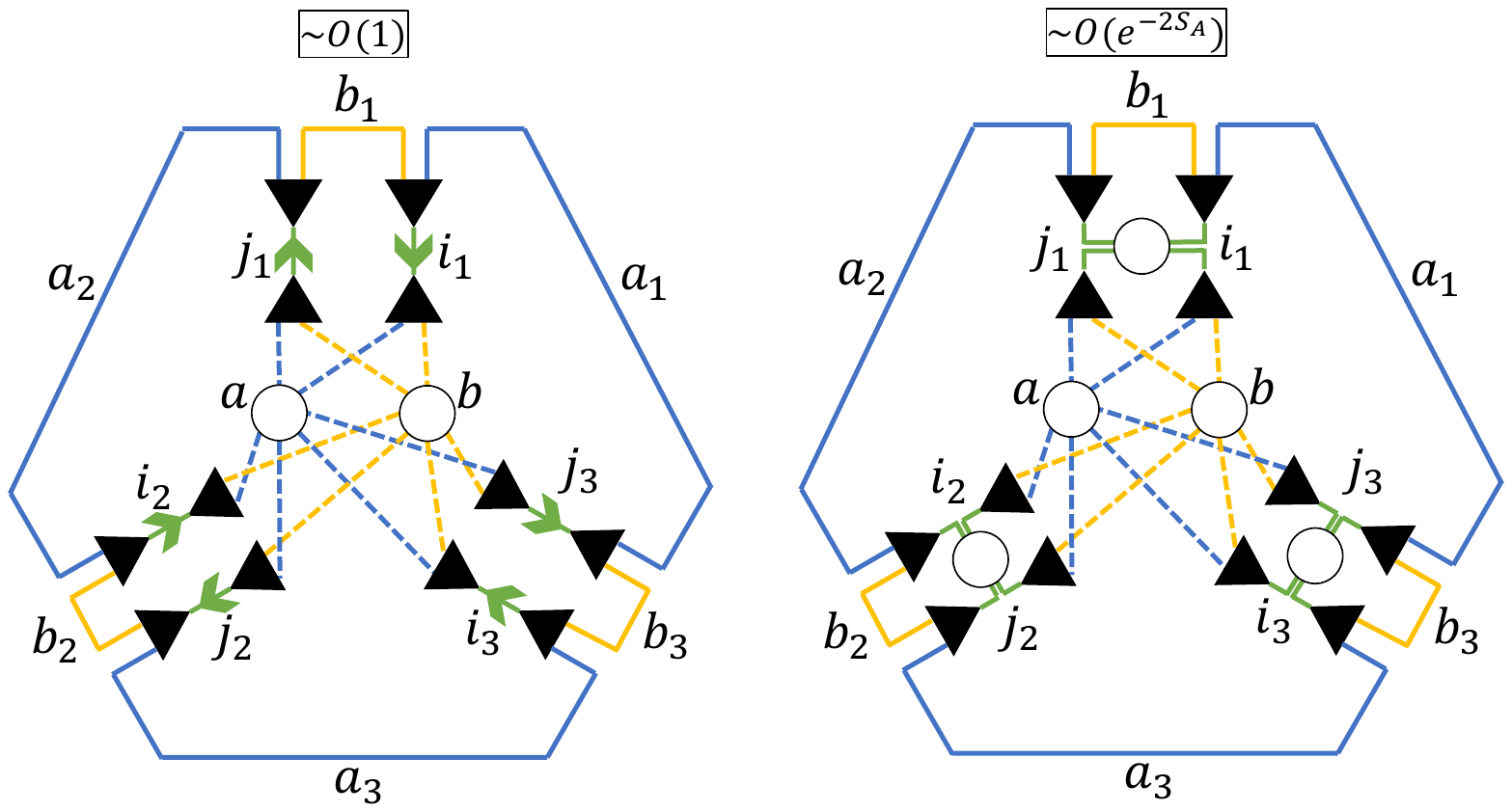}
        \caption{Time dependent 3\textsuperscript{rd} Renyi entropy.}
        \label{subfig:3RenyiNeq}
    \end{subfigure}
    \vspace{0.1cm}
    \centering
    \begin{subfigure}{0.49\textwidth}
        \centering
        \includegraphics[width=\linewidth, viewport=0 50 800 550, clip]{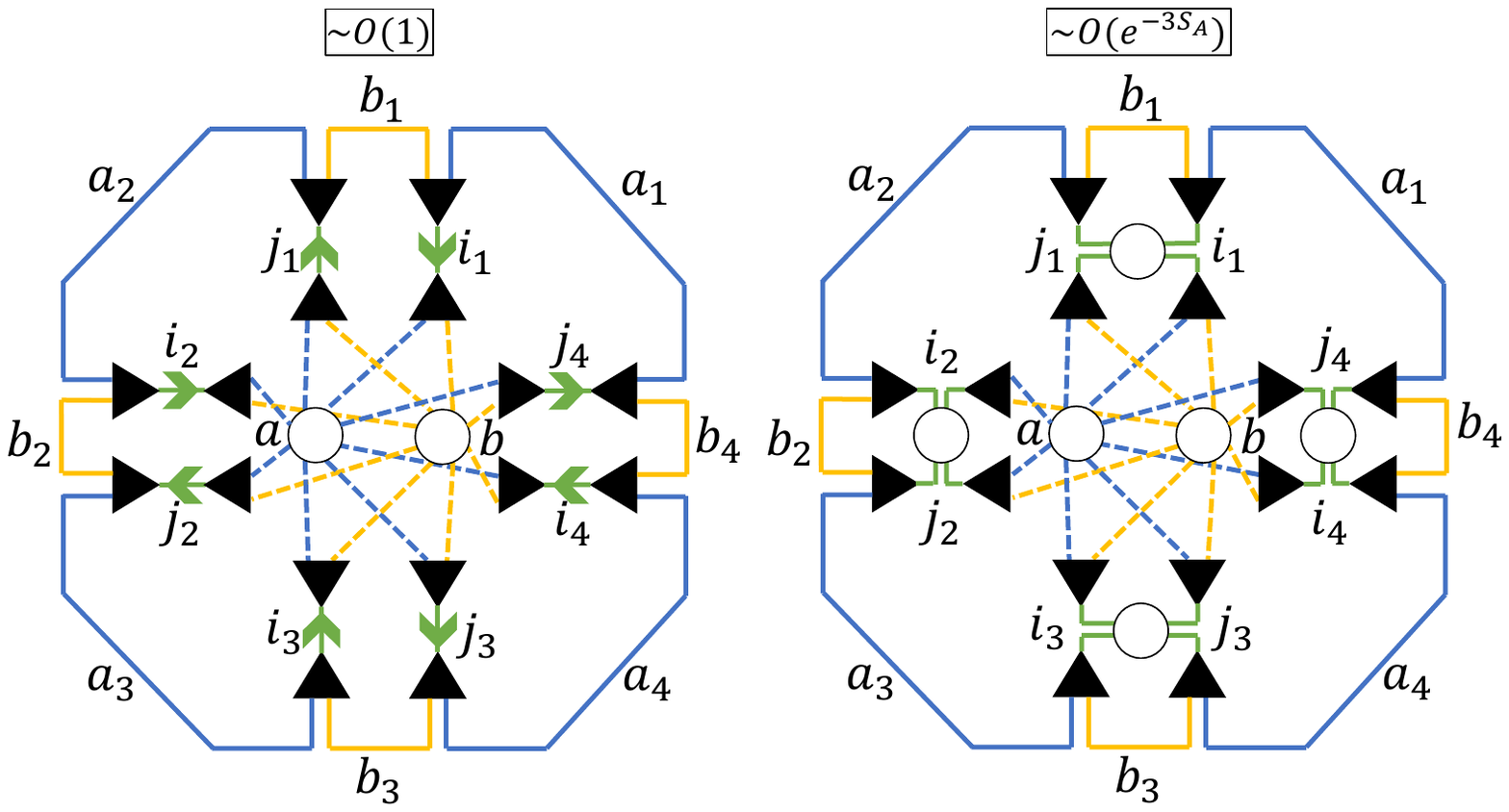}
        \caption{Time dependent 4\textsuperscript{th} Renyi entropy.}
        \label{subfig:4RenyiNeq}
    \end{subfigure}
    \caption{Time-dependent Renyi entropies adjacent to their leading time-independent counterpart. The contractions between $i$ and $j$ indices map one-to-one with the pairings of $i$ for the equilibrium case, represented for the 3-Renyi entropy in figure \ref{fig:PlanarEq}.}
    \label{fig:RenyiNeq}
\end{figure}

Once again, initializing in $\ket{ab}$, we repeat our analysis for the Renyi entropies. The $\alpha$-Renyi entropy is
\begin{align}
\label{eq:Renyi}
    S_\alpha(t) &\equiv \frac{1}{1-\alpha}\operatorname{ln}\left(\operatorname{Tr}_A\left[\left(\operatorname{Tr}_B\left[e^{-iHt}\ket{ab}\!\!\bra{ab}e^{iHt}\right]\right)^{\alpha}\right]\right) \nonumber \\
    &= \frac{1}{1-\alpha}\operatorname{ln}\left(\sum_{a_m'b_m'i_mj_m}c^{a'_1b'_1}_{i_1}c^{i_1}_{ab}c^{ab}_{j_1}c^{j_1}_{b'_1a'_2}\cdots c^{a'_\alpha b'_\alpha}_{i_\alpha}c^{i_\alpha}_{ab}c^{ab}_{j_\alpha}c^{j_\alpha}_{b'_\alpha a'_1}e^{-it\sum_{m}\omega_{i_mj_m}}\right).
\end{align}
Time-independent partitions are those for which $\sum_{m}\omega_{i_mj_m}$ vanishes, which are obtained by contracting $i$ indices with $j$. We have drawn the minimally contracted and the largest time-independent partitions for subsystem $A$ much smaller than $B$ for the $2$-, $3$-, and $4$-Renyi entropies in figure \ref{fig:RenyiNeq}. 

When $S_B>S_A$, the largest time-independent partition is ${(i_mj_m)}$, while for $S_A>S_B$, the largest is $(i_{m+1}j_m)$. When $S_A\approx S_B$, any pairing of $i$ indices with $j$ indices that is non-crossing, as depicted previously in figure \ref{fig:PlanarEq}, contributes. Once again, we can see that the time-independent partitions factor into an inner partition that integrates out and outer partitions that are identical to those considered in the equilibrium case. Since the result is general, we do not repeat our calculations from section \ref{subsec:PageEq}.

\subsection{Entanglement growth}
\label{subsec:entanglementVelocities}
\begin{figure}[htbp]
    \centering
    \begin{subfigure}{0.49\textwidth}
        \centering
        \includegraphics[width=\linewidth, viewport=20 112 770 550, clip]{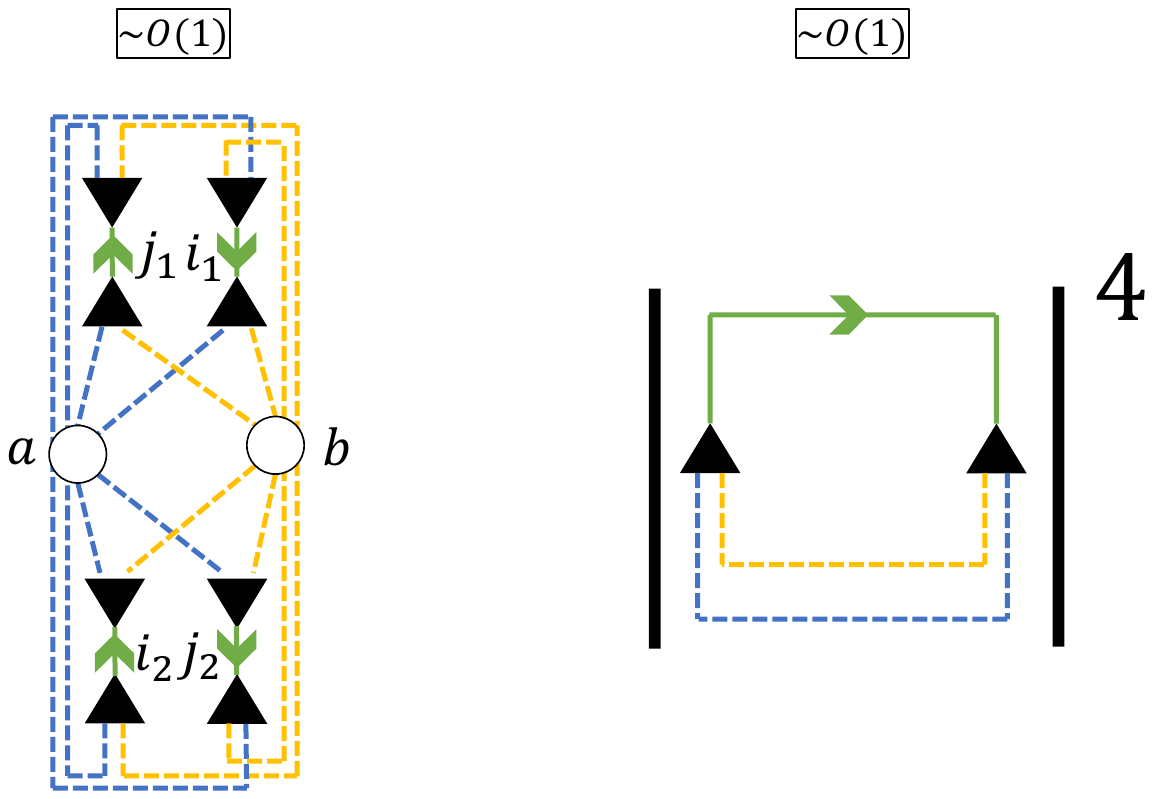}
        \caption{The fourth power of the (modulus) return amplitude is obtained by contracting all summed $a$- and $b$-indices with the initial state. This is the only term which contributes at $t=0$.}
        \label{subfig:eg1}
    \end{subfigure}
    \hfill
    \begin{subfigure}{0.49\textwidth}
        \centering
        \includegraphics[width=\linewidth, viewport=0 112 800 550, clip]{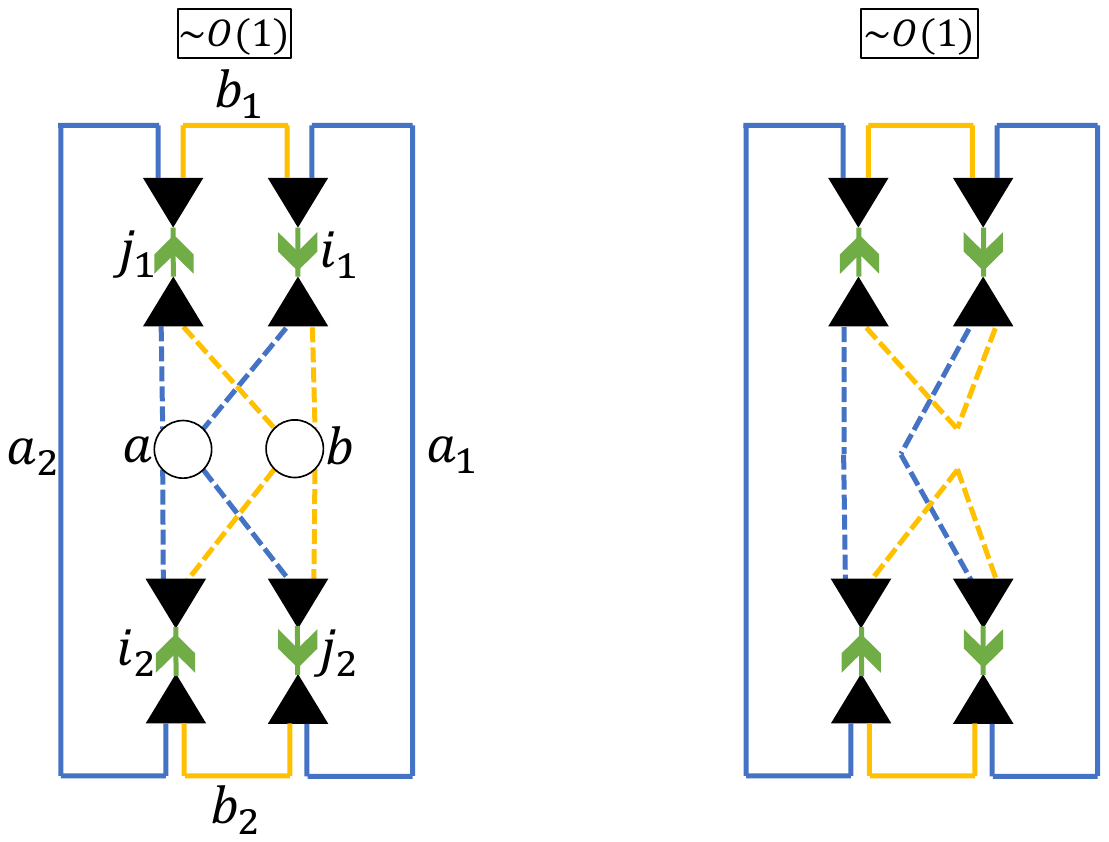}
        \caption{The full, uncontracted partition is expected to govern the ballistic regime of entanglement growth, $t_{\text{eq}}<t<t_{\text{bal}}$, and encode the entanglement velocity.}
        \label{subfig:eg2}
    \end{subfigure}

    \vspace{0.1cm} 

    \begin{subfigure}{0.49\textwidth}
        \centering
        \includegraphics[width=\linewidth, viewport=0 112 800 550, clip]{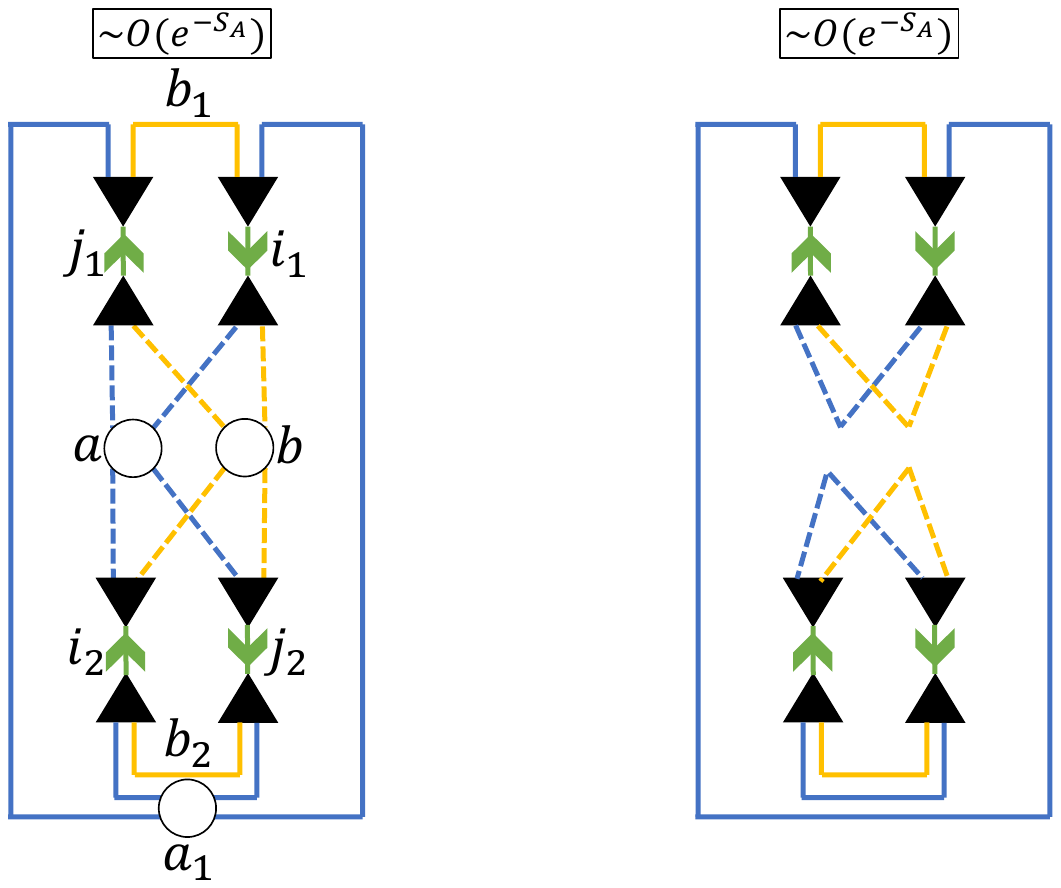}
        \caption{At late times prior to saturation diffusive spreading of energy limits the grwoth of $\alpha>1$-Renyi entropies. This is captured in the relaxation of the diagonal elements of reduced density matrices. Hence, we connect this behavior to the partition that contracts all external (summed) $a$-indices.}
        \label{subfig:eg3}
    \end{subfigure}
    \hfill
    \begin{subfigure}{0.49\textwidth}
        \centering
        \includegraphics[width=\linewidth, viewport=20 112 770 550, clip]{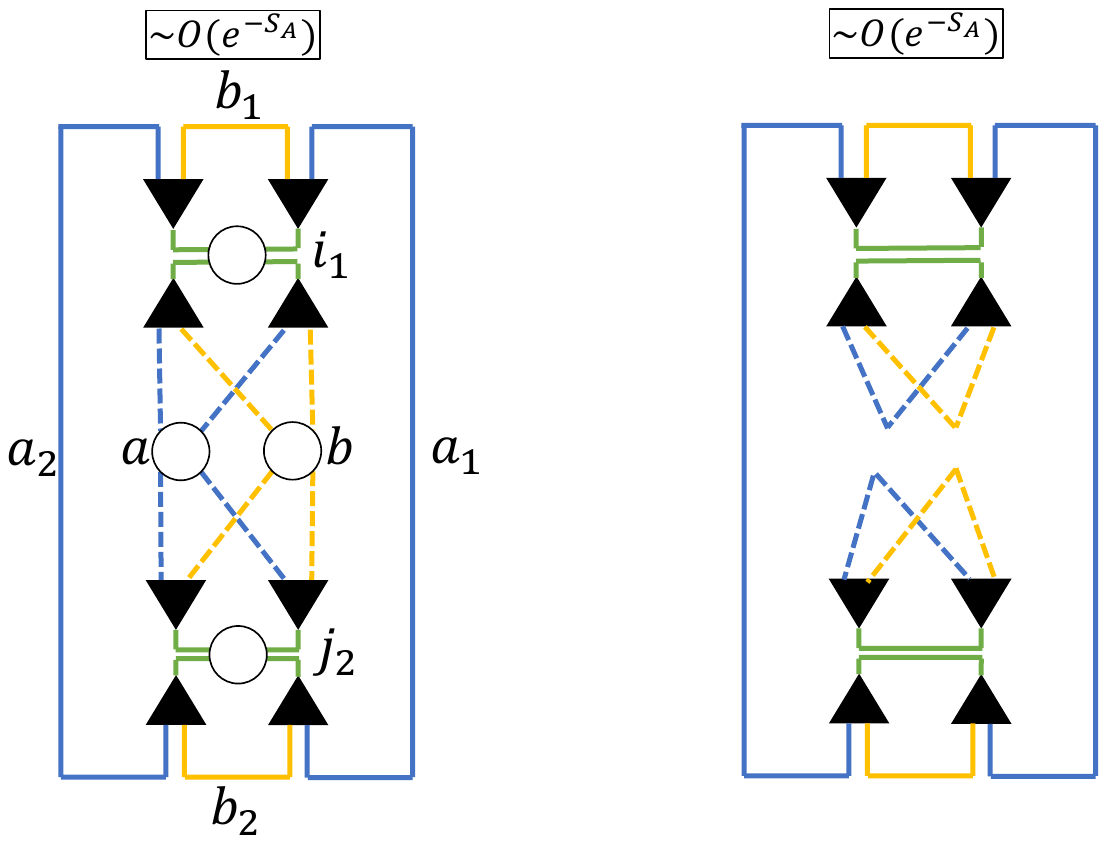}
        \caption{At late times $t>t_{\text{sat}}$, phases cancel against each leaving only partitions that contract forward propogators with backwards propagators.}
        \label{subfig:eg4}
    \end{subfigure}

    \caption{Four partitions associated with the growth of $2$-Renyi entropy adjacent to their leading factorization under the assumption that subsystem $A$ is smaller than subsystem $B$. When the subsystems are of similar size, there will more contributions.}
    \label{fig:eg}
\end{figure}

From an initially unentangled state, the dynamics of Renyi entropies $\alpha>1$ generically exhibit 4 regimes: (i) local equilibration, (ii) ballistic growth, (iii) diffusive growth, and (iv) saturation. At late times, the equilibrium partitions once again map to eigenstate partitions and compute a Page curve. In the previous subsection we associated saturation to equilibrium partitions. In this subsection we will provide similar diagrammatic interpretations to the different regimes of entanglement growth by counting entropic factors (see figure~\ref{fig:eg}). 

Initially, entanglement growth is limited by fidelity to the initial state~\cite{shi_local_2023},
\begin{align}
\label{eq:Retvrn}
    S_\alpha(t\approx 0) \lesssim \frac{1}{1-\alpha}\ln{|F_\text{R}(t)|^{2\alpha}}
\end{align}
where $F_\text{R}(t)$ is the return amplitude $F_\text{R}(t)\equiv \bra{ab}\!e^{-iHt}\!\ket{ab}$ which arises under the contraction $(a_1\dots a_\alpha a)(b_1\dots b_\alpha b)$, represented for $\alpha=2$ in figure~\ref{subfig:eg1}. We discuss functional forms for $F_\text{R}(t)$ in appendix~\ref{subapp:normForm}. At a later time, $t_{\text{eq}}\sim\mathcal{O}(1)$, the system will have reached local equilibrium~\cite{liu_entanglement_2014} and will transition to ballistic entanglement growth. However, as we discuss later in this section, eq.~\eqref{eq:Retvrn} does not provide an adequate accounting of entanglement growth for $t\lesssim t_{\text{eq}}$ and further work is needed to clarify the local equilibration timescale. The problem of modeling the propagator at short times in chaotic systems is closely related to the motivation for the \textit{maximum entropy} approach to the properties of chaotic eigenstates presented in ref.~\cite{hahn_statistical_2023}. Such an approach may be able to shed light on entanglement growth at early times. 

\begin{figure}
    \centering
    \includegraphics[width = 0.75\linewidth, viewport=0 70 800 530, clip]{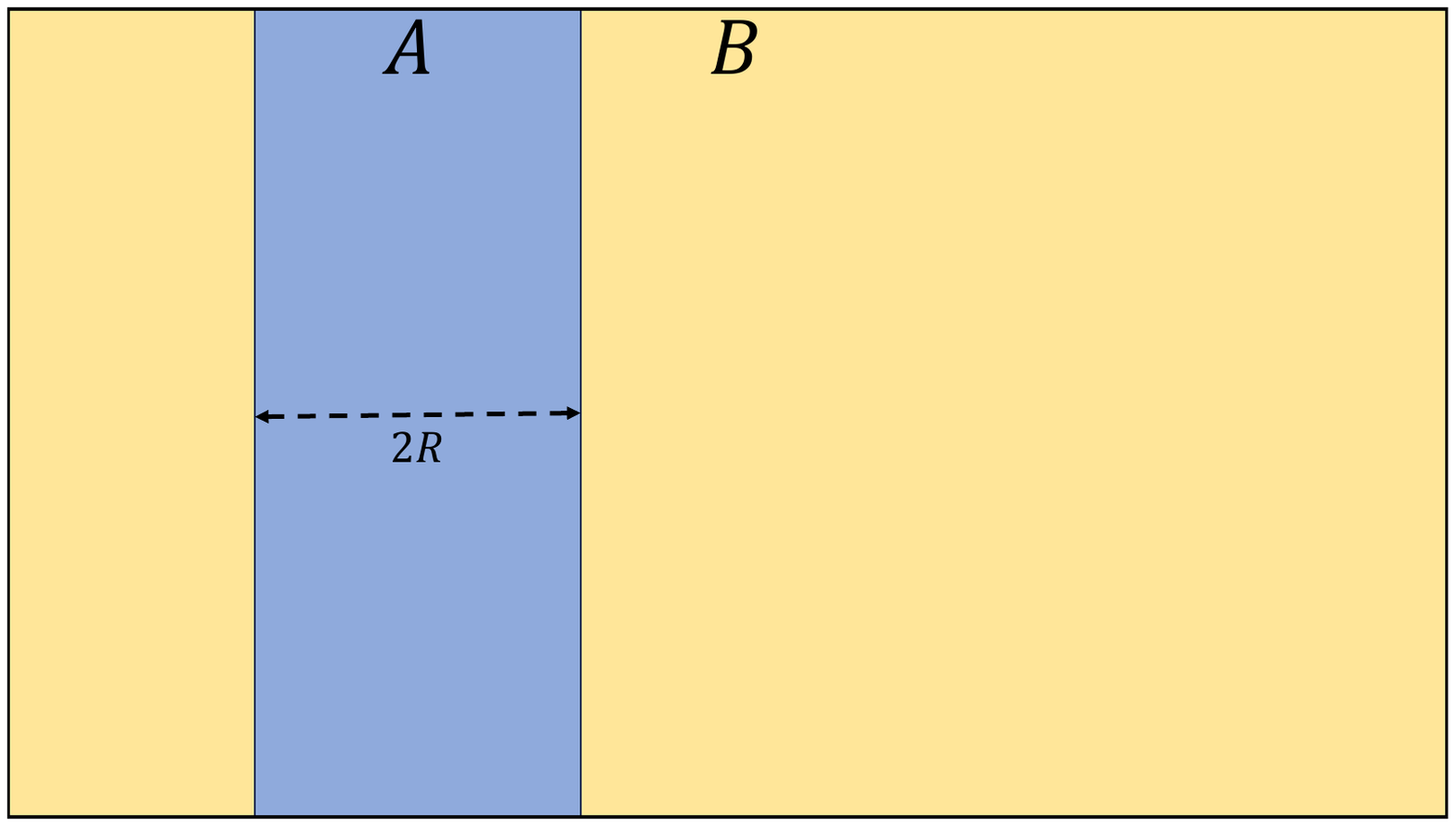}
    \caption{An effective 1D system split into two subsystems $A$ and $B$. The width of $A$ is $2R$.}
    \label{fig:1Dsystem}
\end{figure}

At intermediate times $t_{\text{eq}}\lesssim t\lesssim t_{\text{bal}} \equiv \frac{R}{v_E^{(\alpha)}}$, entanglement growth is expected to be ballistic. For an effective 1D ``strip'' geometry (see figure~\ref{fig:1Dsystem}), this expectation implies a form,
\begin{align}
\label{eq:ballistic}
    S_\alpha(t\lesssim t_{\text{bal}})\approx \frac{v_E^{(\alpha)}t}{R}S_\alpha(\infty)
\end{align}
where $R$ is the radius of the strip. In this regime, $e^{(1-\alpha)S_\alpha}$ is initially $\sim\mathcal{O}(1)$. Whichever partition(s) dominates this regime of entanglement growth will be the slowest decaying partition that contributes at $\mathcal{O}(1)$. The remaining partitions that may contribute at $\mathcal{O}(1)$ are the full, uncontracted partition and partitions generated by contractions of any summed $a_l$- and $b_l$-indices with $a$ and $b$. Performing all such contractions yielded the term composed solely of powers of the return ampltidue, eq.~\eqref{eq:Retvrn}. Any partial such contraction will composed of other metrics of fidelity to the initial state such as $\expval{\rho_{A(B)}(t)}_{a(b)}$, $\expval{\rho_{A(B)}(t)^2}_{a(b)}$, or other powers of the return amplitude. Since each of these terms are highly nonlocal, we expect they should decay at faster rates than any entanglement velocity. Hence, our expectation is that for each $\alpha$-Renyi entropy, only the full uncontracted partition governs the whole regime of ballistic entanglement growth, while other terms $\sim\mathcal{O}(1)$ at best contribute during local equilibration. We've included the case $\alpha=2$ in figure~\ref{subfig:eg2}. Then we assert,
\begin{align}
    p(t/t_{\text{eq}})e^{(1-\alpha)\frac{v^{(\alpha)}_Et}{R}S_\alpha(\infty)} &\approx\sum_{[\cdots]}c^{a'_1b'_1}_{i_1}c^{i_1}_{ab}c^{ab}_{j_1}c^{j_1}_{b'_1a'_2}\cdots c^{a'_\alpha b'_\alpha}_{i_\alpha}c^{i_\alpha}_{ab}c^{ab}_{j_\alpha}c^{j_\alpha}_{b'_\alpha a'_1}e^{-it\sum_{m}\omega_{i_mj_m}} \nonumber \\
    &= \int_{\cdots}e^{2\alpha S(\bar E_i)+\alpha S_A(\bar E_a) + \alpha S_B(\bar E_b)- 3\alpha S(E_a+E_b)}F(\cdots)e^{-it\sum_{m}\omega_{i_mj_m}} \nonumber \\
    &= \int_{\{\omega\}}F(\omega_{i_1j_1},\cdots)e^{-it\sum_{m}\omega_{i_mj_m}}
\end{align}
where the left-hand side contains the exponential decay with a short time suppression, $p(t/t_{\text{eq}})$, that ensures the expression vanishes at $t\ll t_{\text{eq}}$ and the right-hand side contains the only relevant index partition. For convenience, we define $\omega_E^{(\alpha)} \equiv (\alpha - 1){v^{(\alpha)}_E}S_\alpha(\infty)/R$ and study the spectral function,
\begin{align}
    \mathcal F (\omega) &\equiv \int_{\omega_{i_1j_1}\cdots}F(\omega_{i_1j_1},\cdots)\delta\left(\omega - \sum_{m}\omega_{i_mj_m}\right) \nonumber \\ &\approx \int_{-\infty}^\infty p(t/t_{\text{eq}})e^{i(\omega+i\omega_E^{(\alpha)}) t } dt.
\end{align}
The exponential decay ensures that $\mathcal F (\omega)$ analytic within the strip $-\omega_E^{(\alpha)}\leq\text{Im}[\omega]\leq \omega_E^{(\alpha)}$. Furthermore, we expect $p(t/t_{\text{eq}})e^{-\omega_E^{(\alpha)}t}$ to maximize during the crossover from the local equilibration regime to the ballistic regime, $t\approx t_{\text{eq}}$, which requires $\mathcal{F}(\omega)$ to oscillate over all scales larger than $t_{\text{eq}}^{-1}$. Lastly, general arguments based on locality~\cite{arad_connecting_2016} enforce a sharp cutoff at large frequencies $\mathcal{F}(\omega)\lesssim e^{-\frac{|\omega|-V_{AB}}{g}}$, where $V_{AB}$ is an area scaling term determined by the higher moments of $H_{AB}$ and $g$ is an $\mathcal{O}(1)$ effective coupling constant.

At late times still prior to saturation the Renyi entropies for $\alpha>1$ will transition from ballistic to diffusive growth~\cite{rakovszky_sub-ballistic_2019, huang_dynamics_2020} in systems with conservation laws (i.e. energy conservation). Close to saturation, diffusive growth implies a form
\begin{align}
    S_{\alpha}(t\lesssim t_{\text{sat}}) \propto \sqrt{Dt}
\end{align}
where $D$ is the diffusivity of the system, and $t_{\text{sat}}$ is the saturation time for the entanglement entropy. Since we just argued that the Renyi entropies are dominated by the full index partition in the entropic factors, naively, one may expect the same partition to contain both ballistic and diffusive regimes as was suggested in ref.~\cite{shi_local_2023}. However, at late times, the exponential decay of the full partition will diminish its magnitude by the entropy of the smaller subsystem. Then, any other partitions that were only suppressed by the entropy of the smaller subsystem can contribute provided they decay sufficiently slowly. Taking $A$ to be the smaller subsystem, we find precisely one partition to be suitable: the contraction of all summed $a$ indices, $(a_1a_2\dots a_\alpha)$\footnote{analogously for subsystem $B$. When the subsystems are a similar size, there will be a contribution from all non-crossing contractions of summed $a$ and $b$ indices.}. Defining $\rho_A(E_a,t) \equiv \bra{a}\!\operatorname{Tr}_A\!\left[e^{-iHt}\ket{ab}\!\!\bra{ab}e^{iHt}\right]\!\ket{a}$ as the diagonal elements of the reduced density matrix on subsystem $A$ in the subsystem energy eigenbasis, we can write,
\begin{align}
\label{eq:diffusive}
    F_{(a_1a_2\dots a_\alpha)}(t) = \int_{E_a}e^{S_A(E_a)}\left(\rho_A(E_a,t)-\rho_A(E_a,\infty)\right)^\alpha+ \dots \sim \mathcal{O}(e^{(1-\alpha)S_A})
\end{align}
where the error in eq.~\eqref{eq:diffusive} consists of partitions associated with the now negligible return amplitude~\eqref{eq:Retvrn} and partitions that are too suppressed to contribute at leading order in any regime. From eq.~\eqref{eq:diffusive}, it's clear why $(a_1a_2\dots a_\alpha)$ is the right partition to describe diffusive relaxation: $\rho_A(E_a,t)$ encodes the dynamics of the approximately conserved subsystem energy. From an arbitrary unentangled inital state, the diffusive dynamics depend not only on the geometry of the subsystem, but on the temperature dependence of the diffusivity and heat capacity as well. We can conclude, however, that since $F_{(a_1a_2\dots a_\alpha)}(t)$ decays slower than any exponential, but faster than any polynomial, its spectral function $\mathcal{F}_{(a_1a_2\dots a_\alpha)}(\omega)$ is smooth, but non-analytic on the real line\footnote{As an example of an appropriate non-analytic function, the Fourier transform of the standard bump function, $\exp(\frac{1}{x^2-1}),\,|x|\leq 1$, can be evaluated via saddle-points to get the desired asymptotic decay $\sim e^{-\sqrt{t}}$, up to dimensionful constants and other physical features.}. 

We now discuss the $\alpha\rightarrow 1$ limit. Since the saturation regime shares its structure with pure eigenstates, the $\alpha\rightarrow 1$ limit will be the same as in section~\ref{subsec:PageEq}. Interestingly, $S_1$ does not exhibit a diffusive regime and so we should expect $F_{(a_1a_2\dots a_\alpha)}(t)$ to vanish in this limit. Indeed, $\int_{E_a}e^{S_A(E_a)}\left(\rho_A(E_a,t)-\rho_A(E_a,\infty)\right)^1 = 1-1 = 0$. For the ballistic regime, eq.~\eqref{eq:ballistic}, the analytic continuation $\alpha\rightarrow 1$ holds by construction if $v_E^{(\alpha)}$ and $S_\alpha(\infty)$ can be regarded as analytic functions of $\alpha$ predicting a ballistic growth for $S_1$ where one is seen for $S_{\alpha>1}$. In general, for the analytic continuation to exist\footnote{it must since $\rho_A$ is a positive semi-definite operator}, $\alpha = 1 + \epsilon$ must imply that $\operatorname{Tr}[\rho_A(t)^\alpha] = 1-S_1(t)\epsilon+\mathcal{O}(\epsilon^2)$. However, eq.~\eqref{eq:Retvrn} clearly does not have this property, except for $t=0$. Since $|F_{\text{R}}(t>0)|^{2\alpha}$ does not converge to $1$ with $\alpha$, there must exist one or more partitions that exactly cancel the decay of $F_{\text{R}}(t)$ for all $t>0$. We are forced to conclude that decay of the return amplitude does not provide an adequate picture of the local equilibration regime, which in our view remains a mysterious aspect of entanglement growth. Otherwise, our diagrams provide an intriguing structure to the phases of entanglement growth.

\section{Operator-eigenstate correlations}
\label{sec:locality}
In the preceding sections, we assume that both subsystems $A$ and $B$ are thermodynamically large, as parameterized by expansions in $e^{-S_{A(B)}}$. It follows that an operator $X$ obeys the ETH, eq. \eqref{eq:ETH}, with respect to eigenstates both of $H_A$ and of $H$. Furthermore, if $X$ is deep within subsystem $A$ it should have the same properties on the subsystem and the full system, at least over short times. This question was previously considered in ref.~\cite{shi_local_2023} where it was remarked that correlations between $X$ and $c$ are in principle unnecessary to satisfy these conditions based on a calculation that matched factors of the density-of-states. In fact, we show that this is not true and nontrivial correlations between $X$ and $c$ are necessary for physical consistency. We additionally show that these correlations encode the time at which operator $X$ receives information about subsystem $B$, and argue that this timescale is determined by the butterfly velocity. 

\subsection{The decay catastrophe}
\label{subsec:corr}
\begin{figure}
    \centering
    \includegraphics[width = 0.7\linewidth, viewport = 0 110 800 530, clip]{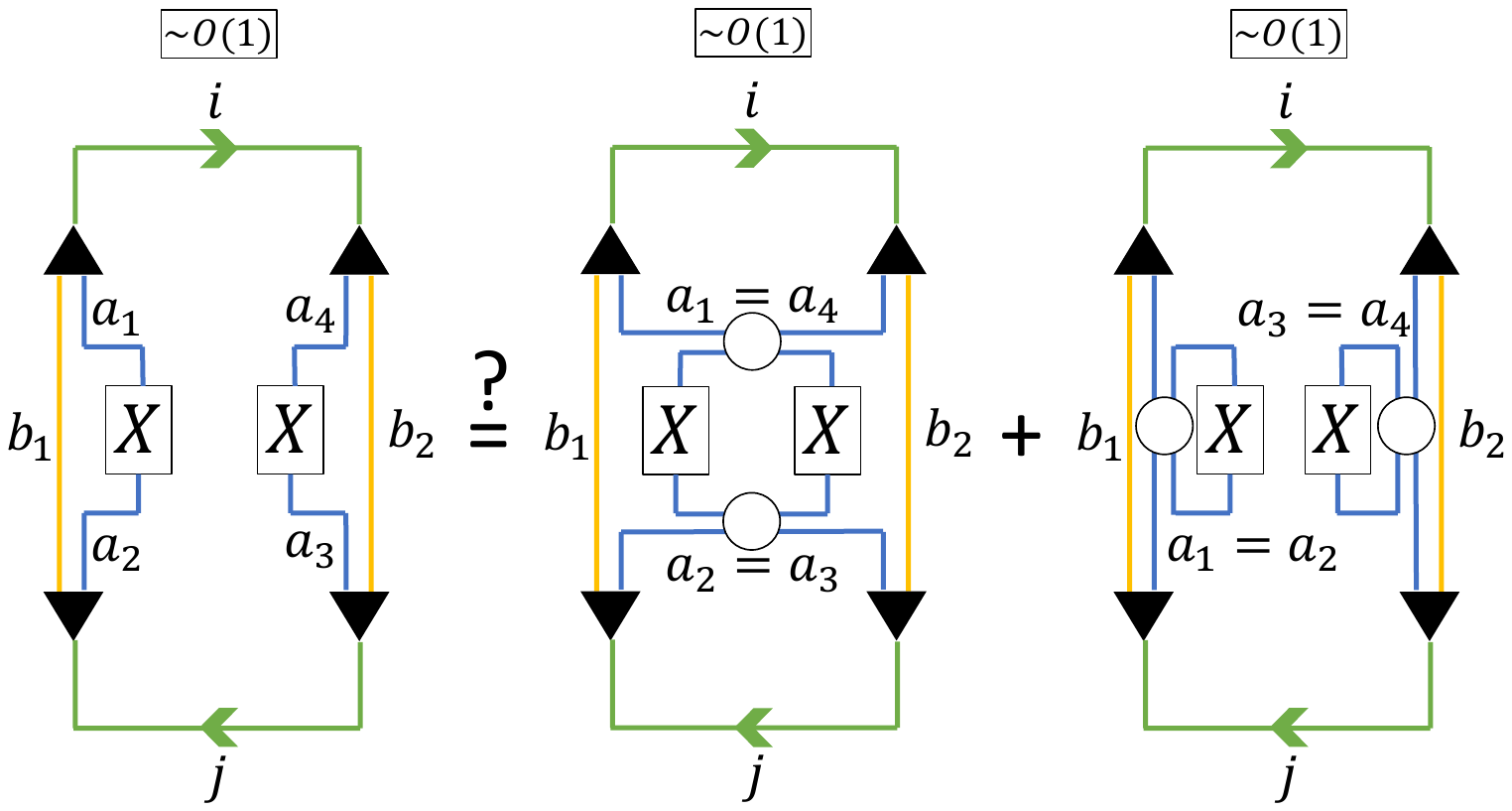}
    \caption{This figure depicts the false hypothesis that operator-eigenstate correlations factor if operators are taken to be far from the boundary between the subsystems.}
    \label{fig:catastrophe}
\end{figure}

When $X$ is deep within subsystem $A$ it may be tempting to assume that its matrix elements become uncorrelated with those of $c$. Consider the 2-point cumulant of $X$,
\begin{align}
\label{eq:2cumulant}
    &f_2(t+i\beta/2)=\expval{X(t)X}_{i}-\expval{X}^2_i=\sum_{[j]a_1a_2a_3a_4b_1b_2} c^{i}_{a_1b_1}X_{a_1a_2}c^{a_2b_1}_jc^{j}_{a_3b_2}X_{a_3a_4}c^{a_4b_2}_{i}e^{i\omega_{ij}t}.
\end{align}
If the elements of $X$ and $c$ are uncorrelated, their combined diagram must factor as in figure \ref{fig:catastrophe}. There are two diagrams that contribute at leading order, but we just focus on one, $(a_1a_4)(a_2a_3)$,  to illustrate the point. For $a_1\neq a_2$, this partition relates the 2-point cumulant on the whole system, $f_2$, to the 2-point cumulant of subsystem $A$, $f_2^{(A)}$. 

The factorization allows us to perform a direct computation,
\begin{align}
\label{eq:convolution}
    f_{(a_1a_4)(a_2a_3)}(t+i\beta/2) &= \sum_{[ja_1a_2b_1b_2]} c^{i}_{a_1b_1}c^{a_2b_1}_jc^{j}_{a_2b_2}c^{a_1b_2}_{i}X_{a_1a_2}X_{a_2a_1}e^{i\omega_{ij}t} \nonumber \\ 
    &\!\!\!\!\!=\int_{E_{a}E_{b_1}E_{b_2}\omega_{ij}\omega_{a}}F_{(a_1a_4)(a_2a_3)}(E_i;\cdots,\omega_{ij},\omega_{a})f_2^{(A)}(E_a;\omega_a)e^{i\omega_{ij}t-\beta\omega_{ij}/2} \nonumber \\
    &\!\!\!\!\!= F_{(a_1a_4)(a_2a_3)}(-t-i\beta/2)f_2^{(A)}(0)
\end{align}
where in the second line, we have defined $F$ to be the relevant eigenstate cumulant and emphasized its dependence on $\omega_{ij}$ and $\omega_a = E_{a_1}-E_{a_2}$. In general, decorrelating $c$ and $X$ causes correlation functions to become convolutions between wavefunction partitions and operator partitions in frequency space and products in time. 

Eq. (\ref{eq:convolution}) implies the obviously false conclusion that the entire dynamics of any local observable in subsystem $A$ far from the boundary are contained in the dynamics of a single eigenstate cumulant that generally depends on the nature of the interaction between the subsystems. This outcome is not a bug associated with the partition we have chosen but an inevitable consequence of decorrelating $c$ with $X$. Thus, we are forced to conclude that $X$ and $c$ are correlated even at short times and far separations. Consequently, the elements of $c$ contain non-local information about the whole system. In section \ref{subsec:butterfly}, we will square this conclusion with causality.
\subsection{The butterfly velocity}
\label{subsec:butterfly}
Existing literature discusses emergent causality in terms of the \textit{butterfly velocity}, which bounds the time at which out-of-time-ordered-correlators (OTOCs) between distant operators can decay~\cite{roberts_lieb-robinson_2016}. The simplest OTOC is $g(t) \equiv \expval{W(t)VW(t)V}$ for distinct operators $V$ and $W$ which commute at time $t=0$. Let $W$ and $V$ be separated by a distance $R$. The decay of $g(t)$ measures the nontriviality of the commutator $\left[W(t),V\right]$. Until information of $W$ has reached $V$, the commutator is approximately zero, and $g(t)$ cannot decay. The butterfly velocity $v_B$ is defined such that the time at which decay first occurs is $t_B = R/v_B$.

We will study an analog of the OTOC between $H_{AB}$ and $X$ to bound when $X$ has received information about subsystem $B$. Consider the Heisenberg evolution of our operator $X$ located a distance $R$ from the boundary between $A$ and $B$. First, we will evolve $X$ forwards in time under $H_{A}$ for a time $t$, then switch on the interaction $H_{AB}$ and reverse evolution back to time $t=0$. For $t<t_B$, $X$ should return approximately to its initial value. Concretely, we define
\begin{align}
    \label{eq:hfunction}
    h(t) \equiv \expval{e^{-iHt}e^{iH_At}Xe^{-iH_At}e^{iHt}X}_i.
\end{align}
For convenience, we will take $\expval{X}_i  = 0$ and $\expval{X^2}_i=1$. 

We justify using $h(t)$ to study the butterfly velocity by defining an interaction propagator $U_{I}(t)\equiv e^{i(H_A+H_B)t}e^{-iHt}$, for which $U_I'(t) = -iH_{AB}(t)U_I(t)$, where we have defined the time dependence of operators in terms of $H_A+H_B$. With this definition, we have:
\begin{align}
    h(t) &= \expval{U_{I}^\dagger(-t)XU_I(-t)X}_i,\nonumber \\
    h'(t) &= -i\expval{U_{I}^\dagger(-t)[H_{AB}(-t), X]U_I(-t)X}_i.
\end{align}
We can understand $h'(t)$ as an inner product on the space of Hermitian operators between $-iU_{I}^\dagger(-t)[H_{AB}(-t), X]U_I(-t)$ and $X$ and utilize Cauchy--Schwarz,
\begin{align}
\label{eq:OTOCInequality}
    \lvert h'(t)\rvert^2\leq -\expval{U_{I}^\dagger(-t) [H_{AB}(-t), X]^2U_I(-t)}_i\expval{X^2}_i = -\expval{ [H_{AB}, X(t)]^2}_i.
\end{align}
Thus, $h'(t)$ is bounded by the squared commutator of $X$ and $H_{AB}$, separated by a time of $t$. This result can be readily adapted into a bound on the decay of $h(t)$ by assuming the growth of the commutator square is monotonic prior to $t_B$,
\begin{align}
    \label{eq:inequalities}
    \left|h(t)-h(0)\right| &\leq \int_0^t\left|h'(t)\right|dt \nonumber \\
    &\leq \int_0^t\sqrt{-\expval{ [H_{AB}, X(t)]^2}_i}dt \nonumber \\
    &\leq t\sqrt{-\expval{ [H_{AB}, X(t)]^2}_i},
\end{align}
hence, the decay of $h(t)$ is bounded by the nontriviality of the squared commutator.

We define $t_{d}$ as the decay timescale of $h(t)$. When $t_B \gg t_{d}$, $h(t)$ will approach a step function. Consequently, for $\omega < t_{d}^{-1}$, $h(\omega)$ will behave as $\sim \operatorname{sinc}(R\omega/v_B)$, while for $\omega > t_{d}^{-1}$, $h(\omega)$ will have some cutoff that in general depends on the decay of $h(t)$. We want to understand correlations in the energy eigenspace of our system, so let us write eq. (\ref{eq:hfunction}) in terms of components of $X$ and $c$,
\begin{align}
\label{eq:htime}
    h(t)=\sum_{ja_1a_2a_3a_4b_1b_2} c^{i}_{a_1b_1}X_{a_1a_2}c^{a_2b_1}_jc^{j}_{a_3b_2}X_{a_3a_4}c^{a_4b_2}_{i}e^{i(\omega_{a_1a_2}-\omega_{ij})t}.
\end{align}
Eq. (\ref{eq:htime}) is nearly identical to eq. (\ref{eq:2cumulant}) and contains the same partitions. From the above considerations,
\begin{align}
\label{eq:hspectrum}
    h(\omega)&=2\pi\sum_{ja_1a_2a_3a_4b_1b_2} c^{i}_{a_1b_1}X_{a_1a_2}c^{a_2b_1}_jc^{j}_{a_3b_2}X_{a_3a_4}c^{a_4b_2}_{i}\delta(\omega_{a_1a_2}-\omega_{ij} - \omega) \nonumber \\ &\sim \operatorname{sinc}(R\omega/v_B)\Lambda_{t_d^{-1}}(\omega)
\end{align}
where $\Lambda_{t_d^{-1}}(\omega)$ is a cutoff function introduced by the finite rate of decay of $h(t)$. 

Eqs. (\ref{eq:htime}) and (\ref{eq:hspectrum}) ensure a nontrivial correlated structure between $X$ and $c$ that is relevant at arbitrarily short times and arbitrarily long distances. However, in contrast to previous quantities we have studied, it is not obvious that the right-hand sides of eqs. (\ref{eq:htime}) and (\ref{eq:hspectrum}) have any particular interpretation in terms of individual generalized free cumulants indicating that our diagrammatic approach does not capture the full picture. Our analysis mirrors that of ref.~\cite{chan_eigenstate_2019}, which considered the eigenbasis representation of the OTOC between spatially separated operators in the context of the ETH and found an analogous sinc-like universal form. More recently, ref.~\cite{hahn_statistical_2023} was able to capture the physics of the butterfly velocity from a pure wavefunction perspective analogous to our own, but considering correlations between distinct partitionings of the system. It would be interesting to connect their approach to our own.

\section{Conclusions and discussion}
The eigenstate thermalization hypothesis was originally introduced to justify the application of statistical mechanics to quantum many-body systems. However, in recent years, especially given the connection to free probability theory, it has become clear that the ETH does more than justify statistical mechanics. Free probability theory has previously emerged in physics in the context of large $N$ limits~\cite{gopakumar_mastering_1994, hruza_coherent_2023}. Then, the ETH is perhaps best understood as a set of phenomena associated with an emergent $e^{-S}$ expansion in quantum many-body systems. This interpretation of the ETH is automatic in the matrix model formulation developed in ref.~\cite{jafferis_matrix_2023}. In this work, we have presented the many-body Berry's conjecture as a reformulation of the ETH, which is traditionally understood in the sense of eq. (\ref{eq:ETH}). However, we may alternately view eq. \eqref{eq:ETH}, the many-body Berry's conjecture, and the various results in this paper and the literature as just aspects of a large $e^{S}$ limit. A mature understanding of this limit remains to be developed.

In this paper, we have formulated the eigenstate thermalization hypothesis and the ergodic bipartition in terms of a many-body Berry's conjecture, the hypothesis that eigenstates of chaotic systems are random vectors up to the symmetry constraints of the system, and we have argued that the MBBC is the natural quantum generalization of the ergodic hypothesis. We showed how this approach naturally leads to a diagrammatic formalism developed in the language of free probability, which is our main result. We demonstrate the power of our formalism by showing that systems relax to a universal reduced density matrix and obey the Page curve at late times, thus establishing irreversible thermalization under reversible unitary evolution. We also establish the subsystem ETH as a consequence of the MBBC. We also discuss the role that locality plays in the ETH and develop connections to butterfly velocities and entanglement growth.

Our considerations in this paper are generic, and our results should apply to any system where ergodicity is not explicitly broken, e.g., through many-body localization. Certainly much more can be done on a similarly generic footing. While we have largely ignored the functional forms of eigenstate cumulants ($F$ functions), they are necessary to understand the time-dependent dynamics of thermalization and entanglement. There is some work that suggests eigenstate cumulants, in some cases, have universal forms that depend on the number of dimensions, locality of interactions, etc but not the model considered~\cite{shi_local_2023}. Additionally, while the operator cumulants ($f$ functions) are generally model dependent, their high-frequency cutoffs appear connected to universal properties of scrambling, and future work may clarify this connection~\cite{murthy_bounds_2019, pappalardi_eigenstate_2022}. Other work may clarify the connection between the butterfly velocity and spectral correlations and uncover correlated structures that exist beyond the ETH~\cite{chan_eigenstate_2019,hahn_statistical_2023}. We hope the formalism we have developed will aid in the classification of generic properties of quantum many-body systems.

On the other hand, future work may focus on concrete realizations of the above results. Exact functional forms of generalized free cumulants may be possible to derive within toy models, such as the SYK model. Applications of the ETH to conformal field theories and to holographic systems~\cite{lashkari_eigenstate_2016} have revealed important structure in those systems and provided insights into the structure of black holes~\cite{kudler-flam_distinguishing_2021}. Additionally, numerical evidence for our work is severely limited by the capabilities of exact diagonalization. Modern techniques and greater computational power could provide clean demonstrations of our results in spin chains or other toy models~\cite{hosur_polynomial_2021, luo_probing_2023}.

Some techniques in this paper are reminiscent of others in the literature. In section \ref{sec:thermalization}, our means of imposing equilibrium on non-equilibrium partitions by matching opposing time dependences reproduces the equilibrated pure state formalism of ref.~\cite{liu_entanglement_2021}. In another case, refs.~\cite{hahn_statistical_2023, de_boer_principle_2023} study the statistical properties of eigenstates via an application of the \textit{principle of maximum entropy}. Ref.~\cite{hahn_statistical_2023} uses this principle to study propagators at early times and studies the behavior of OTOCs in Floquet quantum circuits. Ref.~\cite{de_boer_principle_2023} argues that the various principles of the ETH may each be understood as consequences of a maximum entropy principle and derives a special case of our eq. \eqref{eq:RDMEq} to study saddle-points in the gravitational path integral. In our view, the principle of maximum entropy is a more formal and fundamental approach to the heuristic arguments we present in sections \ref{sec:motivation} and \ref{sec:diagrams} that may even extend the validity of our work. 

\acknowledgments 

This work was supported by the National Science Foundation grant no. DMR 2047193. We are grateful to Xiao-Liang Qi, Chaitanya Murthy, Shinsei Ryu, Philip Crowley, and Roland Speicher for insightful discussions. We thank Mark Srednicki for pointing out the relevance of ref.~\cite{jafferis_matrix_2023} to our work. We are especially grateful to Tarun Grover for useful discussions and comments on the draft.

\appendix
\section{Free probability in quantum chaos}
\label{app:FPT}
In this appendix, we discuss the principles of free probability theory and their emergence in quantum chaos. We do not present rigorous or technical arguments but instead discuss informally how key ideas, namely \textit{freeness}, \textit{free cumulants}, and \textit{random matrices} arise, and obtain analogs in the study of quantum chaos. In particular, we wish to motivate why the decay of out-of-time-ordered correlators (OTOCs), the eigenstate thermalization hypothesis (ETH), and emergent rotational symmetry are, in essence, equivalent definitions of chaos and avatars of free probability theory. Except where another citation is provided, we direct readers to ref.~\cite{mingo_free_2017} for background on free probability.

First, we discuss what it means to have a probability theory in a non-commutative setting\footnote{For an elaboration on the ideas in this paragraph, we direct readers to ref.~\cite{tao_254a_2010}.}. The traditional foundation of classical probability theory starts with a sample space of possible events and a rule for assigning probabilities to subsets of events. Non-commutative probability theory instead starts with an algebra of observables and their expectation values. The rules of non-commutative probability theory generalize the familiar rules of quantum mechanics where we are given a (pure or mixed) density matrix and some (generically non-commuting) observables whose expectation values are of interest to us. Non-commutative probability also reduces to classical probability theory in the limit that observables of interest commute. Often in quantum systems, particularly when $\hbar$ can be considered small, operators are approximately commutative and principles of classical probability theory such as sample spaces and (classical) independence become useful emergent descriptions of statistical physics. Free probability theory concerns itself with, in some sense, an opposite limit in which observables are as non-commutative as possible. We will see that this notion of maximal non-commutativity can be made precise and that quantum mechanics indeed contains such a limit. 

Consider two algebras of observables, $\mathcal X$ and $\mathcal Y$, and a state $\expval{\cdots}$ that satisfies $\expval{ZW} = \expval{WZ}$. As an example, one can take $\mathcal Y$ to be the algebra generated by Pauli operators on a single site of a spin chain, $\mathcal X$ to be the algebra generated by the Pauli operators on a different site evolved far forward in time, and the state to be the conventional thermal regulator discussed in row (i) of table~\ref{tab:mytable}. We will return to this example. $\mathcal X$ and $\mathcal Y$ are considered \textit{freely independent}, or \textit{free}, if
\begin{align}
\label{eq:algebraFreeness}
    \expval{X_1Y_1\cdots X_nY_n} = 0 \quad \text{for all} \quad X_i\in\mathcal X,\, Y_i \in\mathcal Y  \quad\text{such that}\quad \expval{X_i} = \expval{Y_i} = 0.
\end{align}
The expression $\expval{X_1Y_1\cdots X_nY_n}$ is known as an alternating moment and so long as $\mathcal X$ and $\mathcal Y$ are closed algebras any mixed moment of operators between them reduces to such an expression. It is not obvious, but eq.~\eqref{eq:algebraFreeness} encodes maximal non-commutativity between $\mathcal X$ and $\mathcal Y$. If there were any nontrivial algebraic relations between elements of $\mathcal X$ and $\mathcal Y$, e.g. $XY = YX$ for some $X$, $Y$, then the definition~\eqref{eq:algebraFreeness} would imply that at least one of $X$ and $Y$ is a scalar and thus that the algebraic relation is trivial, in contradiction with the assumption of nontriviality. We sketch this proof just for commutators. Assume that $\mathcal X$ and $\mathcal{Y}$ satisfy eq.~\eqref{eq:algebraFreeness} but that $XY = YX$ for some non-scalar $X$, $Y$. Consider the fluctuations $\delta X \equiv X-\expval{X}$, $\delta Y \equiv Y-\expval{Y}$. By~\eqref{eq:algebraFreeness},
\begin{align}
\label{eq:firstStep}
    \expval{\left(\delta X\delta Y\right)^n\left(\delta Y\right)^{m-n}} = \expval{\delta X\delta Y\cdots \delta X\delta Y^{m-n+1}} = 0.
\end{align}
Since $X$ and $Y$ commute, eq.~\eqref{eq:firstStep} reduces to,
\begin{align}
    \expval{\delta X^n\delta Y^m} = 0.
\end{align}
However, eq.~\eqref{eq:algebraFreeness} implies for $n=1$,
\begin{align}
\label{eq:secondFC}
    \expval{\left(X_1 - \expval{X_1}\right)\left(Y_1 - \expval{Y_1}\right)} = 0 \quad\implies\quad \expval{X_1Y_1} = \expval{X_1}\expval{Y_1}
\end{align}
and taking $X_1=\delta X^n$, $Y_1=\delta Y^m$,
\begin{align}
\label{eq:centralMoment}
    \expval{\delta X^n\delta Y^m} = \expval{\delta X^n}\expval{\delta Y^m} = 0.
\end{align}
However, eq.~\eqref{eq:centralMoment} states that the product of arbitrary central moments of $X$ and $Y$ vanish. This can only be the case if at least one of $X$, $Y$ has vanishing central moments and thus is a scalar\footnote{We assume faithfulness: $\expval{X^2} = 0 \implies X = 0$. Freeness is defined with respect to a given state and cannot say much about operators for which the chosen state is not faithful.}. Hence, our hypothesis is contradicted, and no nontrivial element of $\mathcal{X}$ can commute with any nontrivial element of $\mathcal{Y}$. This argument can be extended to any algebraic relation between elements of $\mathcal{X}$ and $\mathcal{Y}$ and thus establishes the idea of maximal non-commutativity. The notion of freeness can also be extended from algebras to pairs of operators as freeness of the subalgebras they generate. $X$ and $Y$ are free if,
\begin{align}
\label{eq:operatorFreeness}
    \expval{\left(X-\expval{X}\right)^{p_1}\left(Y-\expval{Y}\right)^{q_1}\cdots \left(X-\expval{X}\right)^{p_n}\left(Y-\expval{Y}\right)^{q_n}} = 0 \quad \text{for all} \quad n\geq 1,
\end{align}
for positive integer exponents $p_m,\, q_m$. A pair of algebras are free if and only if any pair of their operators are free\footnote{freeness can be further extended to a collection of any number of algebras or operators}.

Freeness has an analogy in quantum chaos. A many-body system is considered chaotic if for simple, few-body observables $X$, $Y$,
\begin{align}
\label{eq:otocFreeness}
    \expval{\left[\left(X(t)-\expval{X(t)}_\beta\right)\left(Y(0)-\expval{Y(0)}_\beta\right)\right]^q}_\beta \to 0 \quad\text{for all}\quad q\geq 1
\end{align}
for $t>t_{\text{scr}}$, where $t_{\text{scr}}$ is known as the scrambling time of the system~\cite{roberts_chaos_2017} and we have used the conventional thermal regulator. ~\eqref{eq:otocFreeness} is formulated to measure the non-commutativity of $X(t)$ with $Y(0)$. Eq.~\eqref{eq:otocFreeness} is essentially the definition of the quantum butterfly effect, in which the decay of OTOCs characterizes the ability for the unitary evolution of a system to scramble quantum information~\cite{hosur_chaos_2016}. In this light, the quantum butterfly effect is a consequence of the asymptotic freeness between simple few-body observables at long time separations. If the example of Pauli operators are to constitute freely independent algebras, then their OTOCs must decay.

The definition of freeness given in eq.~\eqref{eq:algebraFreeness} is the maximally non-commutative analog of the classical definition of independence between two variables. That is, $\mathcal X$ and $\mathcal Y$ are \textit{classically independent}, or simply \textit{independent}, if,
\begin{align}
\label{eq:algebraIndependence}
    \expval{X_1Y_1\cdots X_nY_n} = \expval{X_1\cdots X_n}\expval{Y_1\cdots Y_n} \quad \text{for all} \quad X_i\in\mathcal X,\, Y_i \in\mathcal Y.
\end{align}
Eqs.~\eqref{eq:algebraFreeness} and~\eqref{eq:algebraIndependence} are both rules for computing higher mixed moments from lower moments. Such a rule is obtained from eq.~\eqref{eq:algebraFreeness} by performing the binomial expansion of \\${\expval{(X_1-\expval{X_1})(Y_1-\expval{Y_1})\cdots}=0}$. The analogous rule from~\eqref{eq:algebraIndependence} is simply that mixed moments factorize. Whereas eq.~\eqref{eq:algebraFreeness} implied that elements of $\mathcal X$ and $\mathcal Y$ have no nontrivial algebraic relations, eq.~\eqref{eq:algebraIndependence} implies that all elements of $\mathcal X$ and $\mathcal Y$ satisfy the specific relation that they commute, i.e. $XY = YX$. This result is straightforward to prove. Consider two alternating moments,
\begin{align}
      \expval{X_1Y_1\cdots X_{m-1}Y_{m-1}X_mY_mX_{m+1}Y_{m+1}\cdots X_nY_n} &=  \expval{X_1\cdots X_n}\expval{Y_1\cdots Y_n}, \nonumber \\ 
      \expval{X_1Y_1\cdots X_{m-1}(Y_{m-1}Y_m)(X_mX_{m+1})Y_{m+1}\cdots X_nY_n} &= \expval{X_1\cdots X_n}\expval{Y_1\cdots Y_n}
\end{align}
where both are equal per eq.~\eqref{eq:algebraIndependence}. Then,
\begin{align}
\label{eq:vanishCommute}
    &\expval{\cdots X_{m-1}Y_{m-1}X_mY_mX_{m+1}Y_{m+1}\cdots} - \expval{\cdots X_{m-1}(Y_{m-1}Y_m)(X_mX_{m+1})Y_{m+1}\cdots} = 0 \nonumber \\
    &\quad\implies\quad \expval{\cdots [X_m,Y_m]\cdots}=0.
\end{align}
Since $X_1,Y_1\cdots,X_{m-1},Y_{m-1},X_{m+1},Y_{m+1},\cdots,X_n,Y_n$ are arbitrary, eq.~\eqref{eq:vanishCommute} can only hold if $[X_m,Y_m]=0$ in general. Hence, eq.~\eqref{eq:algebraIndependence} enforces commutativity. 

Many concepts of classical probability have exact analogs in free probability. Eq.~\eqref{eq:operatorFreeness} is analogous to the factorization of mixed moments of classically independent variables $\expval{X^pY^q} = \expval{X^p}\expval{Y^q}$. The classical probability distribution of an operator is instead replaced with its spectrum. The convolution of probability distributions for sums of independent operators is replaced with a free additive convolution of spectra for sums of free operators. Of key importance in classical and free probability is the existence of cumulants, which to a physicist serve as the building blocks of a theory. Cumulants also compactly formulate the rules for computing higher moments from lower moments mentioned above for classically or freely independent variables. Whereas the vanishing of classical cumulants indicates independence, the vanishing of free cumulants indicates freeness. Furthermore, while classical cumulants admit a combinatorial interpretation in terms of set partitions, free cumulants can be interpreted in terms of \textit{non-crossing} partitions. As an example we consider the cumulant decompositions of the moment $\expval{XYXY}$. First, in the classical case,
\begin{align}
\label{eq:crossPart}
    \expval{XYXY} &\equiv \raisebox{-0.77em}{\begin{tikzpicture}
    \begin{scope}[xshift=-0.35em]
    \draw (0em,0) -- ++ (0,-1.0em) -| (0.9em,0);
    \draw (0.9em,-1em) -| (1.8em,0);
    \draw (1.8em,-1em) -| (2.7em,0);
    \end{scope}
    \node[fill=white, inner sep=1pt] at (1em,0) {$\expval{XYXY}$};    
  \end{tikzpicture}} \nonumber \\
  &\quad + \raisebox{-0.77em}{\begin{tikzpicture}
    \begin{scope}[xshift=-0.35em]
    \draw (0em,0) -- ++ (0,-1.1em) -| (0.9em,0);
    \draw (0.9em,-1.1em) -| (1.8em,0);
    \draw (2.7em,-0.9em) -- ++ (0,0.9em);
    \end{scope}
    \node[fill=white, inner sep=1pt] at (1em,0) {$\expval{XYXY}$};    
  \end{tikzpicture}} 
  + \raisebox{-0.77em}{\begin{tikzpicture}
    \begin{scope}[xshift=-0.35em]
    \draw (0em,0) -- ++ (0,-1.1em) -| (0.9em,0);
    \draw (0.9em,-1.1em) -| (2.7em,0);
    \draw (1.8em,-0.9em) -- ++ (0,0.9em);
    \end{scope}
    \node[fill=white, inner sep=1pt] at (1em,0) {$\expval{XYXY}$};    
  \end{tikzpicture}}
  + \raisebox{-0.77em}{\begin{tikzpicture}
    \begin{scope}[xshift=-0.35em]
    \draw (0em,0) -- ++ (0,-1.1em) -| (1.8em,0);
    \draw (1.8em,-1.1em) -| (2.7em,0);
    \draw (0.9em,-0.9em) -- ++ (0,0.9em);
    \end{scope}
    \node[fill=white, inner sep=1pt] at (1em,0) {$\expval{XYXY}$};    
  \end{tikzpicture}}
  + \raisebox{-0.77em}{\begin{tikzpicture}
    \begin{scope}[xshift=-0.35em]
    \draw (0.9em,0) -- ++ (0,-1.1em) -| (1.8em,0);
    \draw (1.8em,-1.1em) -| (2.7em,0);
    \draw (0.0em,-0.9em) -- ++ (0,0.9em);
    \end{scope}
    \node[fill=white, inner sep=1pt] at (1em,0) {$\expval{XYXY}$};    
  \end{tikzpicture}} \nonumber \\
  &\quad + \raisebox{-0.77em}{\begin{tikzpicture}
    \begin{scope}[xshift=-0.35em]
    \draw (0em,0) -- ++ (0,-1.0em) -| (0.9em,0);
    \draw (1.8em,0) -- ++ (0,-1.0em) -| (2.7em,0);
    \end{scope}
    \node[fill=white, inner sep=1pt] at (1em,0) {$\expval{XYXY}$};    
  \end{tikzpicture}}
  + \raisebox{-0.77em}{\begin{tikzpicture}
    \begin{scope}[xshift=-0.35em]
    \draw (0em,0) -- ++ (0,-1.1em) -| (1.8em,0);
    \draw (0.9em,0) -- ++ (0,-0.9em) -| (2.7em,0);
    \end{scope}
    \node[fill=white, inner sep=1pt] at (1em,0) {$\expval{XYXY}$};    
  \end{tikzpicture}}
  + \raisebox{-0.77em}{\begin{tikzpicture}
    \begin{scope}[xshift=-0.35em]
    \draw (0em,0) -- ++ (0,-1.1em) -| (2.7em,0);
    \draw (0.9em,0) -- ++ (0,-0.9em) -| (1.8em,0);
    \end{scope}
    \node[fill=white, inner sep=1pt] at (1em,0) {$\expval{XYXY}$}; 
  \end{tikzpicture}} \nonumber \\
  &\quad + \raisebox{-0.77em}{\begin{tikzpicture}
    \begin{scope}[xshift=-0.35em]
    \draw (0.0em,-1em) -- ++ (0,1em);
    \draw (0.9em,-1em) -- ++ (0,1em);
    \draw (1.8em,-1em) -- ++ (0,1em);
    \draw (2.7em,-1em) -- ++ (0,1em);
    \end{scope}
    \node[fill=white, inner sep=1pt] at (1em,0) {$\expval{XYXY}$}; 
  \end{tikzpicture}} \nonumber \\
  &\equiv k_{X^2Y^2} \nonumber \\
  &\quad + 2k_{X^2Y}k_{Y} + 2k_{XY^2}k_{X} \nonumber \\
  &\quad + k_{X^2}k_{Y^2} + 2k_{XY}^2 \nonumber \\
  &\quad + k_X^2k_Y^2
\end{align}
where we've used $k$ to denote classical cumulants. Then in the free case,
\begin{align}
\label{eq:noncrossPart}
    \expval{XYXY} &\equiv \raisebox{-0.33em}{\begin{tikzpicture}
    \begin{scope}[xshift=-0.25em]
    \draw (0em,0) -- ++ (0,1.0em) -| (0.9em,0);
    \draw (0.9em,1em) -| (1.8em,0);
    \draw (1.8em,1em) -| (2.7em,0);
    \end{scope}
    \node[fill=white, inner sep=1pt] at (1em,0) {$\expval{XYXY}$};    
  \end{tikzpicture}} \nonumber \\
  &\quad + \raisebox{-0.33em}{\begin{tikzpicture}
    \begin{scope}[xshift=-0.25em]
    \draw (0em,0) -- ++ (0,1.1em) -| (0.9em,0);
    \draw (0.9em,1.1em) -| (1.8em,0);
    \draw (2.7em,0.9em) -- ++ (0,-0.9em);
    \end{scope}
    \node[fill=white, inner sep=1pt] at (1em,0) {$\expval{XYXY}$};    
  \end{tikzpicture}} 
  + \raisebox{-0.33em}{\begin{tikzpicture}
    \begin{scope}[xshift=-0.25em]
    \draw (0em,0) -- ++ (0,1.1em) -| (0.9em,0);
    \draw (0.9em,1.1em) -| (2.7em,0);
    \draw (1.8em,0.9em) -- ++ (0,-0.9em);
    \end{scope}
    \node[fill=white, inner sep=1pt] at (1em,0) {$\expval{XYXY}$};    
  \end{tikzpicture}}
  + \raisebox{-0.33em}{\begin{tikzpicture}
    \begin{scope}[xshift=-0.25em]
    \draw (0em,0) -- ++ (0,1.1em) -| (1.8em,0);
    \draw (1.8em,1.1em) -| (2.7em,0);
    \draw (0.9em,0.9em) -- ++ (0,-0.9em);
    \end{scope}
    \node[fill=white, inner sep=1pt] at (1em,0) {$\expval{XYXY}$};    
  \end{tikzpicture}}
  + \raisebox{-0.33em}{\begin{tikzpicture}
    \begin{scope}[xshift=-0.25em]
    \draw (0.9em,0) -- ++ (0,1.1em) -| (1.8em,0);
    \draw (1.8em,1.1em) -| (2.7em,0);
    \draw (0.0em,0.9em) -- ++ (0,-0.9em);
    \end{scope}
    \node[fill=white, inner sep=1pt] at (1em,0) {$\expval{XYXY}$};    
  \end{tikzpicture}} \nonumber \\
  &\quad + \raisebox{-0.33em}{\begin{tikzpicture}
    \begin{scope}[xshift=-0.25em]
    \draw (0em,0) -- ++ (0,1.0em) -| (0.9em,0);
    \draw (1.8em,0) -- ++ (0,1.0em) -| (2.7em,0);
    \end{scope}
    \node[fill=white, inner sep=1pt] at (1em,0) {$\expval{XYXY}$};    
  \end{tikzpicture}}
  + \raisebox{-0.33em}{\begin{tikzpicture}
    \begin{scope}[xshift=-0.25em]
    \draw (0em,0) -- ++ (0,1.1em) -| (2.7em,0);
    \draw (0.9em,0) -- ++ (0,0.9em) -| (1.8em,0);
    \end{scope}
    \node[fill=white, inner sep=1pt] at (1em,0) {$\expval{XYXY}$}; 
  \end{tikzpicture}} \nonumber \\
  &\quad + \raisebox{-0.33em}{\begin{tikzpicture}
    \begin{scope}[xshift=-0.25em]
    \draw (0.0em,1em) -- ++ (0,-1em);
    \draw (0.9em,1em) -- ++ (0,-1em);
    \draw (1.8em,1em) -- ++ (0,-1em);
    \draw (2.7em,1em) -- ++ (0,-1em);
    \end{scope}
    \node[fill=white, inner sep=1pt] at (1em,0) {$\expval{XYXY}$}; 
  \end{tikzpicture}} \nonumber \\
  &\equiv f_{XYXY} \nonumber \\
  &\quad + 2f_{X^2Y}f_{Y} + 2f_{XY^2}f_{X} \nonumber \\
  &\quad + 2f_{XY}^2 \nonumber \\
  &\quad + f_X^2f_Y^2
\end{align}
where we've used $f$ to denote free cumulants and we've distinguished ordinary set partitions from non-crossing partitions by drawing the former below the line and the latter above it. 

Hopefully the rules of the above partitions are clear, but there are a few aspects which should be emphasized. It is straightforward to verify that the vanishing of free cumulants, defined by the non-crossing partitions, implies freeness in the sense of eq.~\eqref{eq:operatorFreeness}. Until 3\textsuperscript{rd} order, free and classical cumulants are identical. At 4\textsuperscript{th} order and above, classical cumulants become undefined for non-commutative operators while free cumulants become sensitive to order (up to cyclic permutations), i.e $f_{XYXY} = f_{YXYX} \neq f_{X^2Y^2} = f_{Y^2X^2}$. The difference is reflected in the fact that eqs.~\eqref{eq:crossPart} and~\eqref{eq:noncrossPart} differ by a single partition. As we discuss in section~\ref{subsec:cactus}, free cumulants are crucial to the definition of the ETH. However, to justify the form given in eq.~\eqref{eq:ETH}, we will turn to a concrete construction of free probability.

Random matrices are known to model free probability in their large $N$ limit. A finite $N\times N$ hermitian matrix can be classically sampled from a distribution over its $N$ real diagonal and $N(N-1)$ complex off-diagonal elements\footnote{One can similarly consider real symmetric matrices and orthogonal rotations or self-adjoint quaternionic matrices and symplectic rotations, but in this appendix we will stick to hermitian matrices and unitary rotations.}. Of particular importance are rotationally invariant probability distributions. That is, for any matrix $X$ and unitary rotation $U$,
\begin{align}
\label{eq:rotationalInvariance}
    p(X) = p(U^\dagger XU)
\end{align}
where $p$ is the probability density. Next, to take the large $N$ limit one has to specify a sequence of $N\times N$ matrices $X^{(N)}$ sampled from a sequence of distributions $p^{(N)}$, $N\in \mathbbm Z$, such that as $N\to\infty$, the spectrum of $X^{(N)}$ converges to a well-defined limit. Then consider two sequences of random matrices with large $N$ limits, $X^{(N)}$ and $Y^{(N)}$. If for each $N$, the elements of $X^{(N)}$ and $Y^{(N)}$ are sampled independently and from rotationally invariant probability distributions, then,
\begin{align}
\label{eq:asymptoticFreeness}
    \lim_{N\to\infty}\expval{\prod_{i=1}^{n}\left(X^{(N)}-\expval{X^{(N)}}\right)^{p_i}\left(Y^{(N)}-\expval{Y^{(N)}}\right)^{q_i}} = 0 \quad \text{for all} \quad n\geq 1.
\end{align}
Eq.~\eqref{eq:asymptoticFreeness} is simply the freeness condition eq.~\eqref{eq:operatorFreeness} asymptotically in the large $N$ limit. Since we have considered rotationally invariant distributions, the specific state $\expval{\cdots}$, is unimportant. 

What we have seen is the remarkable fact that hermitian random matrices whose elements are sampled classically independently from a rotationally invariant ensemble, become freely independent in the large $N$ limit. Rotational invariance is crucial. Asymptotic freeness in the large $N$ limit holds if and only if the ensemble is rotationally invariant. Hence, where in eq.~\eqref{eq:otocFreeness} we found that freeness occurs at long time separations for simple observables, we should expect that at small frequencies the matrix elements of those observables exhibit an emergent rotational invariance.

Lastly, we can introduce the expression for free cumulants in terms of matrix elements. For a sequence of matrices, $X^{(N)}$, that satisfies rotational invariance and has a well-defined limit spectrum, its free cumulants in the large $N$ limit are given by the classical cumulants $k_n$ of its matrix elements~\cite{collins_second_2006},
\begin{align}
\label{eq:matrixFreeCumulants}
    f_{X^n} = \lim_{N\to\infty} N^{n-1}k_{n}\left(X^{(N)}_{i_1i_2}, X^{(N)}_{i_2i_3},\cdots,X^{(N)}_{i_ni_1}\right)
\end{align}
for any choice of distinct indices $i_1,\dots,i_n$. Ref.~\cite{collins_second_2006} presents the result in terms of classical cumulants, but as discussed in ref.~\cite{foini_eigenstate_2019} the only products of matrix elements whose expectations are rotationally invariant are those whose indices are cyclic. Then, instead we have,
\begin{align}
\label{eq:matrixFreeCumulants2}
    f_{X^n} = \lim_{N\to\infty} N^{n-1}\textsc{E}(X^{(N)}_{i_1i_2} X^{(N)}_{i_2i_3}\cdots X^{(N)}_{i_ni_1})
\end{align}
for any choice of distinct indices $i_1,\dots,i_n$, where $\textsc{E}$ represents the classical expectation value. It is clear that eq.~\eqref{eq:ETH} is the chaotic analogy to eq.~\eqref{eq:matrixFreeCumulants2} where the theoretical expectation value has simply been replaced with an empirical average and $N$ with $e^{S}$. 

Hence we have seen how key concepts of quantum chaos: the butterfly effect, the ETH, and emergent rotational invariance are each avatars of free probabilistic limit inherent in chaos of quantum many-body systems. Yet, there are many more tools in the free probability toolbox. We suspect broader awareness of the subject will help those tools find applications in physics.

\section{Saddle-points in the ETH}
\label{app:saddlePoints}

We briefly review the use of saddle-point integration in the context of the ETH. The basic principle of saddle-point integration is that the total integral of a rapidly varying integrand is sharply concentrated around its peak(s) or of a rapidly oscillating integrand around its point(s) of stationary phase. We refer to either peaks or stationary points as saddle-points. In the cases we consider, the contributions to integrals from regions away from the saddle-point will be suppressed by the system size, and thus, saddle-point integration becomes exact in the thermodynamic limit, at least within a logarithm.

We assume that our system has a Hamiltonian $H$ with $\mathcal V$ degrees of freedom (volume). Extensive thermodynamic quantities such as the energy $E = \expval{H}$, microcanonical entropy $S(E) = \ln\left[\sum_i \delta(E_i-E)\right]$, and heat capacity $C(E)\equiv -\beta^2 \left[S''(E)\right]^{-1}$, all scale $\sim\mathcal{O}(\mathcal V)$. Intensive thermodynamic quantities, such as the inverse temperature $\beta \equiv S'(E)$ and the expectation values of local operators $\expval{X}$ are $\sim\mathcal{O}(1)$. 

The diagonal elements of ETH satisfying operators are equal to their microcanonical expectation value,
\begin{equation}
    \label{eq:ETH1}
    X_{ii} = e^{-S(E_i)}\sum_{j}\delta(E_j-E_i)X_j + \mathcal{O}(e^{-S}) \equiv f_1(E_i) + \mathcal{O}(e^{-S}).
\end{equation}
Then eigenstates, by definition, obtain the same equilibrium properties as the microcanonical ensemble. We can also check the expectation values in the canonical ensemble,
\begin{align}
\label{eq:canonical}
    \expval{X}_\beta \equiv \frac{\sum_{i}e^{-\beta E_i}X_{ii}}{\sum_{i}e^{-\beta E_i}} = \frac{\int_E e^{S(E)-\beta E}f_1(E) + \mathcal{O}(e^{-S/2})}{\int_E e^{S(E)-\beta E}}
\end{align}
where we have exploited \eqref{eq:ETH1} and the definition of microcanonical entropy to turn the sums into integrals. We find the saddle-point of the integral by setting the derivative of the integrand to zero and obtain
\begin{align}
\label{eq:condition}
    S'(E_\beta)-\beta +\frac{f_1'(E_\beta)}{f_1(E_\beta)} \implies S'(E_\beta) = \beta + \mathcal{O}(1/{\mathcal V}).
\end{align}
Taking only the saddle-point value of each integral we find,
\begin{align}
\label{eq:ensembleEquivalence}
    \expval{X}_\beta = \frac{e^{S(E_\beta)-\beta E_\beta}f_1(E_\beta)}{e^{S(E_\beta)-\beta E_\beta}} = f_1(E_\beta).
\end{align}
Eq. \eqref{eq:ensembleEquivalence} implies that the properties of the canonical ensemble with temperature $\beta$ are equivalent to the properties of the microcanonical ensemble at a specific energy $E_\beta$. From eq. \eqref{eq:condition}, we can realize that the saddle-point associated to the canonical inverse temperature $\beta$ is equivalent to the microcanonical definition of temperature to leading order in $1/{\mathcal V}$. Thus, we establish ensemble equivalence in the ETH. This result extends readily to the non-equilibrium functions $f_n$ for their slow dependence on the total system energy. 

However, we have not considered the magnitude of corrections away from the saddle-point. To do so, we consider the Taylor expansion of the exponent about the saddle-point, 
\begin{align}
    S(E)-\beta E = S(E_\beta)-\beta E_\beta -\frac{\beta^2(E-E_\beta)^2}{2C} + \sum_{m=3}^{\infty} \frac{S^{(m)}(E_\beta)(E-E_\beta)^m}{m!}.
\end{align}
If we truncate this series to second order, we find that $E$ is Gaussian distributed about $E_\beta$ with a width $\sqrt{C/\beta^2}$ that scales $\sim\mathcal{O}({\mathcal V}^{1/2})$. However, since $f_1'(E)$ is $\sim\mathcal{O}(1/{\mathcal V})$, contributions away from the saddle-point in the Gaussian approximation will only contribute at most at order $\sim\mathcal{O}({\mathcal V}^{-1/2})$. If we assume that $E$ stays close to $E_\beta$ at any order in the expansion, self-consistently including higher derivatives of $S$ results in an asymptotic series about ${\mathcal V}\rightarrow\infty$ which may be evaluated via Feynman diagrams. However, the higher order terms will be suppressed by further powers of ${\mathcal V}$, and going beyond zeroth order will be unnecessary for our purposes.
\section{Non-zero width of the window function}
\label{app:width}
In this appendix we discuss the non-zero width of $F$. In~\ref{subapp:subvariances} we discuss how the width can affect some of the calculations in this paper by focusing on a specific case: subsystem energy fluctuations for both the eigenstates and non-eigenstates. In~\ref{subapp:normForm} we discuss the form of $F$ in eq.~\eqref{eq:EB}. This has been previously discussed in refs.~\cite{murthy_structure_2019} and~\cite{shi_local_2023} which predict a Gaussian and a Lorentzian form, respectively, in tension with one another. We shed some light on this disagreement. 
\subsection{Subsystem energy fluctuations}
\label{subapp:subvariances}

In this subappendix we consider an observable which can distinguish eigenstates from non-eigenstates: subsystem energy fluctuations. As we will see, these fluctuations gain an extra contribution from the energy uncertainty of a non-eigenstate when the subsystem is a finite fraction of the whole system.

To consider the effects of finite $F$ width, we write
\begin{align}
\label{eq:GaussianApprox}
    \ln\left[F(\omega)\right] \approx  \ln\left[F(\omega_{0})\right] - \frac{\left(\omega-\omega_{0}\right)^2}{2\Delta_0^2} + \dots
\end{align}
where $\Delta_0^2$ represents the variance of $F$ and $\omega_0$ maximizes $F$. Eq.~\eqref{eq:GaussianApprox} is in essence a Gaussian approximation for $F$ but the outcomes of convolutions will be exact so long as we are only computing variances\footnote{i.e. Bienaymé's identity}.

First, we consider the eigenstate case, eq.~\eqref{eq:diageq}. The subsystem energy variance is given by, 
\begin{align}
\label{eq:subvarianceeq}
    \expval{\Delta H_A^2}_i\equiv \expval{\left(H_A-E_{iA}\right)^2}_i = \int_{E_aE_b}(E_a-E_{iA})^2e^{-S(E_i)+S_A(E_a)+S_B(E_b)}F(E_i-E_a-E_b).
\end{align}
Per appendix~\ref{app:saddlePoints}, the density of states factors should be well approximated as Gaussians about the saddle-point of eq.~\eqref{eq:subvarianceeq}. $e^{S_{A(B)}}$ obtains a variance of $C_{A(B)}T^2$, where $C_{A(B)}$ is the heat capacity of subsystem $A(B)$. Then, we can write,
\begin{align}
\label{eq:subvarianceeqgauss}
    \expval{\Delta H_A^2}_i= \frac{1}{\mathcal N} \int_{E_aE_b}(E_a-E_{iA})^2e^{-\frac{(E_a-E_{iA})^2}{2C_AT^2}-\frac{(E_b-E_{iB})^2}{2C_BT^2} - \frac{(E_i-E_a -E_b-\omega_0)^2}{2\Delta_0^2}}
\end{align}
where $\mathcal{N}$ represents an overall normalization that ensures $\bra{i}\ket{i} = 1$. The Gaussian integrals in eq.~\eqref{eq:subvarianceeqgauss} can be directly evaluated to find,
\begin{align}
\label{eq:resistorEq}
    \expval{\Delta H_A^2}_i = T^2\left(\frac{1}{C_A}+\frac{1}{C_B+\Delta_0^2/T^2}\right)^{-1}.
\end{align}

The variance of subsystem energies was considered in ref.~\cite{garrison_does_2018} which contrasted the results for eigenstates and the canonical ensemble. For a Gibbs state on subsystem $A$, the variance is 
\begin{align}
\label{eq:one}
    \expval{\Delta H_A^2}_{\text{Gibbs}} = C_AT^2. 
\end{align}
For an eigenstate, one instead expects\footnote{To connect our expression to that of ref.~\cite{garrison_does_2018} one has to assume homogeneity. I.e. that $C_{A(B)} \propto {\mathcal V}_{A(B)}$ where ${\mathcal V}_{A(B)}$ is the volume of subsystem $A(B)$. Then one recovers $\expval{\Delta H_A^2}_{\text{Eigenstate}} = C_AT^2\left(1-{{\mathcal V}_A}/{{\mathcal V}}\right)$ where ${\mathcal V}$ is the volume of the whole system.},
\begin{align}
\label{eq:oneminusf}
    \expval{\Delta H_A^2}_{\text{Eigenstate}} = T^2\left(\frac{1}{C_A}+\frac{1}{C_B}\right)^{-1}.
\end{align}
We can see that eq.~\eqref{eq:resistorEq} interpolates between the two cases. If for some reason $\Delta_0^2/T^2 \gg C_A$ eq.~\eqref{eq:resistorEq} reduces to eq.~\eqref{eq:one}. This situation may be possible in models with long-range interactions and would be interesting to explore further. Such systems may also exhibit some nontrivial scaling with temperature. If $\Delta_0^2$ is vanishing, eq.~\eqref{eq:oneminusf} is recovered. This is more physical because for an eigenstate we should expect $\Delta_0^2$ to scale as the area of the boundary between $A$ and $B$. Thus, the finite width of $F$ adds only an area scaling correction to the subsystem energy variance. Nonetheless, this correction is perceptible in the numerical calculations presented in figure 11 of ref.~\cite{garrison_does_2018}, where the graph of the subsystem energy variance shows a small enhancement with respect to eq.~\eqref{eq:oneminusf} for a single high energy eigenstate when $A$ is larger than $B$. 

Next, we consider an arbitrary initial state $\ket{\psi}$ with initial energy $E_\psi=\expval{H}_\psi$. We consider its eigenstate components,
\begin{align}
    \ket{\psi} \equiv \sum_{i}c^{\psi}_i\ket{i}
\end{align}
and define,
\begin{align}
    \overline{c^{\psi}_ic^{i}_\psi} \equiv e^{-S(E)}F_\psi(E_i-E_\psi)
\end{align}
and
\begin{align}
    \ln\left[F_\psi(\omega)\right] \approx  \ln\left[F_\psi(\omega_\psi)\right] - \frac{(\omega-\omega_\psi)^2}{2\Delta_\psi^2} + \dots
\end{align}
where $\Delta_\psi$ is the energy variance of $F_\psi$ and $\omega_\psi$ maximizes it. Generalizing eq.~\eqref{eq:integrateOut}, in equilibrium, the subsystem energy variance will be given by,
\begin{align}
\label{eq:subvarianceNeq}
    \expval{\Delta H_A^2(t\to\infty)}_\psi = \int_{E_iE_aE_b}(E_a-E_{iA})^2e^{-S(E_i)+S(E_a)+S(E_b)}F(E_i-E_a-E_b)F_\psi(E_\psi-E_i).
\end{align}
In this case we have two more variances to consider: $CT^2$ from $e^{S}$ and $\Delta_\psi^2$ from $F_\psi$. Once again evaluating the Gaussian integrals, we calculate,
\begin{align}
    \expval{\Delta H_A^2(\infty)}_\psi = T^2\left(\frac{1}{C_A}+\frac{1}{C_B+\Delta_0^2/T^2+(T^2/\Delta_\psi^2+C^{-1})^{-1}}\right)^{-1}.
\end{align}
Thus we find some difference between the subsystem energy variance of a system prepared in an eigenstate and of a system that is prepared outside of an eigenstate but allowed to equilibrate. However, such a difference only shows up for a subsystem that is a finite fraction of the whole system.
\subsection{Form of the window function}
\label{subapp:normForm}

We have not yet discussed the form of $F$ in eq.~\eqref{eq:EB}. Where in sections~\ref{subsec:entanglementVelocities} and~\ref{subsec:butterfly} we were able to relate forms of $F$ in more complicated scenarios to known locality-related bounds, for this simplest case the literature provides different answers. 

Ref.~\cite{murthy_structure_2019} conjectured that in this case $F$ obtains an exactly Gaussian form in $D>1$ dimensional systems,
\begin{align}
\label{eq:MSgauss}
    F(\omega) \propto e^{-\frac{(\omega-\omega_0)^2}{2\Delta_0^2}}    
\end{align}
where $\omega_0$ maximizes $F(\omega)$ and $\Delta_0$ is its variance. Their argument relies broadly on two claims: 
\begin{enumerate}
    \item $\int_\omega F(\omega)\omega^n \approx \expval{H_{AB}^n}_i$ up to subleading order corrections that are polynomially suppressed in the system size,
    \item The moments of $H_{AB}$ are those of a Gaussian distribution.
\end{enumerate}
Claim 2 follows from the fact that $H_{AB}$, in $D>1$ spatial dimensions, is a sum of an area scaling number of local terms in a system with a finite correlation length and thus its moments converge weakly towards those of a Gaussian distribution by central limit theorem. Claim 1 is more subtle. We take $H\equiv H_{A}+H_{B}+H_{AB}$, $H_0\equiv H_{A}+H_{B}$ and $\ket{i}$, $\ket{I}$ to label their respective eigenstates. Starting from the identity\footnote{Note that our normalization convention differs from that of ref.~\cite{murthy_structure_2019} but the difference can be absorbed into the definition of $F$.}
\begin{align}
    \bra{I}\!(H-E_I)^n\!\ket{I} = \sum_i |\!\bra{i}\ket{I}\!|^2(E_i-E_I)^n = \int_{\omega}F(\omega)\omega^n
\end{align}
ref.~\cite{murthy_structure_2019} shows that,
\begin{align}
\label{eq:MSmoments}
    \bra{I}\!(H-E_I)^n\!\ket{I} = \bra{I}\!\left(\left[H_0-E_I\right]+H_{AB}\right)^n\!\ket{I} = \bra{I}\!(H_{AB})^n\!\ket{I} + \mathcal{O}(n{\mathcal A}^{\frac{n-1}{2}})
\end{align}
where ${\mathcal A}$ denotes the area of the boundary between the subsystems. Since the moments of $H_{AB}$ are indeed Gaussian, ref.~\cite{murthy_structure_2019} concluded that $F$ must be as well with mean $\omega_0 = \expval{H_{AB}}_I$ and variance $\Delta^2_0=\Delta^2\equiv\expval{H_{AB}^2}_I-\expval{H_{AB}}_I^2$. However, as pointed out in ref.~\cite{shi_local_2023}, eq.~\eqref{eq:MSmoments} only bounds the first $p$ central moments of $F(\omega)$ for some $p\sim\mathcal{O}\left(\sqrt{{\mathcal A}}\right)$ which may not be sufficient to determine the form of $F$.

Ref.~\cite{shi_local_2023} instead argues for a Lorentzian form with an exponential cutoff at large frequencies in any number of dimensions. Their argument follows by approximating the solution to the characteristic equations for the eigenvalues of $H$ after perturbing $H_0$ by $H_{AB}$. Preparing the system in a product state $\ket{I}$, we can decompose the Hilbert space into the direct sum of $\ket{I}$ and its complement, $\mathcal{H} = \mathcal{H}_c\oplus \ket{I}$. Diagonalizing $H$ on the complement Hilbert space yields a set of $d-1$ eigenstates ${\ket{\epsilon_m}}$. Ref.~\cite{shi_local_2023} then mirrors a standard derivation of \textit{Fermi's golden rule
} (FGR)~\cite{santra_fermis_2017}. The key assumption is that the nonzero level spacing and finite bandwidth of the levels $\epsilon$ are irrevelant, that is,
\begin{align}
    \epsilon_{\text{max}} - \epsilon_{\text{min}} \gg \mathcal{O}({\mathcal A}) \gg \delta\epsilon
\end{align}
where $\epsilon_{\text{max}}$ ($\epsilon_{\text{min}}$) is the largest (smallest) value of $\epsilon$ and $\delta\epsilon$ is the typical spacing between consecutive $\epsilon$. Under these assumptions one can approximately solve a set of characteristic equations to yield,
\begin{align}
\label{eq:cauchy}
    F(E_i-E_I) \sim \frac{\Gamma^i_I}{(E_i-E_I)^2 + (\Gamma^i_I)^2}
\end{align}
where $\Gamma^i_I$ gives the width of $F$. So long as $\Gamma^i_I$ has a sufficiently slow dependence on $E_i-E_I$ eq.~\eqref{eq:cauchy} is consistent with a Lorentzian form for $F$. As a consequence, the return amplitude of a system prepared in $\ket I$ and evolved under $H$ exhibits an exponentially decaying return amplitude,
\begin{align}
    e^{iE_It}\bra{I}\! e^{-iHt}\!\ket{I} &= \sum_i |\!\bra{i}\ket{I}\!|^2e^{-i(E_i-E_I)t} \nonumber \\
    &= \int_\omega F(\omega)e^{-i\omega t} \nonumber \\
    &\equiv F(t)\sim e^{-\Gamma^i_I t}
\end{align}
which reflects the physical picture of FGR. Recent work~\cite{crowley_partial_2022, micklitz_emergence_2022, long_beyond_2023} has connected FGR and its breakdown to the \textit{many-body localization} (MBL) to ETH crossover\footnote{We are grateful to Philip Crowley for bringing this work to our attention.}. Ref.~\cite{long_beyond_2023}, in particular, argues that an exponential decay of the (modulus-squared) return amplitude is generic feature of ETH satisfying systems so long as the perturbation is not large enough to modify the entropy function. In contrast the argument of ref.~\cite{murthy_structure_2019} would suggest a Gaussian decay of the return amplitude. Various numerical calculations~\cite{shi_local_2023, crowley_partial_2022, long_beyond_2023} corroborate a regime of exponential decay in 1 dimension, however, we are unaware of any numerical work in higher dimensions where the tension arises. 

To understand the tension, it may be enlightening to study how the moment calculation eq.~\eqref{eq:MSmoments} influences the time dependence of $F(t)$\footnote{neglecting phase factors throughout}. Our analysis complements that in appendix A of ref.~\cite{shi_local_2023} by focusing on the decay of $F(t)$ where they focused on the decay of $F(\omega)$. At late times, the return amplitude saturates to the overlap of two microcanonically random vectors, 
\begin{align}
    e^{iE_It}\bra{I}\! e^{-iHt}\!\ket{I} \rightarrow e^{-S/2}.
\end{align}
The smooth $F$ functions will not capture this saturation but determine what form the return amplitude takes prior to the saturation time. First we estimate the saturation time, $t_{\text{sat}}$, under the Gaussian form implied by eq.~\eqref{eq:MSgauss},
\begin{align}
\label{eq:tsat}
    e^{-\Delta^2 t_{\text{sat}}^2/2} = e^{-S/2} \quad\implies\quad t_{\text{sat}} = \sqrt{\frac{S}{\Delta^2}}\sim\mathcal{O}\left(\sqrt{\frac{{\mathcal V}}{{\mathcal A}}}\right)
\end{align}
where ${\mathcal V}$ is the volume of the system. Next, eq.~\eqref{eq:MSmoments} determines only the first $p$ derivatives of $F(t)$ where $p\sim\mathcal{O}\left(\sqrt{{\mathcal A}}\right)$ is the first (even) central moment for which the error crosses the mean. Then we can estimate the error of the Gaussian description from its Taylor series,
\begin{align}
\label{eq:errorAnal2}
    |F(t)-e^{-\Delta^2 t^2/2}| \approx \left|\frac{(p-1)!!\Delta^{p}t^p}{p!}\right|\approx \frac{1}{\sqrt{\pi}}\left|{\frac{e\Delta^2t^2}{p}}\right|^{\frac{p}{2}}.
\end{align}
The error approximation derived in~\eqref{eq:errorAnal2} implies that $e^{-\Delta^2 t^2/2}$ is only a reliable approximation to $F(t)$ for $t\lesssim t_{\text{early}}$ defined as
\begin{align}
\label{eq:tshort2}
    &e^{-\Delta^2 t_{\text{early}}^2/2} \approx \frac{1}{\epsilon \sqrt{\pi}}\left|{\frac{e\Delta^2t_{\text{early}}^2}{p}}\right|^{\frac{p}{2}} \nonumber \\
    &\implies \frac{\Delta^2t_{\text{early}}^2}{p} \approx W\left(e^{-1}{\left(\epsilon^2\pi\right)^{\frac{1}{p}}}\right) = 0.2785\dots +\mathcal{O}\left(\frac{1}{p}\right) \nonumber \\
    &\implies t_{\text{early}} \sim\mathcal{O}({\mathcal A}^{-\frac{1}{4}})
\end{align}
where $\epsilon$ is a small $\mathcal{O}(1)$ error threshold and $W$ is the Lambert W function. Eq.~\eqref{eq:tshort2} implies that increasing the size of ${\mathcal A}$ will actually shorten the validity of the moment calculation. This is physical since we are scaling $p$ and $\Delta$ simultaneously and $F(t)$ grows sharp as ${\Delta^{-1}}\sim\mathcal{O}\left({\mathcal A}^{-\frac{1}{2}}\right)$ which is faster than the constraint in eq.~\eqref{eq:tshort2} grows tight. Thus, the Gaussian ansatz of ref.~\cite{murthy_structure_2019} can be expected to accurately capture the decay of $F(t)$ for a large number of periods $\Delta^{-1}$ yet only for a vanishing period of time in the thermodynamic limit. Furthermore, eq.~\eqref{eq:tsat} shows that $t_{\text{sat}}$ diverges in the thermodynamic limit. We conclude that the argument of ref.~\cite{murthy_structure_2019} cannot determine the form of $F(t)$, or by extension $F(\omega)$, over all physically relevant scales.

Ref.~\cite{long_beyond_2023} uses a statistical analysis of the Jacobi diagonalization algorithm applied to ETH satisfying systems to develop the following picture for the return amplitude. Initializing the system in an eigenstate of $H_0$ and perturbing it to $H$, the return amplitude is expected to experience an early time Gaussian decay, intermediate exponential decay, and a late time saturation, i.e.
\begin{align}
    e^{iH_0t}\bra{I}\! e^{-iHt}\!\ket{I} = 
    \begin{cases}
      F_{\text{early}}(t) = e^{-J^2 t^2/2}\\
      F_{\text{inter}}(t) \propto e^{-\Gamma t}\\
      e^{-S/2} \quad \text{(saturation)}
    \end{cases} .
\end{align}
For a sufficiently weak perturbation a Gaussian description only holds at very early times with $J^2\approx\Delta^2 \sim \Delta_0^2$. For a sufficiently strong volume-scaling perturbation the Gaussian decay may saturate early and entirely preempt the exponential regime with $J^2\approx \Delta_0^2 \gg\Delta^2$ and. Then, inserting $F_{\text{early}}(t)$ for $F(t)$ into eqs.~\eqref{eq:errorAnal2} and~\eqref{eq:tshort2}, we get,
\begin{align}
    &e^{\left(J^2-\Delta^2\right)t_{\text{early}}^2/2} - 1 = \epsilon \implies t_{\text{early}} = \sqrt{\frac{2\ln(1+\epsilon)}{J^2-\Delta^2}} \nonumber \\
    &\implies J^2 \approx \left(1+\frac{2\epsilon}{W\left(e^{-1}{\left(\epsilon^2\pi\right)^{\frac{1}{p}}}\right)p}\right)\Delta^2 = \Delta^2 + \mathcal{O}\left(\sqrt{{\mathcal A}}\right)
\end{align}
and our estimate of $t_{\text{early}}$ indicates that the area-scaling perturbation, $H_{AB}$, should be considered weak with $J^2\approx\Delta^2\approx \Delta_0^2$. Returning to the frequency space picture, the exponential cut-off of $F(t)$ after $t_{\text{early}}$ will constrain the form of $F(\omega)$ on scales below $t_{\text{early}}^{-1}\sim\mathcal{O}(\mathcal{A}^{1/4})$. However, the width of $F(\omega)$, $\Delta_0 \approx \Delta\sim\mathcal{O}(\mathcal{A}^{1/2})$ is parametrically larger than this scale in $D>1$ dimensions. Thus, $F(\omega)$ should appear Gaussian in $D>1$ dimensions over scales comparable to $\Delta\sim\mathcal{O}(\mathcal{A}^{1/2})$. $F(\omega)$ must also respect an $\mathcal{O}(1)$ cutoff at very large scales $\omega\sim\mathcal{O}(\mathcal{A})$~\cite{arad_connecting_2016}.

We conclude that the tension between the Gaussian and Lorentzian forms in $D>1$ dimensions boils down to relevant timescales. The $F$ function in eq.~\eqref{eq:EB} captures the return amplitude of a system prepared in an eigenstate of $H_A+H_B$ and evolved under the Hamiltonian $H_A+H_B+H_{AB}$. We expect that the return amplitude will vanish rapidly as a Gaussian at very early times, but continue to decay as an exponential for a much longer period until saturation. Future work should clarify how the various arguments in the literature capture different aspects of this picture.

\section{Operator thermalization}
\label{app:OFSR}
We wish to show in this section that the ETH, eq. (\ref{eq:ETH}), is a consequence of the MBBC. Consider a local Hermitian operator $X$ supported on a local Hilbert space of dimension $d_X$ and an associated rotation, 
\begin{align}
\label{eq:rotation}
    U_X(s)\equiv e^{isX} = \sum_{m=0}^{d_X-1}p_m(s)X^m
\end{align}
where the second equality states that $U_X$ is a finite polynomial in $X$ and follows from the Cayley--Hamilton theorem. For $\ket{i}$ an eigenstate of a Hamiltonian $H$, $\ket{i(s)}\equiv U_X(s)\ket{i}$ is an eigenstate of Hamiltonian $H(s)\equiv U_X(s)HU_X(-s)$. We can decompose $H(s)$ as
\begin{align}
    H(s) = H + (H(s) - H) \equiv H + V(s)
\end{align}
where $V(s)$ is a small $O(1)$ perturbation so long as $X$ is a local operator and $H$ and sum of local terms. Then, by eq. 
(\ref{eq:stateCumulant}) we can assert the following,
\begin{align}
    \overline{\bra{i_1(s)}\ket{i_2(2s)}\!\!\bra{i_2(2s)}\cdots \ket{i_n(ns)}\!\!\bra{i_n(ns)}\ket{i_1(s)}} = e^{-(n-1)S(\bar E)}F_X(s;\bar E; \Vec \omega)
\end{align}
where we have considered $n$ distinct eigenstates of $H$ and rotated them each by a different angle $s$. Since distinct eigenstates are orthogonal, $F_X(0;\bar E; \Vec \omega) = 0$. In fact, the first $n-1$ derivatives of $F_X$ with respect to $s$ at $s=0$ must vanish as well,
\begin{align}
\label{eq:powerSeries}
   e^{-(n-1)S(\bar E)}F_X(s;\cdots) &= \sum_{m=0}^{n(d_X-1)}\!\!\sum_{\sum_l m_l = m}\!\!q_{\{m\}}(s)\overline{(X^{m_1})_{i_1i_2}\cdots (X^{m_n})_{i_ni_1}} \nonumber \\ &= 0 \cdot s^0 + \dots + 0 \cdot s^{n-1} +  q_{\{1\cdots 1\},n} s^n\overline{X_{i_1i_2}\cdots X_{i_ni_1}} + \mathcal{O}(s^{n+1}) 
\end{align}
where $q_{\{m\}}$ collects the sum of products of $p_m$ and we have defined the expansion $q_{\{m\}}(s) \equiv \sum_{m'}q_{\{m\},m'}s^{m'}$. In general, the terms $p_m$ and $q_{\{m\}}$ will depend only on the characteristic polynomial of $X$ and are finite for any $d_X$. Thus we can extract correlations of $X$ from the following non singular limit,
\begin{align}
    \overline{X_{i_1i_2}\cdots X_{i_ni_1}} = \lim_{s\rightarrow 0}q_{\{1\cdots 1\},n}^{-1}s^{-n}e^{-(n-1)S(\bar E)}F_X(s;\cdots) \equiv e^{-(n-1)S(\bar E)}f(\bar E;\Vec \omega)
\end{align}
and we arrive at our desired result. The approach of this section suggests a sufficient criteria for operators to satisfy the ETH in the sense of eq. (\ref{eq:ETH}): For a given operator $X$,  $U_X(s)HU_X(-s)$ is close to $H$ even as $X$ mixes a large number of eigenstates. 
\section*{}
\bibliographystyle{JHEP}
\bibliography{WETH}
\end{document}